\documentclass[iop,apj]{emulateapj}
\usepackage{epsfig}
\usepackage{graphics,graphicx}
\usepackage{tikz}
\usepackage{subfigure}
\usepackage{multirow}
\usepackage{amssymb}
\usepackage{natbib}
\usepackage{rotating}
\usepackage{setspace}
\usepackage[bookmarks=false, colorlinks=true, citecolor=blue, backref=true]{hyperref}
\usepackage{apjfonts}

\shorttitle{Herschel Observations of Major Merger Pairs at z=0}
\shortauthors{C. Cao et al.}

\newcommand{\lsim}{\, \lower2truept\hbox{${< \atop\hbox{\raise4truept\hbox{$\sim$}}}$}\,}
\newcommand{\gsim}{\, \lower2truept\hbox{${> \atop\hbox{\raise4truept\hbox{$\sim$}}}$}\,}

\newcommand{\ch}{\colhead}

\newcommand{\noprint}[1]{}
\newcommand{\figsetstart}{{\bf Fig. Set} }
\newcommand{\figsetend}{}
\newcommand{\figsetgrpstart}{}
\newcommand{\figsetgrpend}{}
\newcommand{\figsetnum}[1]{{\bf #1.}}
\newcommand{\figsettitle}[1]{ {\bf #1} }
\newcommand{\figsetgrpnum}[1]{\noprint{#1}}
\newcommand{\figsetgrptitle}[1]{\noprint{#1}}
\newcommand{\figsetplot}[1]{\noprint{#1}}
\newcommand{\figsetgrpnote}[1]{\noprint{#1}}

\begin{document}

\slugcomment{{\bf Draft 2.0}; \today}

\title{Herschel Observations of Major Merger Pairs at z=0: Dust Mass and Star Formation}

\author{
Chen Cao \altaffilmark{1,2,3},
Cong Kevin Xu \altaffilmark{3},
Donovan Domingue \altaffilmark{4},
Veronique Buat \altaffilmark{5},
Yi-Wen Cheng \altaffilmark{6},
Yu Gao \altaffilmark{7},
Jiasheng Huang \altaffilmark{8},
Thomas H. Jarrett \altaffilmark{9},
Ute Lisenfeld \altaffilmark{10},
Nanyao Lu \altaffilmark{3},
Joe Mazzarella \altaffilmark{3},
Wei-Hsin Sun \altaffilmark{11},
Hong Wu \altaffilmark{12},
Min S. Yun \altaffilmark{13},
Joseph Ronca \altaffilmark{4},
Allison Jacques \altaffilmark{4}
}
\altaffiltext{1}{School of Space Science and Physics, Shandong University,
Weihai, Weihai, Shandong 264209, China; caochen@sdu.edu.cn}
\altaffiltext{2}{Shandong Provincial Key Laboratory of Optical Astronomy $\&$
Solar-Terrestrial Environment, Weihai, Shandong 264209, China}
\altaffiltext{3}{Infrared Processing and Analysis Center,
California Institute of Technology 100-22, Pasadena, CA 91125, USA; cxu@ipac.caltech.edu}
\altaffiltext{4}{Georgia College $\&$ State University, CBX 82, Milledgeville, GA 31061, USA}
\altaffiltext{5}{Laboratoire d'Astrophysique de Marseille - LAM, Universit\'e d'Aix-Marseille
\& CNRS, UMR7326, 38 rue F. Joliot-Curie, 13388 Marseille Cedex 13, France}
\altaffiltext{6}{Institute of Astronomy, National Central University, Chung-Li 32054, Taiwan}
\altaffiltext{7}{Purple Mountain Observatory, Chinese Academy of Sciences, 2 West Beijing Road, Nanjing 210008, China}
\altaffiltext{8}{Harvard-Smithsonian Center for Astrophysics, 60 Garden Street, Cambridge, MA 02138}
\altaffiltext{9}{Astronomy Department, University of Cape Town, Rondebosch 7701, South Africa}
\altaffiltext{10}{Departamento de F\'isica Te\'orica y del Cosmos, Universidad de Granada, Spain}
\altaffiltext{11}{Institute of Astrophysics, National Taiwan University and
The National Museum of Natural Science, Taiwan}
\altaffiltext{12}{National Astronomical Observatories, Chinese Academy of Sciences, Beijing, China}
\altaffiltext{13}{Department of Astronomy, University of Massachusetts, Amherst, MA 01003, USA}

\begin{abstract}
We present Herschel PACS $\&$ SPIRE far-infrared (FIR) and sub-mm
imaging observations for a large K-band selected sample of 88 close
major-merger pairs of galaxies (H-KPAIRs) in 6 photometric bands (70,
100, 160, 250, 350, and 500 $\mu m$). Among 132 spiral galaxies in the
44 spiral-spiral (S$+$S) pairs and 44 spiral-elliptical (S$+$E) pairs,
113 are detected in at least one Herschel band. Star formation rate
(SFR) and dust mass ($M_{\rm dust}$) are derived from the IR SED
fitting. Mass of total gas ($M_{\rm gas}$) is estimated by assuming a
constant dust-to-gas mass ratio of 0.01. Star forming spiral galaxies
(SFGs) in S$+$S pairs show significant enhancements in both specific
star formation rate (sSFR) and star formation efficiency (SFE), while
having nearly the same gas mass, compared to control galaxies. On the
other hand, for SFGs in S$+$E pairs, there is no significant sSFR
enhancement and the mean SFE enhancement is significantly lower
than that of SFGs in S$+$S pairs. This suggests an important role for
the disc-disc collision in the interaction induced star formation. The
$M_{\rm gas}$ of SFGs in S$+$E pairs is marginally lower than that of
their counterparts in both S$+$S pairs and the control sample.
Paired galaxies with and without interaction signs do not differ
significantly in their mean sSFR and SFE. As found in previous works, 
this much larger sample confirms the primary and secondary spirals 
in S+S pairs follow a Holmberg effect correlation on sSFR.
\end{abstract}

\keywords{galaxies: interactions --- galaxies: evolution ---
galaxies: starburst --- galaxies: general}

\vskip3truecm

\section{Introduction}

Galaxy interactions and mergers are important external mechanisms for
triggering galactic evolution \citep{Kormendy2004}. Both observations
and numerical simulations have shown that dynamical instabilities induced
by galaxy interactions can cause cold gas inflow and massive star
formation in the central region of disk galaxies (see e.g.,
\citealt{Bahcall1995, Hopkins2006, Dasyra2006}).
Merger induced star formation was first predicted by
\citet{Toomre1972}, and confirmed by \citet{Larson1978} in a study of
optical colors of Arp galaxies. Many subsequent studies of the
$H_{\rm \alpha}$ emission and Far-Infrared (FIR) emission in Arp galaxies
and in paired galaxies provided further support to this theory (see
\citealt{Kennicutt1996} for a review). On the other hand,
\citet{Haynes1988} found little or no enhanced FIR emission in a
sample of optically selected pairs compared to a control sample of
single galaxies.  In a more influential paper, \citet{Bergvall2003}
reported a multi-wavelength study in which they found no significant
star formation rate (SFR) enhancement for a sample of morphologically
selected merger candidates.
Apparently, only some merging galaxies have significantly
enhanced SFR, with ultra-luminous infrared galaxies (ULIRGs) as the
extreme examples, and the others do not.  Therefore, whether the mean
SFR of a merger sample shows significant enhancement depends very much
on how it is selected. \citet{Kennicutt1987} and \citet{Bushouse1988}
found that merger candidates which show strong signs of tidal
interactions have significantly stronger SFR enhancement than
optically selected paired galaxies, the latter being only marginally
enhanced (a factor of $\sim 2$) compared to single galaxies.
\citet{Telesco1988} found a strong tendency for pairs with the highest
FIR color temperatures to have the smallest
separation. \citet{Xu1991} showed that the enhancement of the FIR
emission of close spiral-spiral (S$+$S) pairs with separation less than
the size of the primary and with signs of interaction is significantly
stronger than that of wider pairs and pairs without interaction signs.
\citet{Sulentic1989} found that elliptical-elliptical (E+E) pairs are
equally quiet in the FIR emission as single ellipticals. Very few E's
in S$+$E pairs are FIR bright, possibly cross-fueled by their S
companions \citep{Domingue2003}.

More recently, large digitized surveys (e.g. SDSS, 2MASS, 2df, etc)
enabled large and homogeneously selected pair samples. A clear
anti-correlation between the specific SFR (sSFR=SFR/$M_{\rm star}$)
and the pair separation has been well established \citep{Barton2000,
  Lambas2003, Alonso2004, Nikolic2004, Li2008, Ellison2008,
  Scudder2012, Patton2013}. In particular, close pairs with projected
separation $r_{\rm proj} \leq 20\, h^{-1} kpc$ have relatviely strong
sSFR enhancement of a factor of $\gsim 2$ \citep{Ellison2008}, while
the sSFR enhancement of wider pairs is significantly weaker
\citep{Scudder2012, Patton2013}. On the other hand, not all star
forming galaxies (SFGs) in close pairs have enhanced
sSFR. \citet{Xu2010} studied the sSFR enhancement in a sample of
K-band selected close major-merger pairs (primary-to-secondary mass
ratio $m_{\rm pri}/m_{\rm 2nd} \leq 2.5$) using Spitzer FIR observations,
and found that (1) on average, SFGs in S$+$E pairs do not show any sSFR
enhancement compared to their counterparts in a mass-matched control
sample; (2) the sSFR enhancement in S$+$S major-merger pairs is mass
dependent in the sense that significant sSFR enhancement is confined
to massive SFGs while no enhancement is found in low mass SFGs with
nearly equal (low) mass companions. Using data obtained in Herschel
(PEP/ HerMES) and Spitzer surveys of the COSMOS field
\citep{Scoville2007}, \citet{Xu2012b} studied the cosmic evolution of
the sSFR enhancement in close major-merger pairs since z=1, and
found that the sSFR enhancement in massive S$+$S pairs decreases with
increasing redshift, while there is no significant sSFR enhancement in
massive S$+$E pairs at any redshift. Other authors also found evidence
for the lack of SFR enhancement in S$+$E pairs \citep{Hwang2011, Moon2015}
or low SFR activity in spiral galaxies with early type close neighbors
\citep{Park2005, Park2008, Park2009}. The literature results on the
sSFR enhancement in close minor-merger pairs are controversial.
\citet{Woods2007} found that in close minor mergers, only secondary
companions are sSFR enhanced while there is no significant enhancement in
primary companions. This is opposite to the results of \citet{Lambas2003}
that show stronger sSFR enhancement in the primaries. More recently,
\citet{Scudder2012} found that both the primaries and the secondaries
have significantly enhanced sSFRs in close minor-merger pairs.
\citet{Lanz2013} studied SEDs for 31 interacting galaxies in 14 major
$\&$ minor merger systems, and found increases in SFR but not sSFR
as the interaction sequences progress from non-interacting to strongly
interacting.

In this paper, we present new FIR imaging observations in the 6
photometry bands (70, 100, 160, 250, 350, and 500 $\mu m$) of Herschel
PACS \citep{Poglitsch2010} and SPIRE \citep{Griffin2010}, for a large
and complete sample of close major-merger pairs of SFGs (including
both S$+$S and S$+$E pairs). The focus of this new study is on the
dependence of the SFR enhancement on the dust mass, which is a good
proxy of gas mass. Star formation activity in galaxies is fueled by
cold gas which dominates the gas mass, and all SFGs are gas rich.
Does the SFR enhancement depend on the gas content of a paired SFG?
Both in simulations \citep{Hopkins2009a, Perret2014, Scudder2015} and
in observations \citep{Xu2012b} there have been indications that the
SFR enhancement may decrease with increasing gas fraction. According
to \citet{Hopkins2009a} (see also \citet{Mihos1996}), this is because
the gravitational torque imposed by the stellar disk to the gas disk
is less effective when gas fraction is high, therefore less disk gas
can sink to the nuclear region to fuel the merger-induced starburst by
losing angular momentum to stars. Maps of the dust emission in the six
Herschel bands enabled us to estimate accurately the dust mass (and its
distribution) for individual SFGs. All galaxies in our Herschel sample,
a subset of the KPAIR sample of K-band selected major-merger pairs
\citep{Domingue2009, Xu2012a}, have $M_{\rm star} > 10^{10}\; M_{\sun}$ and
normal metallicity. Therefore their gas mass is likely to be related
to dust mass with a rather constant
gas-to-dust ratio of $\sim 100$ \citep{DL07}, and the sSFR vs.
$M_{\rm gas}$ relation for these paired SFGs can be explored using the
Herschel observations. It is worth noting that it is difficult to
measure directly the gas mass in individual galaxies in close
major-merger pairs. The angular resolutions of single dish HI 21~cm
line observations are too coarse ($\rm beam \gsim 5'$) to resolve
pairs into individual galaxies, while the interferometry observations
using the VLA are very expensive in terms of the integration time.

The rest of the paper is arranged as following: Section 2 describes
the selection of the H-KPAIR sample, details of Herschel observation
$\&$ photometry are given in Section 3. Section 4 shows the images $\&$
catalogs of the H-KPAIR galaxies, and describes the estimation of dust
mass and star formation rates using SED fittings, while we describe
the selection, data reduction $\&$ photometry on control sample galaxies
in Section 5. Sections 6 $\&$ 7 show statistical comparison results on
sSFRs, total gas masses, and star formation efficiencies (SFEs) in
H-KPAIRs and the control sample. Discussions will be given in Section
8. We summarize our results and briefly introduce future plans in Section 9.

Throughout this paper, we adopt the $\Lambda$-cosmology with
$\Omega_{\rm m}=0.3$ and $\Omega_{\rm \Lambda} = 0.7$, and $H_{\rm 0}= 70\;
(km~sec^{-1} Mpc^{-1})$.

\section{The Herschel KPAIR (H-KPAIR) Sample}
The local galaxy pairs sample used in this work (hereafter Herschel
KPAIR sample, or H-KPAIR) was constructed from the KPAIR which is a
complete and unbiased K-band selected sample of 170 close major-merger
galaxy pairs (see details in \citealt{Domingue2009},
\citealt{Xu2010}). Pairs in the KPAIR with following characteristics
are excluded from H-KPAIR: (1) Elliptical+Elliptical (E+E) pairs; (2)
pairs with only one measured redshift; (3) pairs with recession
velocities $<$ 2000km/s.  The resulting sample contains 88 galaxy
pairs (176 paired galaxies), of which 44 are Spiral+Spiral (S$+$S) pairs
and 44 are Spiral+Elliptical (S$+$E) pairs. Interaction morphology of
these pairs was visually checked by three of us (C.C., C.K.X. and D.D).
Accordingly, pairs are classified into three types: (1) INT
(with signs of interacting, e.g., morphological distortions, tidal
tails and bridges, plumes etc.), (2) MER (merging systems) and (1) JUS
('just' pairs without clear signs of interaction).  All galaxies have
$\rm z < 0.1$ with a median of z$=$0.04.

\section{Herschel Observations}

\subsection{PACS $\&$ SPIRE images}
Pairs in H-KPAIR were observed successfully using PACS and
SPIRE photometers in the three PACS bands (70, 100, 160$\mu$m) and
three SPIRE bands (250, 350, 500$\mu$m).  Observations of 83 pairs
were carried within our own proposal (OT2$\_$cxu$\_$2).  All but two
of these pairs are smaller than 2', while the sizes of two large pairs
(J1406+5043 and J2047+0019) are between 3' to 5'.

PACS observations were done in the scanmap mode with medium scan speed
(20"/s).  Each pair was observed with four concatenated PACS
observations.  The first two are for the 70 and 160 $\mu$m bands with
orientation angles of $+$45deg (nominal) and $-$45deg (orthogonal),
respectively, and the other two for the 100 and 160 $\mu$m bands of
the same cross-scan configuration. For small pairs ($<$2'), each scan
map has 6 scans with separation of 30" and scan length $=$ 3. This
yields nearly uniform coverage of a 3' $\times$ 3' region, leaving
enough margin outside the pairs for background determination. The
average 4-$\sigma$ sensitivity limits are 36, 42, and 56 mJy for the
three PACS bands. For large pairs we used scan number $=$ 10, separation
$=$ 36'' and scan length $=$ 6', yielding a uniform coverage of a
6'$\times$ 6' region and the average 4-$\sigma$ sensitivity limits of
30, 36, and 48 mJy. Data reduction was carried out using Herschel
Interactive Processing Environment (HIPE; \citealt{Ott2010}), with the
mapmaking done using UNIMAP \citep{Traficante2011}.  The mean beam
size (full width at half maximum: FWHM) are 5.7", 6.8", 12.0" for 70,
100, 1600$\mu$m images, respectively.

SPIRE observations of small pairs were done in the small-map mode,
providing uniformly covered cross-scan maps of 4' $\times$ 4' in 3
SPIRE bands. For large pairs, the observations were done in the
large-map mode with nominal scan speed (30"/s), both scan-length and
scan-height equal to 6', and in A$\&$B directions. For both the small
and large maps, each observation has 4 repeats, yielding confusion
limite 4-$\sigma$ sensitivities of 29, 30, and 34 mJy in the 3 SPIRE
bands, respectively. The SPIRE data were reduced using HIPE 10.0.0
which uses the de-striper for the mapmaking.  Bad pixels (PSWF8,
PSWE9, PSWB5, PSWD11) were masked. The mean beam FWHM are 18.2",
24.9", 36.3" for 250, 350, 500$\mu$m images, respectively.

The remaining 5 pairs in H-KPAIR were observed by other OT and KPGT
projects, and their Herschel data were taken from the archives.  SPIRE
images of two pairs (J1101+5720 and J1429+3534) were taken directly
from HerMES data release v2. In Table~1 we present the pair $\&$
galaxy ID, RA $\&$ Decl, redshift (z), Ks band magnitude (from 2MASS),
morphological types (spiral or elliptical), interaction morphological
types (as described in Sect.2), and Herschel proposal ID for 176
paired galaxies in the H-KPAIR sample. SDSS, PACS, $\&$ SPIRE images
of all pairs are shown in Figure~\ref{fig1:colorimages}.

\figsetstart
\figsetnum{1}
\figsettitle{SDSS, Herschel PACS, $\&$ SPIRE three-color images of H-KPAIR galaxy pairs}

\figsetgrpstart
\figsetgrpnum{1.1}
\figsetgrptitle{Images of H-KPAIRs J0020+0049, J0118-0013, J0211-0039, J0338+0110}
\figsetplot{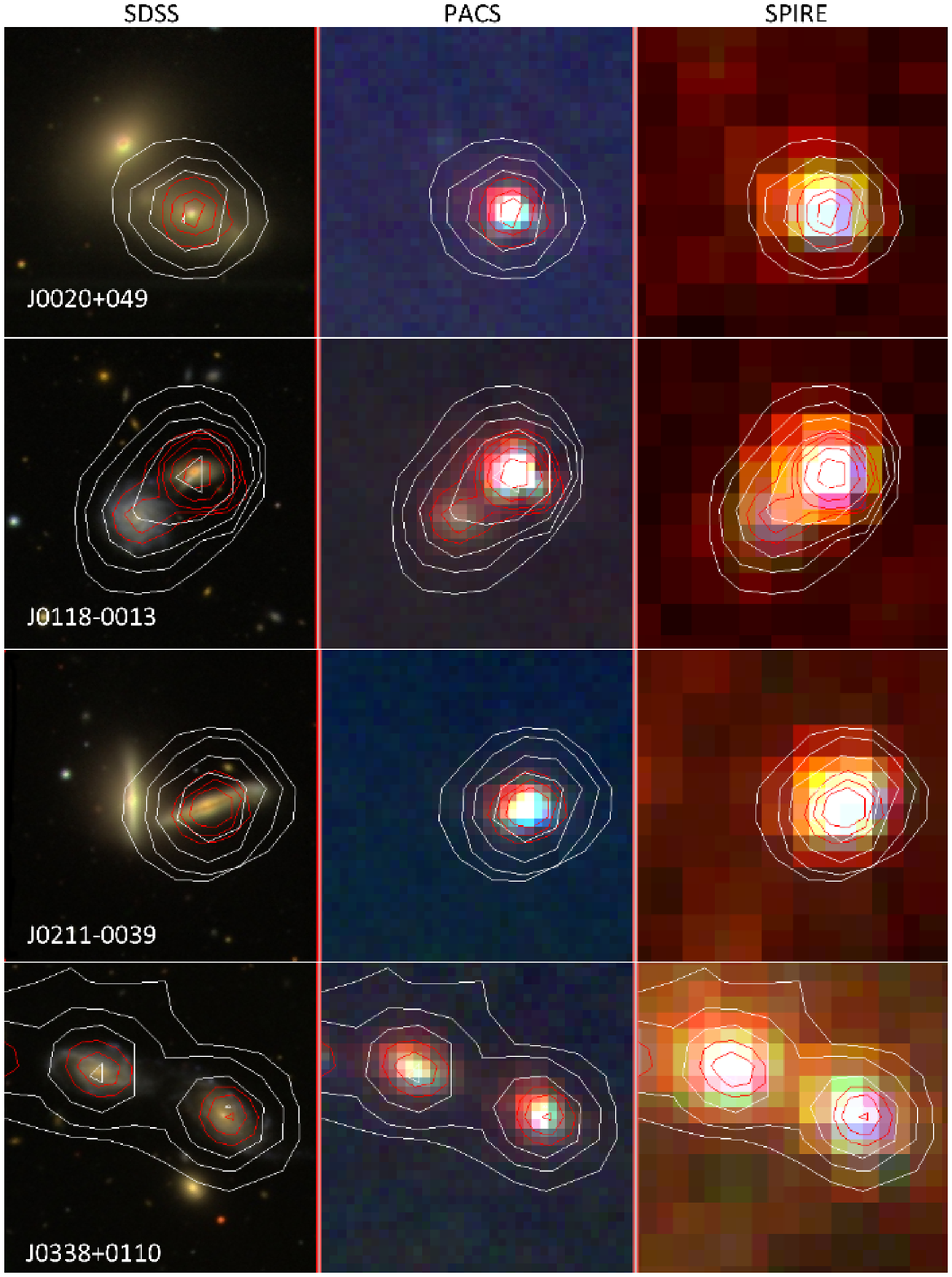}
\figsetgrpnote{SDSS, Herschel PACS, $\&$ SPIRE three-color images of H-KPAIR galaxy pairs. All images are overlaid by the same two sets of contours of 70$\mu$m (red) and 250$\mu$m (white) images.}
\figsetgrpend

\figsetend

\begin{figure}[!htbp]
\includegraphics[width=0.45\textwidth,angle=0]{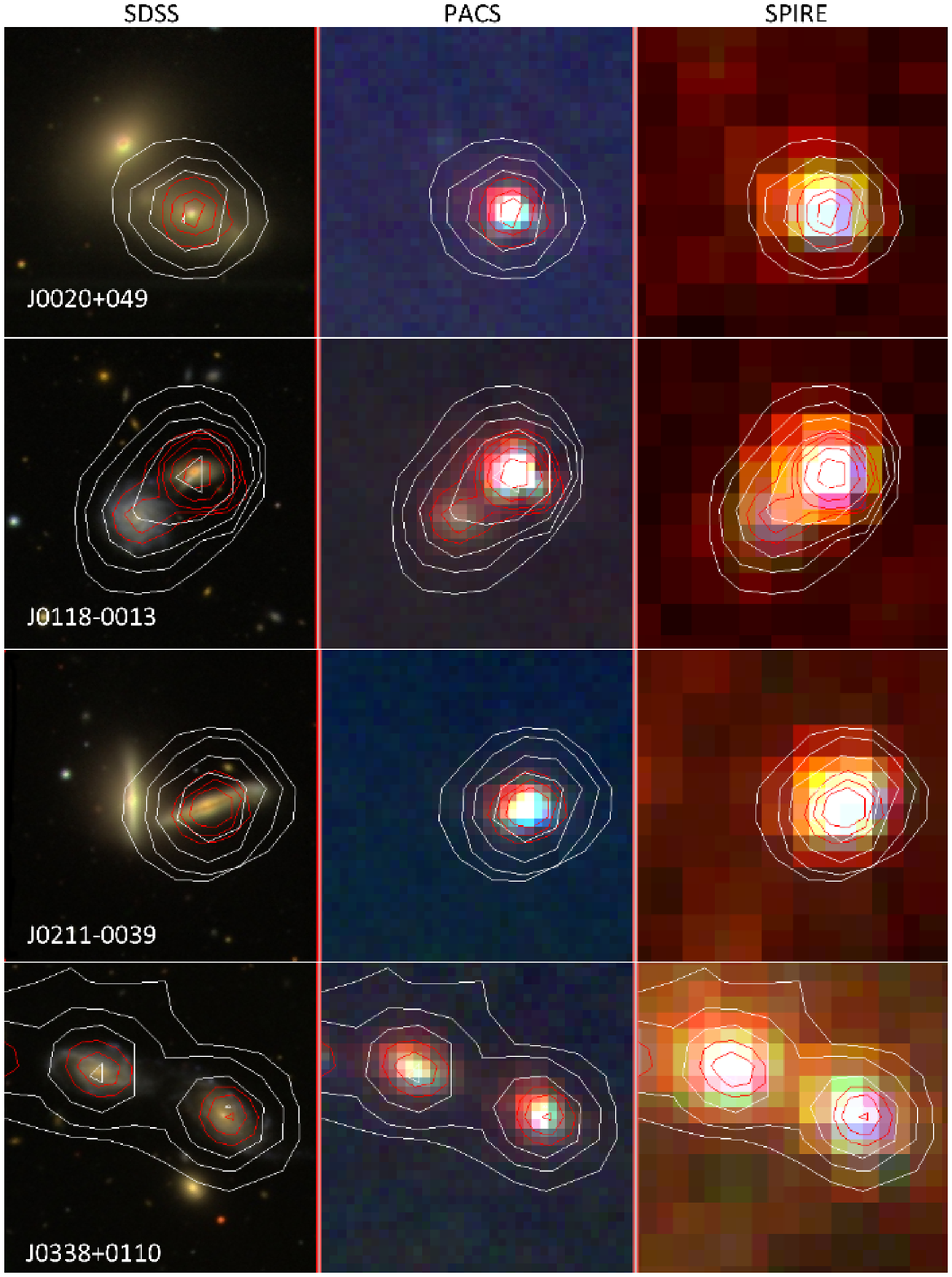}
\caption{SDSS, Herschel PACS, $\&$ SPIRE three-color images of H-KPAIR galaxy pairs. All images
are overlaid by the same two sets of contours of 70$\mu$m (red) and 250$\mu$m (white) images.}
\label{fig1:colorimages}
\setcounter{figure}{1}
\end{figure}

\begin{deluxetable*}{ccrrccccccc}
\label{tbl:hkpair}
\tabletypesize{\scriptsize}
\setlength{\tabcolsep}{0.05in} 
\tablenum{1}
\tablewidth{0.95\linewidth}
\tablecaption{H-KPAIR Galaxy Sample}
\tablehead{
\ch{(1)}   &\ch{(2)}    &\ch{(3)}    &\ch{(4)}&\ch{(5)}&\ch{(6)} &\ch{(7)}
             &\ch{(8)}         &\ch{(9)}    &\ch{(10)}   &\ch{(11)}     \\
\ch{Pair ID} & \ch{Galaxy ID} & \ch{RA}  & \ch{Dec} &
\ch{z} & \ch{K$_s$} & \ch{Type} & \ch{Int-type} & \ch{Herschel-ID} & \ch{s(p)} & \ch{$\delta$(Vz)} \\
\ch{(KPAIR)}  & \ch{(2MASX)} & \ch{(J2000)} &
\ch{(J2000)} & \ch{redshift} & \ch{(mag)} & & &  & \ch{(kpc)} & \ch{(km s$^{-1}$)}
}
\startdata
J0020$+$0049 &      J00202580$+$0049350 &      5.107446   &      0.826346    &    0.014864 &     10.98 &     S     &     JUS   &      OT2$\_$cxu$\_$2   &      11.44  &      402.0   \\
              &      J00202748$+$0050009 &      5.114637   &      0.833554    &    0.016204 &     10.5  &     E     &     JUS   &                  &     &      \\
  J0118$-$0013 &      J01183417$-$0013416 &      19.642259  &      $-$0.228261 &    0.045344 &     12.05 &     S     &     JUS   &      OT2$\_$cxu$\_$2   &      27.23  &      54.0    \\
              &      J01183556$-$0013594 &      19.648352  &      $-$0.233075 &    0.045524 &     12.88 &     S     &     JUS   &                  &     &      \\
  J0211$-$0039 &      J02110638$-$0039191 &      32.776675  &      $-$0.655108 &    0.017746 &     11.42 &     S     &     JUS   &      OT2$\_$cxu$\_$2   &      10.63  &      96.0    \\
              &      J02110832$-$0039171 &      32.784577  &      $-$0.654777 &    0.018066 &     10.9  &     S     &     JUS   &                  &     &      \\
  J0338$+$0109 &      J03381222$+$0110088 &      54.550919  &      1.16912     &    0.039201 &     12.42 &     S     &     INT   &      OT2$\_$cxu$\_$2   &      24.9   &      432.9   \\
              &      J03381299$+$0109414 &      54.554136  &      1.161523    &    0.040644 &     12.77 &     E     &     INT   &                  &     &      \\
  J0754$+$1648 &      J07543194$+$1648214 &      118.633109 &      16.805959   &    0.045869 &     12.03 &     S     &     MER   &      OT2$\_$cxu$\_$2   &      13.89  &      104.1   \\
              &      J07543221$+$1648349 &      118.634232 &      16.809722   &    0.046216 &     11.55 &     S     &     MER   &                  &     &      \\
  J0808$+$3854 &      J08083377$+$3854534 &      122.14073  &      38.914841   &    0.040173 &     12.37 &     S     &     JUS   &      OT2$\_$cxu$\_$2   &      18.71  &      23.7    \\
              &      J08083563$+$3854522 &      122.148487 &      38.914504   &    0.040094 &     11.88 &     E     &     JUS   &                  &     &      \\
  J0823$+$2120 &      J08233266$+$2120171 &      125.886114 &      21.338096   &    0.018076 &     12.15 &     S     &     JUS   &      OT2$\_$cxu$\_$2   &      15.44  &      0.3     \\
              &      J08233421$+$2120515 &      125.892554 &      21.34764    &    0.018075 &     11.52 &     S     &     JUS   &                  &     &      \\
  J0829$+$5531 &      J08291491$+$5531227 &      127.312147 &      55.522992   &    0.025059 &     11.73 &     S     &     JUS   &      OT2$\_$cxu$\_$2   &      27.75  &      55.2    \\
              &      J08292083$+$5531081 &      127.336797 &      55.518931   &    0.025243 &     11.42 &     S     &     JUS   &                  &     &      \\
  J0836$+$4722 &      J08364482$+$4722188 &      129.186777 &      47.371907   &    0.05256  &     12.28 &     S     &     JUS   &      OT2$\_$cxu$\_$2   &      15.8   &      0.0     \\
              &      J08364588$+$4722100 &      129.191208 &      47.369469   &    0.05256  &     11.9  &     S     &     JUS   &                  &     &      \\
  J0838$+$3054 &      J08381759$+$3054534 &      129.573312 &      30.914861   &    0.047559 &     12.65 &     S     &     INT   &      OT2$\_$cxu$\_$2   &      9.09   &      152.7   \\
              &      J08381795$+$3055011 &      129.5748   &      30.916973   &    0.048068 &     11.94 &     S     &     INT   &                  &     &      \\
  J0839$+$3613 &      J08385973$+$3613164 &      129.748888 &      36.221234   &    0.055618 &     12.04 &     E     &     JUS   &      OT2$\_$cxu$\_$2   &      26.58  &      231.6   \\
              &      J08390125$+$3613042 &      129.755226 &      36.217846   &    0.054846 &     12.55 &     S     &     JUS   &                  &     &      \\
  J0841$+$2642 &      J08414959$+$2642578 &      130.456662 &      26.716073   &    0.084834 &     12.27 &     S     &     JUS   &      OT2$\_$cxu$\_$2   &      30.64  &      300.0   \\
              &      J08415054$+$2642475 &      130.460609 &      26.713209   &    0.085834 &     12.99 &     E     &     JUS   &                  &     &      \\
  J0906$+$5144 &      J09060283$+$5144411 &      136.511812 &      51.744702   &    0.02912  &     11.68 &     E     &     JUS   &      OT2$\_$cxu$\_$2   &      24.29  &      3.3     \\
              &      J09060498$+$5144071 &      136.520782 &      51.735305   &    0.029131 &     11.95 &     S     &     JUS   &                  &     &      \\
  J0912$+$3547 &      J09123636$+$3547180 &      138.151523 &      35.788345   &    0.023532 &     11.53 &     E     &     JUS   &      OT2$\_$cxu$\_$2   &      14.24  &      6.0     \\
              &      J09123676$+$3547462 &      138.153186 &      35.796182   &    0.023512 &     12.34 &     S     &     JUS   &                  &     &      \\
  J0913$+$4742 &      J09134461$+$4742165 &      138.435884 &      47.704599   &    0.051226 &     11.91 &     E     &     INT   &      OT2$\_$cxu$\_$2   &      24.32  &      450.9   \\
              &      J09134606$+$4742001 &      138.441955 &      47.700051   &    0.052729 &     12.1  &     S     &     INT   &                  &     &      \\
  J0915$+$4419 &      J09155467$+$4419510 &      138.977827 &      44.33086    &    0.039568 &     12.23 &     S     &     MER   &      OT1$\_$dsanders$\_$1  &      9.69   &      2.4     \\
              &      J09155552$+$4419580 &      138.981364 &      44.332785   &    0.039576 &     11.31 &     S     &     MER   &                  &     &      \\
  J0926$+$0447 &      J09264111$+$0447247 &      141.671319 &      4.790211    &    0.089132 &     13.16 &     S     &     MER   &      OT2$\_$cxu$\_$2   &      8.13   &      483.6   \\
              &      J09264137$+$0447260 &      141.672412 &      4.790559    &    0.090744 &     12.48 &     S     &     MER   &                  &     &      \\
  J0937$+$0245 &      J09374413$+$0245394 &      144.433951 &      2.760819    &    0.024156 &     10.01 &     S     &     INT   &      OT2$\_$cxu$\_$2   &      25.64  &      197.7   \\
              &      J09374506$+$0244504 &      144.437634 &      2.74737     &    0.023497 &     10.45 &     E     &     INT   &                  &     &      \\
  J1010$+$5440 &      J10100079$+$5440198 &      152.503332 &      54.672181   &    0.046004 &     12.2  &     S     &     MER   &      OT2$\_$cxu$\_$2   &      13.92  &      83.7    \\
              &      J10100212$+$5440279 &      152.508859 &      54.674434   &    0.046283 &     12.34 &     S     &     MER   &                  &     &      \\
  J1015$+$0657 &      J10155257$+$0657330 &      153.969058 &      6.959172    &    0.029931 &     12.28 &     S     &     JUS   &      OT2$\_$cxu$\_$2   &      13.02  &      244.5   \\
              &      J10155338$+$0657495 &      153.972451 &      6.963758    &    0.029116 &     11.3  &     E     &     JUS   &                  &     &      \\
  J1020$+$4831 &      J10205188$+$4831096 &      155.216286 &      48.519406   &    0.052968 &     13.26 &     S     &     INT   &      OT2$\_$cxu$\_$2   &      25.94  &      45.0    \\
              &      J10205369$+$4831246 &      155.223658 &      48.523383   &    0.053118 &     12.27 &     E     &     INT   &                  &     &      \\
  J1022$+$3446 &      J10225647$+$3446564 &      155.73532  &      34.78235    &    0.05537  &     13.18 &     S     &     INT   &      OT2$\_$cxu$\_$2   &      11.56  &      301.5   \\
              &      J10225655$+$3446468 &      155.735626 &      34.779681   &    0.056375 &     12.39 &     S     &     INT   &                  &     &      \\
  J1023$+$4220 &      J10233658$+$4220477 &      155.902453 &      42.346583   &    0.045621 &     12.36 &     S     &     INT   &      OT2$\_$cxu$\_$2   &      15.98  &      53.1    \\
              &      J10233684$+$4221037 &      155.903524 &      42.351041   &    0.045444 &     13.01 &     S     &     INT   &                  &     &      \\
  J1027$+$0114 &      J10272950$+$0114490 &      156.872904 &      1.246742    &    0.023592 &     11.79 &     S     &     INT   &      OT2$\_$cxu$\_$2   &      14.21  &      64.8    \\
              &      J10272970$+$0115170 &      156.873782 &      1.254605    &    0.023376 &     10.9  &     E     &     INT   &                  &     &      \\
  J1032$+$5306 &      J10325316$+$5306536 &      158.22153  &      53.114916   &    0.06403  &     12.62 &     S     &     INT   &      OT2$\_$cxu$\_$2   &      8.32   &      44.7    \\
              &      J10325321$+$5306477 &      158.221747 &      53.113263   &    0.063881 &     12.15 &     E     &     INT   &                  &     &      \\
  J1033$+$4404 &      J10332972$+$4404342 &      158.373843 &      44.076176   &    0.052303 &     11.91 &     S     &     JUS   &      OT2$\_$cxu$\_$2   &      27.42  &      64.2    \\
              &      J10333162$+$4404212 &      158.381778 &      44.072572   &    0.052089 &     12.37 &     S     &     JUS   &                  &     &      \\
  J1036$+$5447 &      J10364274$+$5447356 &      159.178124 &      54.79323    &    0.045841 &     12.47 &     S     &     INT   &      OT2$\_$cxu$\_$2   &      16.9   &      17.1    \\
              &      J10364400$+$5447489 &      159.183338 &      54.796929   &    0.045784 &     11.63 &     E     &     INT   &                  &     &      \\
  J1039$+$3904 &      J10392338$+$3904501 &      159.847442 &      39.080611   &    0.043464 &     12.49 &     S     &     JUS   &      OT2$\_$cxu$\_$2   &      20.35  &      43.2    \\
              &      J10392515$+$3904573 &      159.854824 &      39.082589   &    0.04332  &     12.29 &     E     &     JUS   &                  &     &      \\
  J1043$+$0645 &      J10435053$+$0645466 &      160.960666 &      6.76296     &    0.028694 &     11.96 &     S     &     INT   &      OT2$\_$cxu$\_$2   &      23.23  &      173.7   \\
              &      J10435268$+$0645256 &      160.969501 &      6.75702     &    0.028115 &     12.2  &     S     &     INT   &                  &     &      \\
  J1045$+$3910 &      J10452478$+$3910298 &      161.353285 &      39.174955   &    0.026834 &     11.63 &     S     &     JUS   &      OT2$\_$cxu$\_$2   &      22.72  &      340.5   \\
              &      J10452496$+$3909499 &      161.354008 &      39.163873   &    0.025699 &     11.41 &     E     &     JUS   &                  &     &      \\
  J1051$+$5101 &      J10514368$+$5101195 &      162.93181  &      51.022155   &    0.025027 &     10.27 &     E     &     JUS   &      OT2$\_$cxu$\_$2   &      6.99   &      366.0   \\
              &      J10514450$+$5101303 &      162.935331 &      51.025075   &    0.023807 &     10.97 &     S     &     JUS   &                  &     &      \\
  J1059$+$0857 &      J10595869$+$0857215 &      164.994547 &      8.955983    &    0.063163 &     12.01 &     E     &     JUS   &      OT2$\_$cxu$\_$2   &      21.62  &      140.1   \\
              &      J10595915$+$0857357 &      164.996465 &      8.95992     &    0.062696 &     12.95 &     S     &     JUS   &                  &     &      \\
  J1101$+$5720 &      J11014357$+$5720058 &      165.431565 &      57.334969   &    0.046903 &     12.46 &     E     &     JUS   &      KPGT$\_$soliver$\_$1  &      28.03  &      276.0   \\
              &      J11014364$+$5720336 &      165.431837 &      57.342686   &    0.047823 &     13.17 &     S     &     JUS   &                  &     &      \\
  J1106$+$4751 &      J11064944$+$4751119 &      166.706021 &      47.853312   &    0.064324 &     12.66 &     S     &     INT   &      OT2$\_$cxu$\_$2   &      17.92  &      334.8   \\
              &      J11065068$+$4751090 &      166.711183 &      47.85252    &    0.06544  &     12.44 &     S     &     INT   &                  &     &      \\
  J1120$+$0028 &      J11204657$+$0028142 &      170.194061 &      0.470612    &    0.025534 &     11.78 &     S     &     JUS   &      OT2$\_$cxu$\_$2   &      12.33  &      25.2    \\
              &      J11204801$+$0028068 &      170.200062 &      0.468579    &    0.025618 &     11.0  &     S     &     JUS   &                  &     &      \\
  J1125$+$0226 &      J11251704$+$0227007 &      171.321025 &      2.450198    &    0.050379 &     12.75 &     S     &     INT   &      OT2$\_$cxu$\_$2   &      12.99  &      84.0    \\
              &      J11251716$+$0226488 &      171.321518 &      2.446913    &    0.050659 &     12.14 &     S     &     INT   &                  &     &      \\
  J1127$+$3604 &      J11273289$+$3604168 &      171.887066 &      36.071335   &    0.035053 &     11.67 &     S     &     JUS   &      OT2$\_$cxu$\_$2   &      27.43  &      25.2    \\
              &      J11273467$+$3603470 &      171.894471 &      36.063083   &    0.035137 &     10.82 &     S     &     JUS   &                  &     &      \\
  J1137$+$4728 &      J11375476$+$4727588 &      174.478176 &      47.466349   &    0.034334 &     11.28 &     E     &     JUS   &      OT2$\_$cxu$\_$2   &      26.63  &      137.7   \\
              &      J11375801$+$4728143 &      174.491711 &      47.470652   &    0.033875 &     11.86 &     S     &     JUS   &                  &     &      \\
\enddata
\end{deluxetable*}

\begin{deluxetable*}{ccrrccccccc}
\label{tbl:hkpair}
\tabletypesize{\scriptsize}
\setlength{\tabcolsep}{0.05in} 
\tablenum{1}
\tablewidth{0.95\linewidth}
\tablecaption{(Continued)}
\tablehead{
\ch{(1)}   &\ch{(2)}    &\ch{(3)}    &\ch{(4)}&\ch{(5)}&\ch{(6)} &\ch{(7)}
             &\ch{(8)}         &\ch{(9)}    &\ch{(10)}   &\ch{(11)}     \\
\ch{Pair ID} & \ch{Galaxy ID} & \ch{RA}  & \ch{Dec} &
\ch{z} & \ch{K$_s$} & \ch{Type} & \ch{Int-type} & \ch{Herschel-ID} & \ch{s(p)} & \ch{$\delta$(Vz)} \\
\ch{(KPAIR)}  & \ch{(2MASX)} & \ch{(J2000)} &
\ch{(J2000)} & \ch{redshift} & \ch{(mag)} & & &  & \ch{(kpc)} & \ch{(km s$^{-1}$)}
}
\startdata
  J1144$+$3332 &      J11440335$+$3332062 &      176.013986 &      33.535072   &    0.031799 &     11.74 &     E     &     INT   &      OT2$\_$cxu$\_$2   &      20.48  &      100.5   \\
              &      J11440433$+$3332339 &      176.018049 &      33.54277    &    0.031464 &     12.66 &     S     &     INT   &                  &     &      \\
  J1148$+$3547 &      J11484370$+$3547002 &      177.182095 &      35.783414   &    0.064099 &     12.81 &     S     &     INT   &      OT2$\_$cxu$\_$2   &      29.11  &      144.0   \\
              &      J11484525$+$3547092 &      177.188549 &      35.785896   &    0.063619 &     11.94 &     S     &     INT   &                  &     &      \\
  J1150$+$3746 &      J11501333$+$3746107 &      177.555564 &      37.769659   &    0.055042 &     12.7  &     S     &     INT   &      OT2$\_$cxu$\_$2   &      25.36  &      149.4   \\
              &      J11501399$+$3746306 &      177.558293 &      37.775167   &    0.05554  &     12.44 &     S     &     INT   &                  &     &      \\
  J1150$+$1444 &      J11505764$+$1444200 &      177.740186 &      14.738912   &    0.057202 &     12.69 &     S     &     INT   &      OT2$\_$cxu$\_$2   &      17.17  &      300.0   \\
              &      J11505844$+$1444124 &      177.743517 &      14.736805   &    0.056202 &     11.73 &     E     &     INT   &                  &     &      \\
  J1154$+$4932 &      J11542299$+$4932509 &      178.595801 &      49.547493   &    0.070154 &     13.0  &     S     &     INT   &      OT2$\_$cxu$\_$2   &      8.23   &      308.4   \\
              &      J11542307$+$4932456 &      178.596149 &      49.546019   &    0.071182 &     12.02 &     E     &     INT   &                  &     &      \\
  J1202$+$5342 &      J12020424$+$5342317 &      180.517927 &      53.708769   &    0.064651 &     12.97 &     S     &     INT   &      OT2$\_$cxu$\_$2   &      27.14  &      210.3   \\
              &      J12020537$+$5342487 &      180.522233 &      53.713478   &    0.06395  &     12.43 &     E     &     INT   &                  &     &      \\
  J1205$+$0135 &      J12054066$+$0135365 &      181.419422 &      1.59349     &    0.022002 &     12.1  &     S     &     JUS   &      OT2$\_$cxu$\_$2   &      30.83  &      332.4   \\
              &      J12054073$+$0134302 &      181.419711 &      1.575058    &    0.020894 &     11.21 &     E     &     JUS   &                  &     &      \\
  J1211$+$4039 &      J12115507$+$4039182 &      182.97949  &      40.655067   &    0.022854 &     11.82 &     S     &     INT   &      OT2$\_$cxu$\_$2   &      7.73   &      195.3   \\
              &      J12115648$+$4039184 &      182.98535  &      40.655124   &    0.023505 &     11.98 &     S     &     INT   &                  &     &      \\
  J1219$+$1201 &      J12191719$+$1200582 &      184.821646 &      12.01619    &    0.027303 &     11.83 &     E     &     INT   &      OT2$\_$cxu$\_$2   &      13.18  &      160.8   \\
              &      J12191866$+$1201054 &      184.827781 &      12.018189   &    0.026767 &     12.8  &     S     &     INT   &                  &     &      \\
  J1243$+$4405 &      J12433887$+$4405399 &      190.911978 &      44.094436   &    0.041791 &     12.09 &     S     &     JUS   &      OT2$\_$cxu$\_$2   &      22.6   &      249.9   \\
              &      J12433936$+$4406046 &      190.914009 &      44.101295   &    0.040958 &     11.94 &     E     &     JUS   &                  &     &      \\
  J1252$+$4645 &      J12525011$+$4645272 &      193.208797 &      46.757577   &    0.061331 &     12.42 &     S     &     JUS   &      OT2$\_$cxu$\_$2   &      27.67  &      104.7   \\
              &      J12525212$+$4645294 &      193.21717  &      46.758185   &    0.060982 &     12.2  &     E     &     JUS   &                  &     &      \\
  J1301$+$4803 &      J13011662$+$4803366 &      195.319279 &      48.06018    &    0.030278 &     12.2  &     S     &     INT   &      OT2$\_$cxu$\_$2   &      11.84  &      129.6   \\
              &      J13011835$+$4803304 &      195.326491 &      48.058471   &    0.029846 &     12.67 &     S     &     INT   &                  &     &      \\
  J1308$+$0422 &      J13082737$+$0422125 &      197.114112 &      4.370179    &    0.025476 &     13.39 &     S     &     JUS   &      OT2$\_$cxu$\_$2   &      18.52  &      66.0    \\
              &      J13082964$+$0422045 &      197.123418 &      4.367971    &    0.025696 &     12.44 &     S     &     JUS   &                  &     &      \\
  J1313$+$3910 &      J13131429$+$3910360 &      198.309546 &      39.176686   &    0.071586 &     12.24 &     E     &     INT   &      OT2$\_$cxu$\_$2   &      8.29   &      0.0     \\
              &      J13131470$+$3910382 &      198.311265 &      39.177305   &    0.071586 &     13.1  &     S     &     INT   &                  &     &      \\
  J1315$+$4424 &      J13151386$+$4424264 &      198.807791 &      44.407356   &    0.03586  &     12.01 &     S     &     INT   &      OT1$\_$rmushotz$\_$1  &      27.84  &      35.1    \\
              &      J13151726$+$4424255 &      198.821937 &      44.407105   &    0.035743 &     11.23 &     S     &     INT   &                  &     &      \\
  J1315$+$6207 &      J13153076$+$6207447 &      198.878026 &      62.129135   &    0.030566 &     11.99 &     S     &     INT   &      OT1$\_$dsanders$\_$1  &      22.39  &      6.0     \\
              &      J13153506$+$6207287 &      198.896085 &      62.124613   &    0.030586 &     11.54 &     S     &     INT   &                  &     &      \\
  J1332$-$0301 &      J13325525$-$0301347 &      203.230301 &      $-$3.026349 &    0.049306 &     12.95 &     S     &     INT   &      OT2$\_$cxu$\_$2   &      21.45  &      297.0   \\
              &      J13325655$-$0301395 &      203.235759 &      $-$3.027679 &    0.048316 &     12.19 &     S     &     INT   &                  &     &      \\
  J1346$-$0325 &      J13462001$-$0325407 &      206.583483 &      $-$3.428114 &    0.024781 &     11.19 &     S     &     JUS   &      OT2$\_$cxu$\_$2   &      25.03  &      222.0   \\
              &      J13462215$-$0325057 &      206.592397 &      $-$3.418283 &    0.025521 &     11.66 &     E     &     JUS   &                  &     &      \\
  J1400$-$0254 &      J14003661$-$0254327 &      210.152556 &      $-$2.909102 &    0.025579 &     11.49 &     S     &     INT   &      OT2$\_$cxu$\_$2   &      12.23  &      391.5   \\
              &      J14003796$-$0254227 &      210.15818  &      $-$2.906313 &    0.026884 &     11.73 &     S     &     INT   &                  &     &      \\
  J1400$+$4251 &      J14005783$+$4251203 &      210.240969 &      42.85566    &    0.032741 &     11.87 &     S     &     INT   &      OT2$\_$cxu$\_$2   &      27.27  &      234.3   \\
              &      J14005879$+$4250427 &      210.244986 &      42.845198   &    0.033522 &     12.18 &     S     &     INT   &                  &     &      \\
  J1405$+$6542 &      J14055079$+$6542598 &      211.461632 &      65.716632   &    0.03063  &     12.8  &     S     &     JUS   &      OT2$\_$cxu$\_$2   &      23.27  &      39.6    \\
              &      J14055334$+$6542277 &      211.472286 &      65.707721   &    0.030762 &     12.0  &     E     &     JUS   &                  &     &      \\
  J1406$+$5043 &      J14062157$+$5043303 &      211.589913 &      50.725106   &    0.006456 &     9.7   &     S     &     JUS   &      OT2$\_$cxu$\_$2   &      25.21  &      240.6   \\
              &      J14064127$+$5043239 &      211.671969 &      50.723327   &    0.007258 &     9.56  &     E     &     JUS   &                  &     &      \\
  J1407$-$0234 &      J14070703$-$0234513 &      211.779296 &      $-$2.580923 &    0.058555 &     12.44 &     S     &     INT   &      OT2$\_$cxu$\_$2   &      14.43  &      300.0   \\
              &      J14070720$-$0234402 &      211.780016 &      $-$2.577849 &    0.057555 &     12.96 &     E     &     INT   &                  &     &      \\
  J1423$+$3400 &      J14234238$+$3400324 &      215.926602 &      34.009016   &    0.013553 &     11.48 &     S     &     JUS   &      OT2$\_$cxu$\_$2   &      16.14  &      294.0   \\
              &      J14234632$+$3401012 &      215.943004 &      34.017002   &    0.012573 &     11.04 &     S     &     JUS   &                  &     &      \\
  J1424$-$0304 &      J14245831$-$0303597 &      216.242996 &      $-$3.066606 &    0.052517 &     12.37 &     S     &     JUS   &      OT2$\_$cxu$\_$2   &      13.91  &      300.0   \\
              &      J14245913$-$0304012 &      216.246381 &      $-$3.067027 &    0.053517 &     11.9  &     S     &     JUS   &                  &     &      \\
  J1425$+$0313 &      J14250552$+$0313590 &      216.272925 &      3.233119    &    0.037083 &     11.98 &     E     &     INT   &      OT2$\_$cxu$\_$2   &      22.37  &      114.0   \\
              &      J14250739$+$0313560 &      216.280712 &      3.232055    &    0.037463 &     12.97 &     S     &     INT   &                  &     &      \\
  J1429$+$3534 &      J14294766$+$3534275 &      217.448585 &      35.574313   &    0.028996 &     10.93 &     S     &     JUS   &      KPGT$\_$soliver$\_$1  &      22.07  &      180.3   \\
              &      J14295031$+$3534122 &      217.459663 &      35.570065   &    0.029597 &     11.9  &     S     &     JUS   &                  &     &      \\
  J1433$+$4004 &      J14334683$+$4004512 &      218.445139 &      40.080911   &    0.026047 &     10.78 &     S     &     INT   &      OT2$\_$cxu$\_$2   &      28.01  &      114.0   \\
              &      J14334840$+$4005392 &      218.451475 &      40.094155   &    0.026427 &     11.17 &     S     &     INT   &                  &     &      \\
  J1444$+$1207 &      J14442055$+$1207429 &      221.085649 &      12.128593   &    0.030445 &     11.51 &     S     &     MER   &      OT2$\_$cxu$\_$2   &      8.26   &      299.4   \\
              &      J14442079$+$1207552 &      221.086655 &      12.132      &    0.031443 &     10.72 &     S     &     MER   &                  &     &      \\
  J1500$+$4317 &      J15002374$+$4316559 &      225.098935 &      43.282197   &    0.031088 &     11.34 &     E     &     JUS   &      OT2$\_$cxu$\_$2   &      14.58  &      147.0   \\
              &      J15002500$+$4317131 &      225.10417  &      43.286997   &    0.031578 &     11.73 &     S     &     JUS   &                  &     &      \\
  J1505$+$3427 &      J15053137$+$3427534 &      226.380709 &      34.464851   &    0.074528 &     12.98 &     S     &     INT   &      OT2$\_$cxu$\_$2   &      9.55   &      300.0   \\
              &      J15053183$+$3427526 &      226.38266  &      34.464635   &    0.073528 &     12.36 &     E     &     INT   &                  &     &      \\
  J1506$+$0346 &      J15064391$+$0346364 &      226.682879 &      3.776844    &    0.036278 &     11.48 &     S     &     JUS   &      OT2$\_$cxu$\_$2   &      25.24  &      315.0   \\
              &      J15064579$+$0346214 &      226.690883 &      3.772574    &    0.035228 &     11.6  &     S     &     JUS   &                  &     &      \\
  J1510$+$5810 &      J15101587$+$5810425 &      227.566059 &      58.178518   &    0.030343 &     11.77 &     S     &     JUS   &      OT2$\_$cxu$\_$2   &      10.54  &      402.0   \\
              &      J15101776$+$5810375 &      227.574164 &      58.176979   &    0.031683 &     12.35 &     S     &     JUS   &                  &     &      \\
  J1514$+$0403 &      J15144544$+$0403587 &      228.689368 &      4.06633     &    0.0386   &     11.89 &     S     &     INT   &      OT2$\_$cxu$\_$2   &      18.81  &      186.0   \\
              &      J15144697$+$0403576 &      228.695709 &      4.066007    &    0.03922  &     12.06 &     S     &     INT   &                  &     &      \\
  J1523$+$3748 &      J15233768$+$3749030 &      230.907038 &      37.817522   &    0.023365 &     12.55 &     S     &     INT   &      OT2$\_$cxu$\_$2   &      20.09  &      78.3    \\
              &      J15233899$+$3748254 &      230.912489 &      37.807073   &    0.023626 &     12.43 &     E     &     INT   &                  &     &      \\
  J1526$+$5915 &      J15264774$+$5915464 &      231.698953 &      59.262907   &    0.044712 &     12.45 &     S     &     INT   &      OT2$\_$cxu$\_$2   &      8.77   &      224.1   \\
              &      J15264892$+$5915478 &      231.703856 &      59.263304   &    0.045459 &     12.13 &     E     &     INT   &                  &     &      \\
  J1528$+$4255 &      J15281276$+$4255474 &      232.053241 &      42.929924   &    0.018839 &     10.02 &     S     &     INT   &      OT2$\_$cxu$\_$2   &      26.53  &      243.0   \\
              &      J15281667$+$4256384 &      232.069607 &      42.944107   &    0.018029 &     10.59 &     S     &     INT   &                  &     &      \\
  J1552$+$4620 &      J15523258$+$4620180 &      238.135769 &      46.338356   &    0.059385 &     12.07 &     E     &     INT   &      OT2$\_$cxu$\_$2   &      19.46  &      489.0   \\
              &      J15523393$+$4620237 &      238.141398 &      46.339933   &    0.061015 &     12.75 &     S     &     INT   &                  &     &      \\
\enddata
\end{deluxetable*}

\begin{deluxetable*}{ccrrccccccc}
\label{tbl:hkpair}
\tabletypesize{\scriptsize}
\setlength{\tabcolsep}{0.05in} 
\tablenum{1}
\tablewidth{0.95\linewidth}
\tablecaption{(Continued)}
\tablehead{
\ch{(1)}   &\ch{(2)}    &\ch{(3)}    &\ch{(4)}&\ch{(5)}&\ch{(6)} &\ch{(7)}
             &\ch{(8)}         &\ch{(9)}    &\ch{(10)}   &\ch{(11)}     \\
\ch{Pair ID} & \ch{Galaxy ID} & \ch{RA}  & \ch{Dec} &
\ch{z} & \ch{K$_s$} & \ch{Type} & \ch{Int-type} & \ch{Herschel-ID} & \ch{s(p)} & \ch{$\delta$(Vz)} \\
\ch{(KPAIR)}  & \ch{(2MASX)} & \ch{(J2000)} &
\ch{(J2000)} & \ch{redshift} & \ch{(mag)} & & &  & \ch{(kpc)} & \ch{(km s$^{-1}$)}
}
\startdata
  J1556$+$4757 &      J15562191$+$4757172 &      239.091225 &      47.954829   &    0.019103 &     12.1  &     S     &     JUS   &      OT2$\_$cxu$\_$2   &      22.96  &      240.0   \\
              &      J15562738$+$4757302 &      239.114264 &      47.958429   &    0.019903 &     12.16 &     E     &     JUS   &                  &     &      \\
  J1558$+$3227 &      J15583749$+$3227379 &      239.656243 &      32.460535   &    0.049368 &     13.16 &     S     &     INT   &      OT2$\_$cxu$\_$2   &      10.8   &      261.0   \\
              &      J15583784$+$3227471 &      239.657667 &      32.463088   &    0.048498 &     12.29 &     S     &     INT   &                  &     &      \\
  J1602$+$4111 &      J16024254$+$4111499 &      240.677392 &      41.197267   &    0.033536 &     11.69 &     S     &     JUS   &      OT2$\_$cxu$\_$2   &      18.66  &      69.0    \\
              &      J16024475$+$4111589 &      240.686475 &      41.199706   &    0.033306 &     12.5  &     S     &     JUS   &                  &     &      \\
  J1608$+$2529 &      J16080559$+$2529091 &      242.023328 &      25.485866   &    0.041547 &     11.94 &     S     &     MER   &      OT2$\_$cxu$\_$2   &      10.96  &      224.1   \\
              &      J16080648$+$2529066 &      242.02704  &      25.485182   &    0.042294 &     11.32 &     S     &     MER   &                  &     &      \\
  J1608$+$2328 &      J16082261$+$2328459 &      242.094219 &      23.479425   &    0.04092  &     13.17 &     S     &     JUS   &      OT2$\_$cxu$\_$2   &      22.25  &      31.8    \\
              &      J16082354$+$2328240 &      242.098125 &      23.473347   &    0.040814 &     12.44 &     S     &     JUS   &                  &     &      \\
  J1614$+$3711 &      J16145418$+$3711064 &      243.72577  &      37.185123   &    0.058169 &     12.13 &     S     &     INT   &      OT2$\_$cxu$\_$2   &      9.13   &      0.0     \\
              &      J16145421$+$3711136 &      243.725893 &      37.187131   &    0.058169 &     12.04 &     E     &     INT   &                  &     &      \\
  J1628$+$4109 &      J16282497$+$4110064 &      247.104069 &      41.168463   &    0.033017 &     11.47 &     S     &     JUS   &      OT2$\_$cxu$\_$2   &      27.9   &      370.8   \\
              &      J16282756$+$4109395 &      247.114849 &      41.161      &    0.031781 &     11.52 &     S     &     JUS   &                  &     &      \\
  J1635$+$2630 &      J16354293$+$2630494 &      248.928908 &      26.513727   &    0.070061 &     12.29 &     S     &     INT   &      OT2$\_$cxu$\_$2   &      14.99  &      378.0   \\
              &      J16354366$+$2630505 &      248.931925 &      26.514045   &    0.071321 &     12.22 &     E     &     INT   &                  &     &      \\
  J1637$+$4650 &      J16372583$+$4650161 &      249.357631 &      46.837824   &    0.057817 &     11.8  &     S     &     JUS   &      OT2$\_$cxu$\_$2   &      25.82  &      299.7   \\
              &      J16372754$+$4650054 &      249.364787 &      46.834858   &    0.056818 &     12.44 &     S     &     JUS   &                  &     &      \\
  J1702$+$1859 &      J17020320$+$1900006 &      255.513366 &      19.000181   &    0.057322 &     12.39 &     E     &     INT   &      OT2$\_$cxu$\_$2   &      17.14  &      462.6   \\
              &      J17020378$+$1859495 &      255.515763 &      18.997088   &    0.05578  &     13.18 &     S     &     INT   &                  &     &      \\
  J1704$+$3448 &      J17045089$+$3448530 &      256.212001 &      34.814721   &    0.057213 &     13.06 &     S     &     INT   &      OT2$\_$cxu$\_$2   &      11.74  &      270.0   \\
              &      J17045097$+$3449020 &      256.212288 &      34.817342   &    0.056313 &     12.4  &     S     &     INT   &                  &     &      \\
  J2047$+$0019 &      J20471908$+$0019150 &      311.829434 &      0.320801    &    0.012971 &     9.08  &     S     &     JUS   &      OT2$\_$cxu$\_$2   &      28.98  &      414.0   \\
              &      J20472428$+$0018030 &      311.851263 &      0.300826    &    0.011591 &     9.74  &     E     &     JUS   &                  &     &      \\
\enddata
\tablecomments{Descriptions of Columns: (1) Pair ID. The designations are ``KPAIR J0020+0049'',
etc. (2) Galaxy ID, taken from 2MASS. (3) RA (deg, J2000). (4) Dec (deg, J2000). (5) redshift z
taken from SDSS or other telescopes. (6) $K_{rm s}$ ($K_{\rm 20}$) magnitude taken from 2MASS.
(7) Morphological type (S: Spirals, E: Ellipticals). (8) Interacting morphological type (INT:
with interaction signs; MER: mergers; JUS: without interacting signs, 'just' pairs). (9)
Herschel proposal ID. (10) projected separation (s(p)) between two galaxies in pairs, in kpc.
(11) the difference in radial velocity ($\delta$(Vz)) between two galaxies in pairs, in km s$^{-1}$.
}
\end{deluxetable*}

\subsection{PACS $\&$ SPIRE photometry}
We made extended-source photometry on individual galaxies in each
pair to get their integrated FIR-submm fluxes. The following two
methods were used:
\begin{description}
\setlength{\itemsep}{-2pt}
\item{1.} {\bf Aperture photometry}: for pairs with large separation
  compared with the beam size of a given band, elliptical/circular
  apertures were selected and aperture photometry was performed. The
  photometric error is the quadratic sum of background subtraction
  error and the rms error as calculated in \citet{Dale2012}.
  For PACS data, photometry
  in aperture matching annuli were used to determine the values for
  sky background subtraction.  For SPIRE data, the sky background was
  estimated using the mean of photometric measurements on 8
  elliptical/circular regions (of the same size as the photometric
  aperture) surrounding the pair.  Aperture corrections were applied
  according to measurements on the PSF images.
\item{2.} {\bf IMFIT model fitting photometry}: for blended pairs with
  small separation, IMFIT \citep{Erwin2014} was used to do the
  2-component simultaneous model fitting to get the deblended flux for
  each galaxy. The PACS photometry fits simultaneously a combination
  of sky level, gaussians and exponential disks for both galaxy
  components. Initial parameters include the peak intensity,
  coordinates and appropriate scale. The IMFIT uses $\chi^2$
  minimization and the residual images were examined and adjusted
  manually. Subsequent initial parameter adjustments were performed to
  minimize the residual images. The relative flux of each galaxy is
  fairly robust to the parameter adjustments. Total aperture flux of
  the pair with division based on relative fits was therefore the
  chosen photometric method. Errors are based on the area dependent
  background error estimation for this large aperture with a
  conservative approach based on half the area of the flux assigned to
  each galaxy. The pair J0926$+$0447
  is too blended for the fits to converge to a useful residual and the
  relative flux could not be determined. The reported errors for this
  pair are large but constrained by the total flux.  The SPIRE
  photometry uses two types of models in IMFIT: (1) PSF model for
  point-like galaxies; (2) 2-D gaussian
  for extended galaxies. Both models also include an uniform background. The
  photometric errors were estimated using the rms error measured on
  the residual images: error$_{rms}$ = rms (residual,per pixel)
  $\times$ N / $\sqrt{n}$ (N: number of pixels in the aperture; n:
  number of beams in the aperture).
\end{description}

Color corrections were made for 3 SPIRE bands with multiplicative correcting
factors of 0.95385, 0.95632, 0.97215 for 250, 350, 500$\mu$m fluxes, respectively.
No calibration errors are included in PACS and SPIRE photometric errors.
4-$\sigma$ upper-limits were given for non-detections in both aperture
$\&$ model fitting photometry methods.

\section{Dust Mass, $L_{\rm IR}$ and stellar mass}

Dust mass $M_{\rm dust}$ and integrated infrared luminosity $L_{\rm IR}$
(8 -- 1000$\mu$m) of individual galaxies are estimated via FIR-submm
SED (Spectral Energy Distributions) fitting using the dust emission
model of \citep{DL07} (hereafter DL07). The DL07 model
includes emissions from polycyclic aromatic hydrocarbon (PAH)
molecules and graphite and silicate grains, covering the entire IR
wavelength range from the mid infrared (MIR) though the submm.  It is
consistent with observations of a variety of infrared continuum and PAH
features in local galaxies (e.g., \citealt{Draine2007}, see also
\citealt{Dale2012} for more detailed descriptions).  In the SED
fittings, the Milky way extinction curve (MW3.1) is assumed and the
maximum interstellar radiation field intensity is fixed at $\rm Umax =
10^{6}$.  The best-fit values of parameters $M_{\rm dust}$, gamma, qPAH
(between 0.5 and 4.5 with the step of 0.1), and Umin (among values of
0.10, 0.15, 0.20, 0.30, 0.40, 0.50, 0.70, 0.80, 1.00, 1.20, 1.50,
2.00, 2.50, 3.00, 4.00, 5.00, 7.00, 8.00, 12.0, 15.0, 20.0, 25.0) are
then found via a simple $\chi^{2}$ minimization. The errors of
$M_{\rm dust}$ includes two terms: (1) error associated with model
fitting, estimated using Bayesian analysis \citep{daCunha2008};
(2) error caused by observational uncertainties, estimated using a
Monte-Carlo method: it equals to the standard deviation of the values
of $M_{\rm dust}$ resulting from fittings of 300 simulated IR-submm
SEDs. In each SED, the flux in a given Herschel band is generated
randomly assuming a gaussian probability distribution with the mean
equal to the observed flux and the $\sigma$ equal to the photometric
error. The mean errors of $M_{\rm dust}$ for spirals in H-KPAIRs are
estimated to be: 0.086$\pm$0.020 dex (log$_{10}$(M$_{\odot}$)),
0.101$\pm$0.044 dex, and 0.213$\pm$0.191 dex for those with all
6-bands Herschel detections, 5-bands Herschel detections and 4-bands
Herschel detections, respectively.

We carried out a test to check whether the lack of MIR broadband fluxes
in the SED fittings introduces any bias, using $\sim$25 H-KPAIR spirals
observed by Spitzer \citep{Xu2010}.  As shown in Figure~\ref{fig:comp_mdfmir},
the $M_{\rm dust}$ obtained using SED fittings with and without MIR fluxes
(IRAC 8$\mu$m $\&$ MIPS 24$\mu$m) agree very well with each other. The
scatter is quite small and much lower than other uncertainties. We also
found good agreement between L$_{IR}$ and L$_{TIR}$ (estimated from 24, 70,
160$\mu$m luminosities; \citealt{Dale2005}). Thus, both $M_{\rm dust}$ and
L$_{IR}$ estimated using DL07 model SED fittings from FIR-submm bands are
reliable even though the MIR data points are missing. On the other hand,
in trials that used two-graybody model (2GB; \citealt{Dunne2001}) in the
SED fittings of H-KPAIR galaxies, we found that the resulting $M_{\rm dust}$
and $L_{\rm IR}$ are both systematically lower than those obtained using
DL07 model (Figure~\ref{fig:sedfitting}). This is mainly due to the fact
that the 2GB model does not include the MIR emission at $\lambda < 20\mu m$,
therefore it underestimates $M_{\rm dust}$ and $L_{\rm IR}$ \citep{Dale2012}.

It should be noted that there might be a systematic underestimation (up
to 20$\%$) of the L$_{IR}$ for galaxies with very warm MIR-to-FIR colors
(f(25$\mu$m)/f(60$\mu$m)$>$0.2). These sources are generally associated
with bright AGNs \citep{Surace1998}. As discussed in Section 8.1, very
few galaxies in our samples may be associated with bright AGNs. Therefore
our results shall not be affected significantly by this possible bias.
Indeed, Figure~\ref{fig:comp_ircolor} shows the dependence of L$_{IR}$ estimations
using DL07 SED fittings without or with MIR (Spitzer IRAC $\&$ MIPS 24$\mu$m)
data points on Spitzer MIR-to-FIR colors: f(24$\mu$m)/f(70$\mu$m), for 30
spirals in H-KPAIRs which were also included in \citet{Xu2010}. There is
only one spiral in pairs (J20471908$+$0019150) with warm MIR-to-FIR color
(f(24$\mu$m)/f(70$\mu$m)$>$0.2), and there are no obvious underestimations
of L$_{IR}$ without the use of MIR points.

\begin{figure}[!htb]
\includegraphics[width=0.5\textwidth,angle=90]{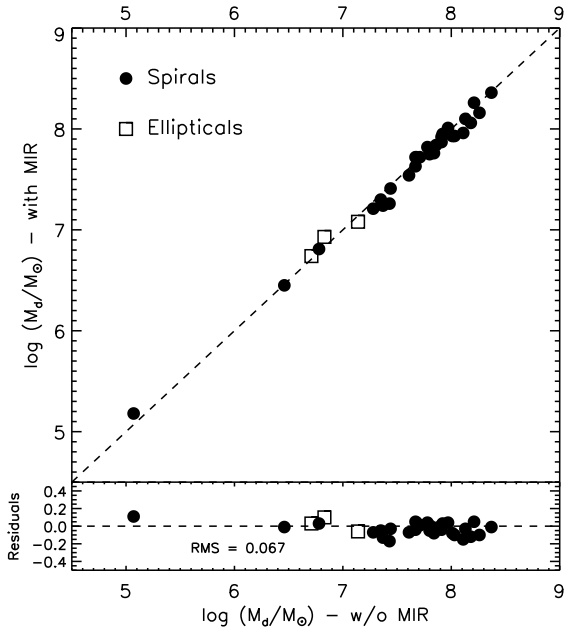}
\caption{Comparison of dust masses using DL07 SED fitting with and without MIR fluxes
for H-KPAIR galaxies with Spitzer observations by \citet{Xu2010}. Spirals are shown
as black dots, and ellipticals are shown as open squares. Residuals are shown in the
bottom panel.
}
\label{fig:comp_mdfmir}
\setcounter{figure}{2}
\end{figure}

\begin{figure*}[!htb]
\centering
\begin{minipage}[!htbp]{0.48\linewidth}
\centerline{\includegraphics[width=0.7\textwidth,angle=90]{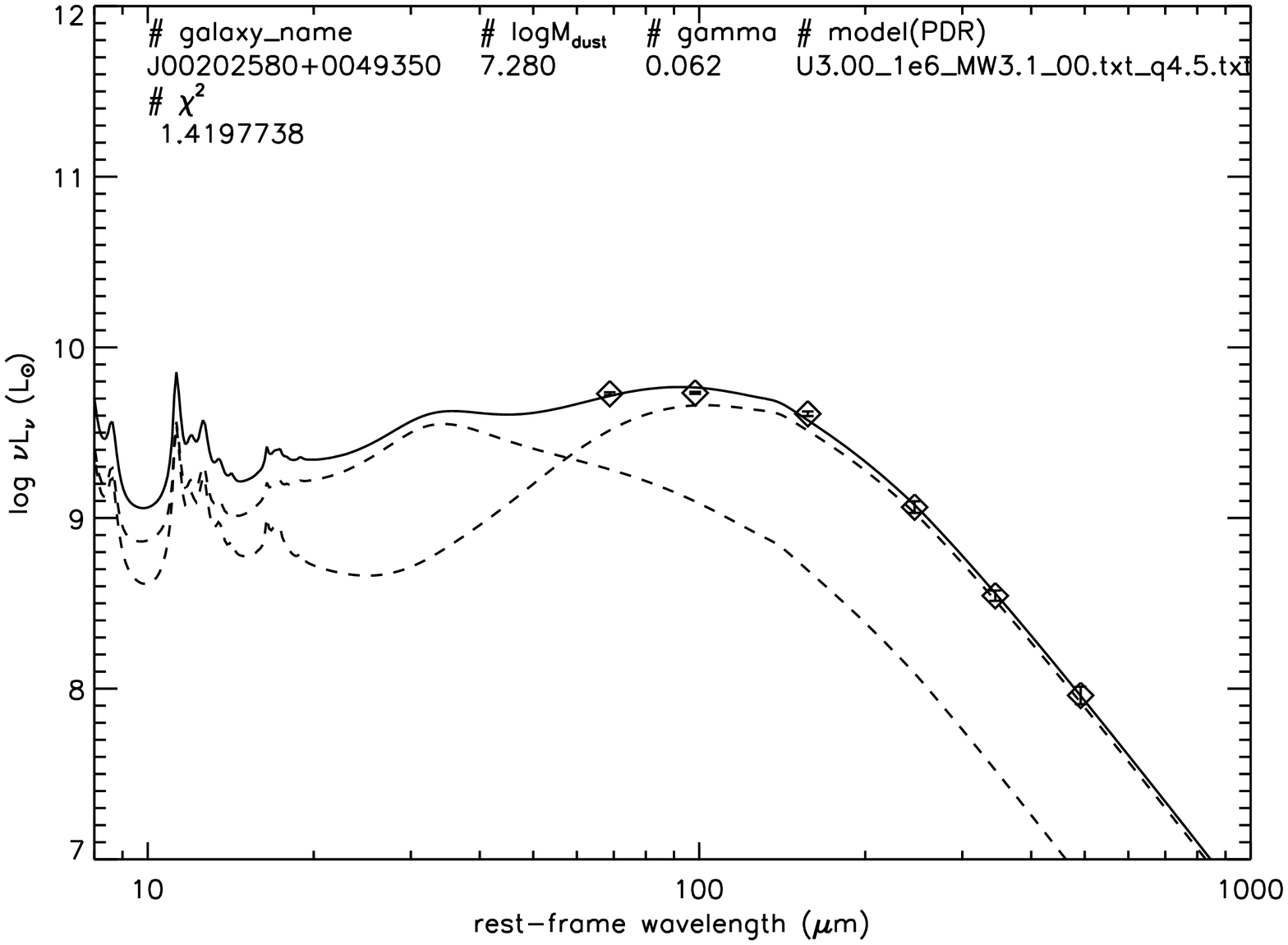}}
\end{minipage}
\hfill
\begin{minipage}[!htbp]{0.48\linewidth}
\centerline{\includegraphics[width=0.7\textwidth,angle=90]{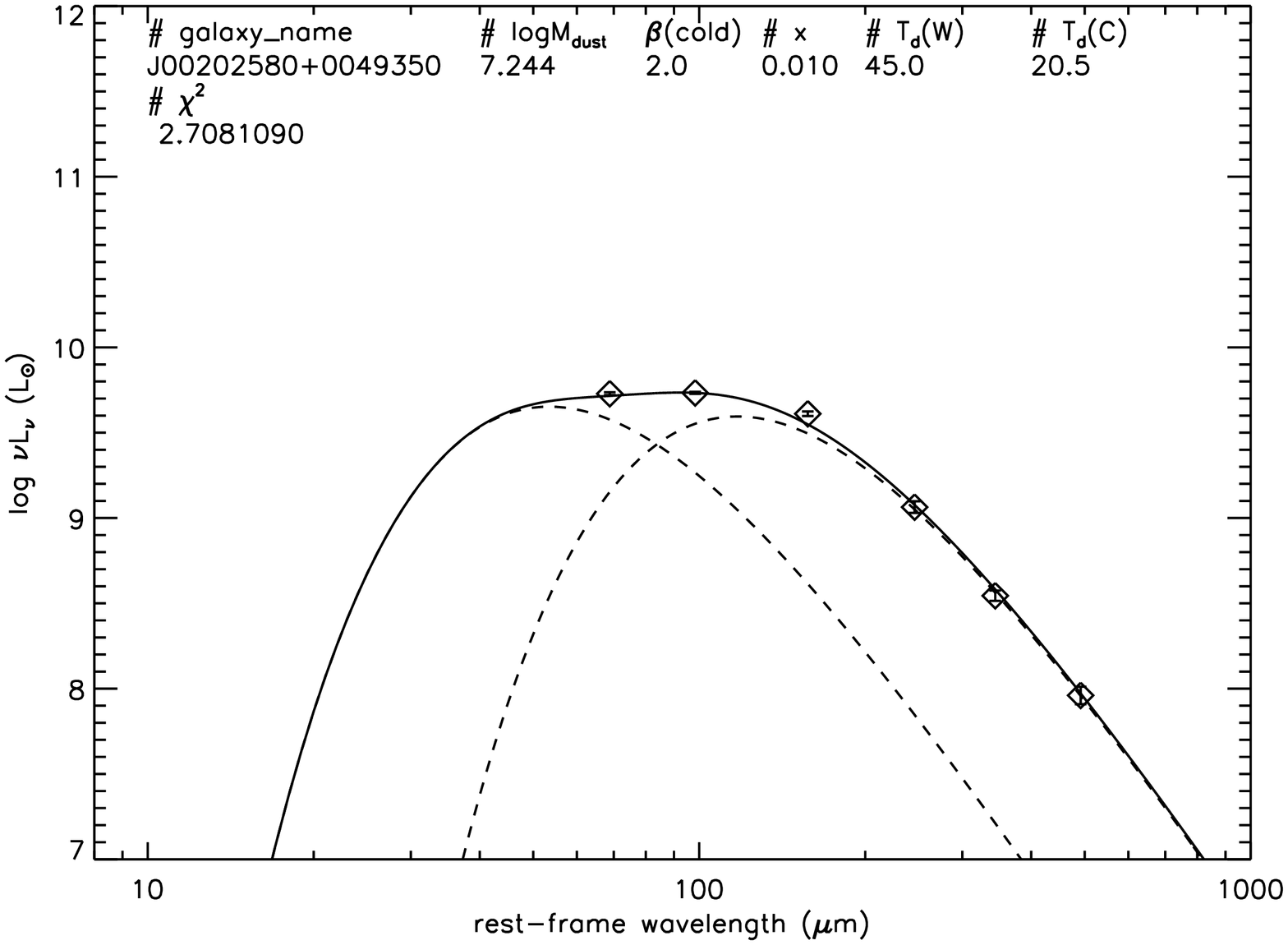}}
\end{minipage}
\hfill
\begin{minipage}[!htbp]{0.48\linewidth}
\centerline{\includegraphics[width=0.7\textwidth,angle=90]{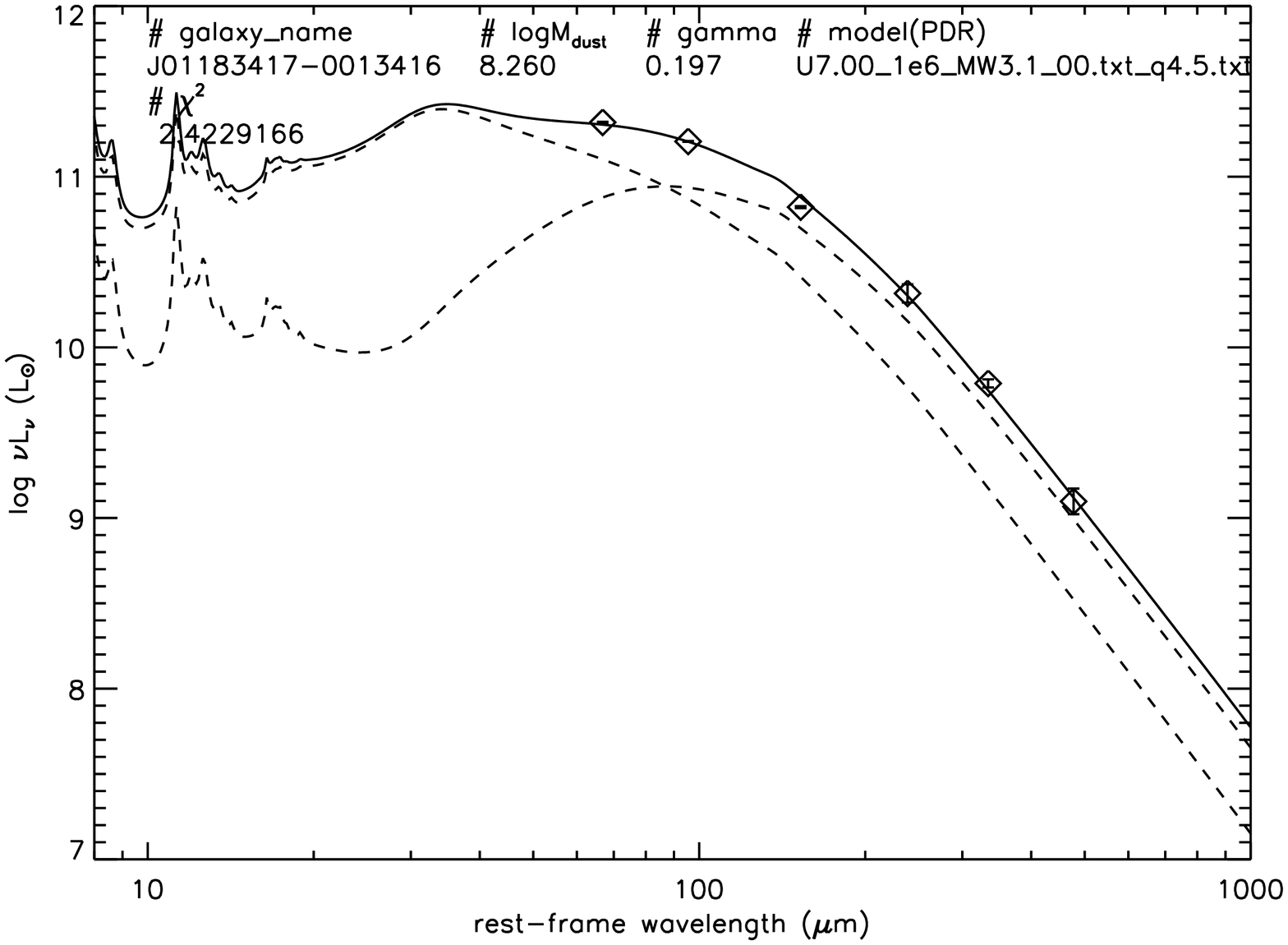}}
\end{minipage}
\hfill
\begin{minipage}[!htbp]{0.48\linewidth}
\centerline{\includegraphics[width=0.7\textwidth,angle=90]{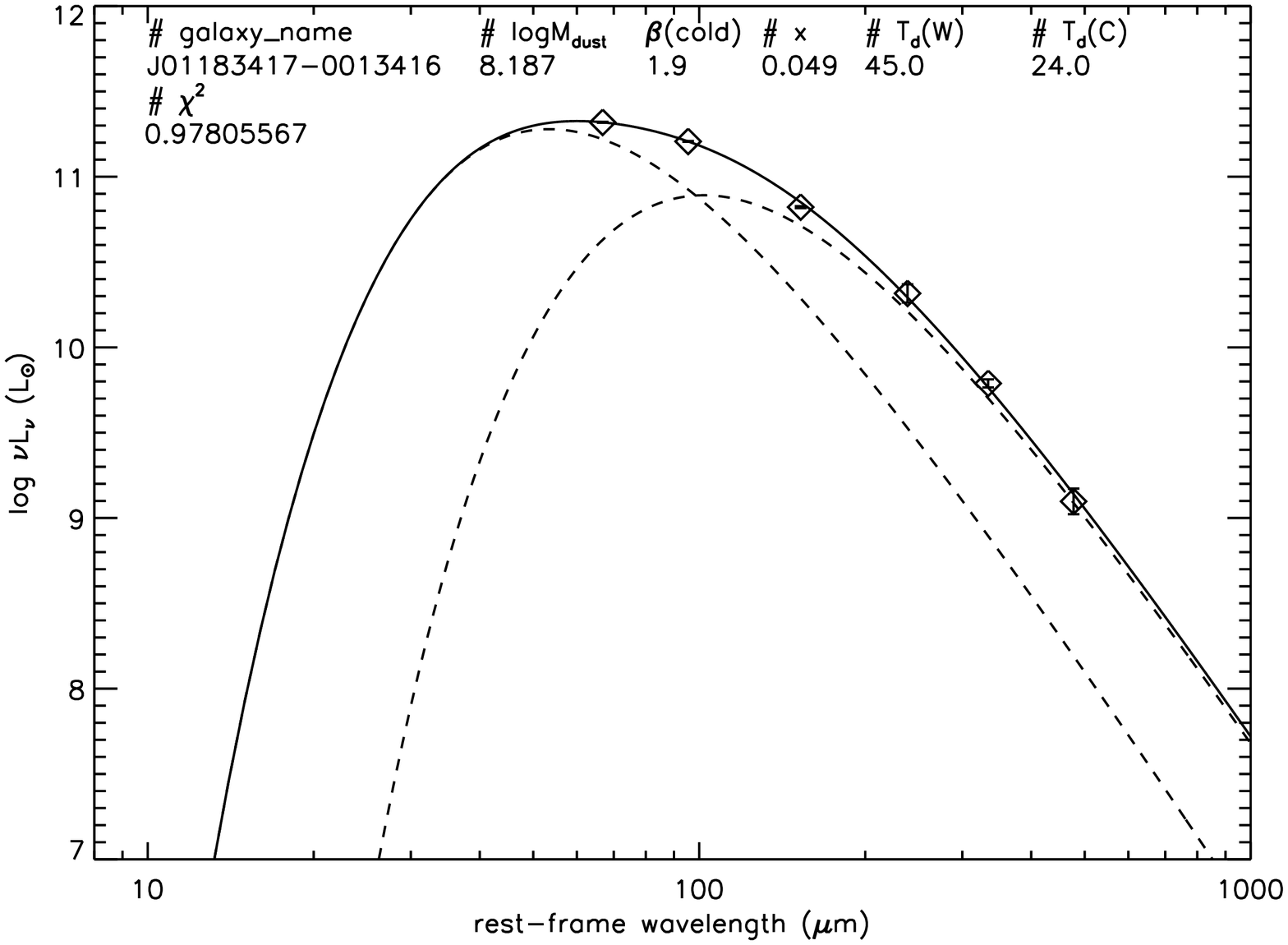}}
\end{minipage}
\hfill
\begin{minipage}[!htbp]{0.48\linewidth}
\centerline{\includegraphics[width=0.7\textwidth,angle=90]{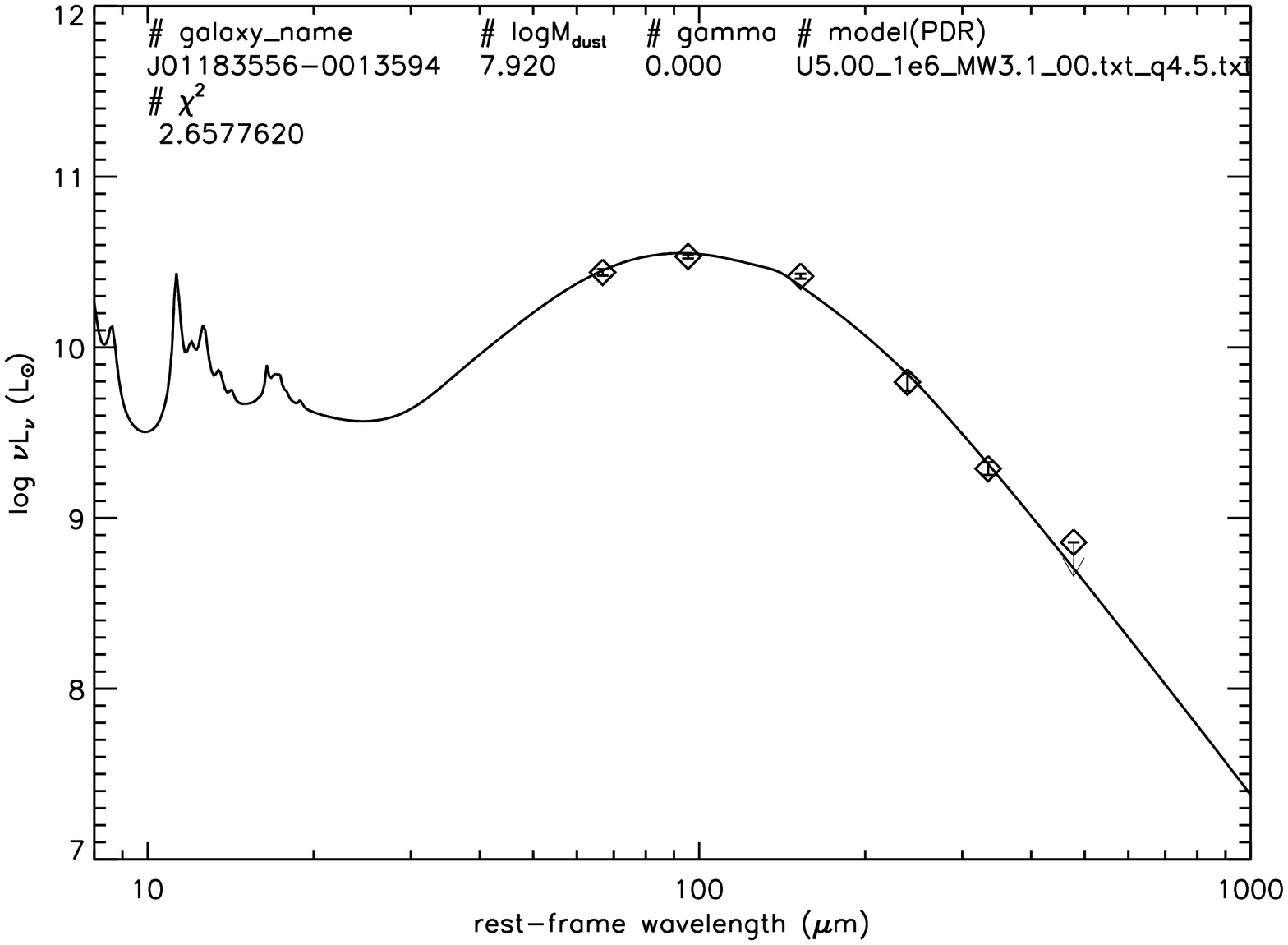}}
\end{minipage}
\hfill
\begin{minipage}[!htbp]{0.48\linewidth}
\centerline{\includegraphics[width=0.7\textwidth,angle=90]{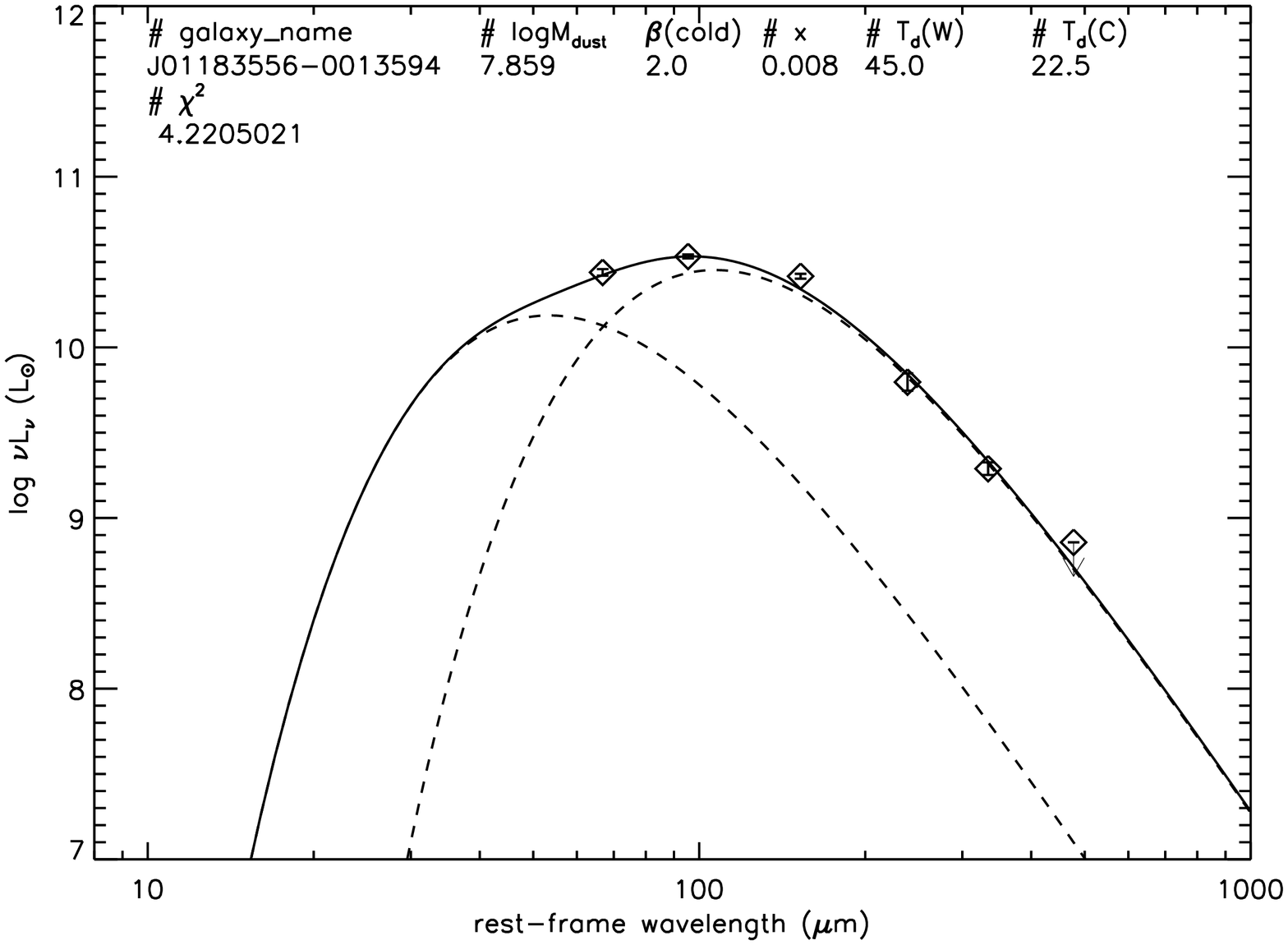}}
\end{minipage}
\caption{Examples of DL07 (left panels) and 2GB (right panels) SED fittings for 3 spirals in H-KPAIR.}
\label{fig:sedfitting}
\setcounter{figure}{3}
\end{figure*}

\begin{figure}[!htb]
\includegraphics[width=0.35\textwidth,angle=90]{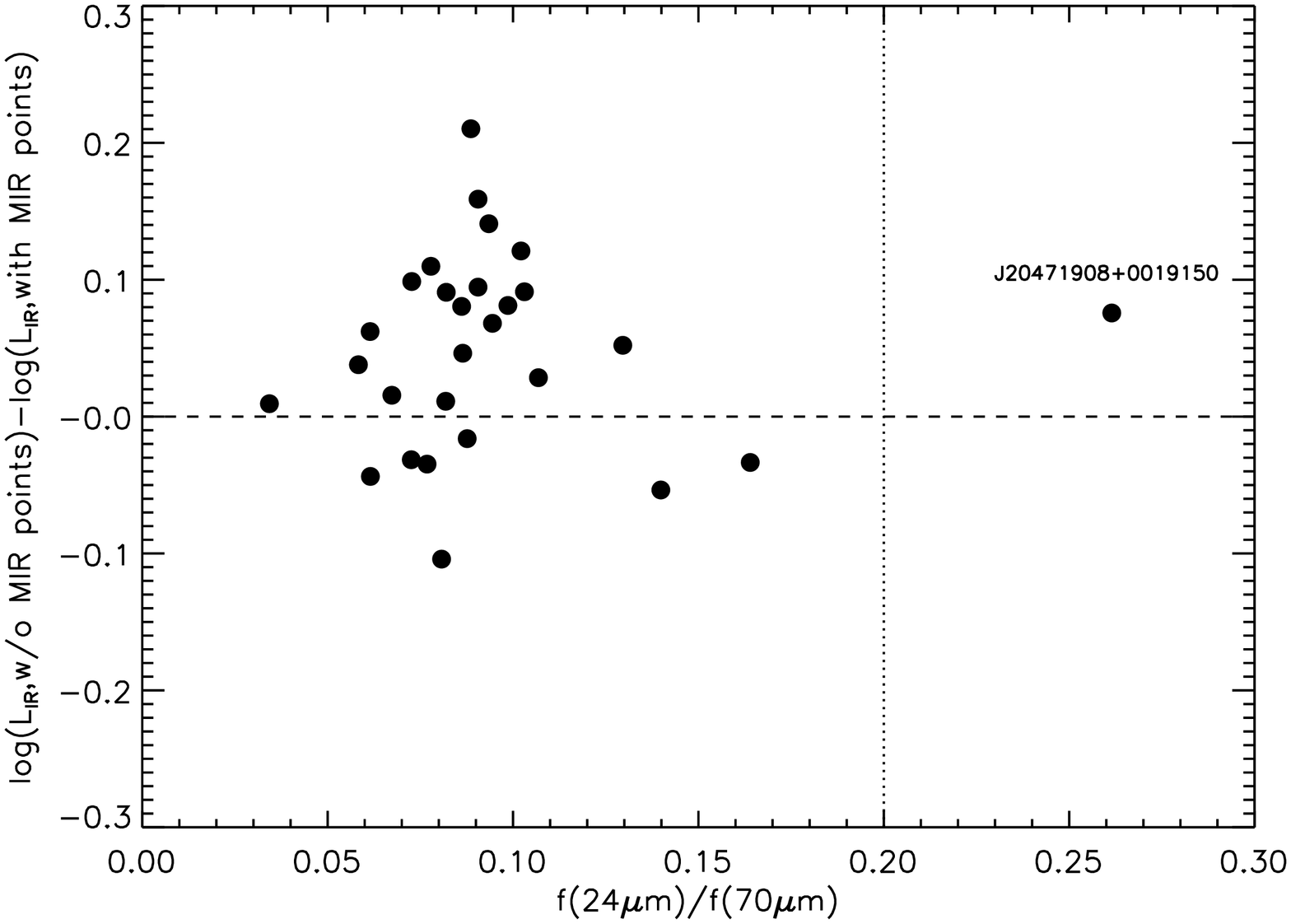}
\caption{The differences between L$_{IR}$ using DL07 SED fitting without
and with MIR fluxes vs. MIR-to-FIR color: f(24$\mu$m)/f(70$\mu$m), for
H-KPAIR spiral galaxies with Spitzer observations by \citet{Xu2010}.
}
\label{fig:comp_ircolor}
\setcounter{figure}{4}
\end{figure}

For H-KPAIR and control sample galaxies (see Section~\ref{section:Control})
with the number of Herschel band of detections less than four, we used an
empirical relation between $M_{\rm dust}$ and $L_{\rm 250\mu m}$ (the
monochromatic luminosity $\nu L_{\rm \nu}$ at 250 $\mu m$) to estimate
the dust mass (or its upper-limit) from the 250$\mu$m flux (or the upper-limit).
The adopted relation is $log(M_{\rm dust}/M_{\rm \odot}) = -0.13 + 0.83 \times
log(L_{\rm 250\mu m}/L_{\rm \odot})$. The $L_{\rm IR}$ (or its upper-limit) of these
galaxies was estimated from the the 100$\mu$m flux (or the upper-limit)
according to another empirical relation: $log(L_{\rm IR}/L_{\rm \odot}) = 0.84
+ 0.95\times log(L_{\rm 100\mu m}/L_{\rm \odot})$. Both relations were derived using
H-KPAIR spirals with all 6-band or 5-band detections. Comparing results
obtained from SED fittings to those calculated using these relations,
the RMS errors of the two empirical relations are 0.27 dex (for $M_{\rm dust}$)
and 0.11 dex (for $L_{\rm IR}$), respectively.

In Table~2 and Table~3 we listed the Herschel fluxes, stellar mass
M$_{star}$, the SFR, and total gas masses $M_{\rm gas}$ ($M_{\rm gas}
= 100 \times M_{\rm dust}$) of paired spiral and elliptical galaxies,
respectively. The SFR is derived from $L_{\rm IR}$ using the formula
of \citet{Kennicutt1998a}, with an additional
correction factor of 10$^{-0.20}$ for the conversion from the Salpeter
IMF to the Kroupa IMF \citep{Calzetti2013}. It is worth noting that
this formalism misses the contribution from unobscured UV radiation,
which is on the order of 20\% for KPAIR galaxies \citep{Yuan2012}.
It is also contaminated by the dust emission powered by the radiation
of old stars \citep{Buat1996}. However, since the same formalism is
applied to both H-KPAIR galaxies and control galaxies, these possible
biases shall not affect the results on the star formation enhancement
of paired galaxies. M$_{\rm star}$ is estimated using 2MASS Ks band luminosity:
M$_{\rm star}$/L$_{\rm K}$ = 1.32 M$_{\rm \odot}$/L$_{\rm \odot}$ \citep{Xu2004, Domingue2009},
with an additional correction factor of 10$^{-0.39}$ \citep{Xu2012a} in order
to match those given in SDSS value-added catalogs (\citealt{Kauffmann2003};
with Kroupa IMF).

{\begin{turnpage}
\begin{deluxetable*}{cccccccccccc}
\label{tbl:hkpair_table2}
\tabletypesize{\scriptsize}
\setlength{\tabcolsep}{0.05in} 
\tablenum{2}
\tablewidth{0.95\linewidth}
\tablecaption{Herschel 6-band fluxes and physical parameters of H-KPAIR spirals}
\tablehead{
\ch{(1)}   &\ch{(2)}    &\ch{(3)}    &\ch{(4)}&\ch{(5)}&\ch{(6)} &\ch{(7)}
             &\ch{(8)}         &\ch{(9)}  &\ch{(10)}  &\ch{(11)}  &\ch{(12)}       \\
\ch{Galaxy ID} & \ch{F70$\mu$m} & \ch{F100$\mu$m}  & \ch{F160$\mu$m} & \ch{P-PACS} &
\ch{F250$\mu$m} & \ch{F350$\mu$m} & \ch{F500$\mu$m} & \ch{P-SPIRE} &
\ch{logM$_{star}$} & \ch{SFR} & \ch{log$M_{\rm gas}$} \\
\ch{(2MASX)}  & \ch{(mJy)} & \ch{(mJy)} & \ch{(mJy)} & & \ch{(mJy)} & \ch{(mJy)} & \ch{(mJy)} & &
\ch{(M$_{\odot}$)} & \ch{(M$_{\odot}$/yr)} & \ch{(M$_{\odot}$)}
}
\startdata
J00202580+0049350 &   738.90$\pm$14.03 &  1067.90$\pm$15.14 &  1288.50$\pm$40.76 & AAA &   571.35$\pm$46.82 &   242.06$\pm$17.53 &    90.04$\pm$11.28 & APP &    10.49 &     1.39 &     9.28 \\
J01183417-0013416 &  3507.90$\pm$20.73 &  3874.90$\pm$22.54 &  2555.50$\pm$33.61 & MMM &  1246.09$\pm$ 163.57 &   517.78$\pm$28.79 &   150.56$\pm$28.41 & PPP &    10.98 &    55.91 &    10.26 \\
J01183556-0013594 &   459.70$\pm$20.73 &   814.30$\pm$22.54 &   997.10$\pm$33.61 & MMM &   373.62$\pm$46.31 &   162.56$\pm$14.32 &  $<$86.01 & PPP &    10.65 &     5.92 &     9.92 \\
J02110638-0039191 &  1126.50$\pm$15.75 &  1822.70$\pm$17.70 &  1839.40$\pm$51.51 & AAA &   989.35$\pm$83.06 &   401.95$\pm$20.05 &   135.12$\pm$11.49 & APP &    10.45 &     2.95 &     9.61 \\
J02110832-0039171 &  $<$37.65 &  $<$42.34 &  $<$60.49 & AAA &    52.50$\pm$9.65 &  $<$25.81 &  $<$35.17 & APP &    10.68 &   $<$0.09 &     8.67 \\
J03381222+0110088 &   368.04$\pm$9.08 &   607.68$\pm$9.90 &   801.85$\pm$16.88 & AAA &   513.69$\pm$47.72 &   200.00$\pm$31.08 &  $<$92.45 & APP &    10.70 &     5.02 &    10.13 \\
J07543194+1648214 &  1365.84$\pm$10.40 &  1934.76$\pm$11.60 &  2078.56$\pm$29.24 & MMM &  1024.48$\pm$50.21 &   361.63$\pm$19.41 &   101.26$\pm$21.93 & PPP &    10.96 &    18.00 &    10.22 \\
J07543221+1648349 &   769.56$\pm$10.40 &  1485.14$\pm$11.60 &  1434.74$\pm$29.24 & MMM &   790.32$\pm$63.32 &   422.92$\pm$16.90 &   167.38$\pm$21.69 & PPP &    11.15 &    12.02 &    10.42 \\
J08083377+3854534 &   121.91$\pm$5.59 &   223.42$\pm$6.56 &   167.75$\pm$12.23 & AAA &   127.53$\pm$15.64 &    63.36$\pm$14.90 &  $<$51.08 & PPP &    10.71 &     1.35 &     9.36 \\
J08233266+2120171 &   732.66$\pm$7.06 &  1037.70$\pm$11.49 &   928.07$\pm$14.95 & AAA &   538.00$\pm$42.87 &   193.89$\pm$10.74 &    76.23$\pm$12.43 & AGP &    10.08 &     1.71 &     9.26 \\
J08233421+2120515 &   933.60$\pm$8.62 &  1229.60$\pm$16.68 &  1115.50$\pm$21.58 & AAA &   643.17$\pm$43.90 &   223.94$\pm$22.02 &    83.05$\pm$13.54 & AGP &    10.34 &     2.27 &     9.33 \\
J08291491+5531227 &   344.84$\pm$19.45 &   674.10$\pm$17.78 &   957.97$\pm$26.71 & AAA &   614.63$\pm$45.86 &   295.64$\pm$26.82 &   160.67$\pm$9.75 & AAG &    10.56 &     1.93 &    10.06 \\
J08292083+5531081 &   368.50$\pm$12.42 &   697.51$\pm$14.41 &   927.75$\pm$19.50 & AAA &   691.67$\pm$45.51 &   363.71$\pm$28.46 &   120.42$\pm$11.07 & AAG &    10.69 &     2.05 &    10.00 \\
J08364482+4722188 &    43.04$\pm$8.44 &    55.95$\pm$8.26 &    69.16$\pm$14.98 & AAA &  $<$65.06 &  $<$29.21 &  $<$37.30 & PPP &    10.98 &     0.82 &   $<$9.49 \\
J08364588+4722100 &  $<$33.75 &  $<$33.05 &  $<$59.90 & AAA &  $<$35.67 &  $<$31.45 &  $<$30.70 & PPP &    11.13 &   $<$0.49 &   $<$9.27 \\
J08381759+3054534 &   232.06$\pm$5.87 &   365.96$\pm$5.92 &   581.73$\pm$12.75 & MMM &   137.05$\pm$11.63 &    69.74$\pm$10.84 &  $<$36.50 & GPP &    10.74 &     2.90 &     9.61 \\
J08381795+3055011 &    91.52$\pm$5.87 &   144.27$\pm$5.92 &  $<$59.49 & MMM &   161.38$\pm$11.29 &    77.64$\pm$11.02 &    38.19$\pm$8.99 & GPP &    11.03 &     1.31 &     8.28 \\
J08390125+3613042 &    62.86$\pm$8.09 &   104.81$\pm$4.79 &   221.31$\pm$14.71 & AAA &   161.63$\pm$12.42 &    74.07$\pm$12.79 &    33.46$\pm$7.98 & PPP &    10.91 &     1.91 &    10.13 \\
J08414959+2642578 &    25.88$\pm$5.11 &    70.29$\pm$8.46 &   106.84$\pm$15.80 & AAA &    56.00$\pm$8.24 &  $<$25.91 &  $<$33.25 & PPP &    11.41 &     1.44 &     9.69 \\
J09060498+5144071 &   155.62$\pm$8.21 &   270.74$\pm$8.46 &   492.79$\pm$20.23 & AAA &   338.81$\pm$24.19 &   157.22$\pm$16.16 &    69.62$\pm$10.39 & AAP &    10.60 &     1.26 &     9.84 \\
J09123676+3547462 &  $<$14.80 &  $<$13.55 &  $<$16.17 & AAA &  $<$28.35 &  $<$20.63 &  $<$41.44 & PPP &    10.24 &   $<$0.04 &   $<$8.57 \\
J09134606+4742001 &   208.14$\pm$10.16 &   406.01$\pm$8.89 &   589.73$\pm$21.83 & AAA &   308.54$\pm$11.21 &   148.80$\pm$8.83 &    62.91$\pm$8.28 & GGP &    11.05 &     4.27 &    10.13 \\
J09155467+4419510 &  2426.67$\pm$14.10 &  2992.48$\pm$12.48 &  2994.09$\pm$28.97 & MMM &  1085.36$\pm$92.00 &   342.20$\pm$31.64 &   248.93$\pm$27.21 & PPP &    10.74 &    26.18 &    10.19 \\
J09155552+4419580 &  4541.43$\pm$14.10 &  5820.14$\pm$12.48 &  5134.21$\pm$28.97 & MMM &  2254.33$\pm$ 100.27 &  1026.76$\pm$31.55 &   218.35$\pm$26.69 & PPP &    11.11 &    43.22 &    10.41 \\
J09264111+0447247 &    16.36$\pm$16.36 &    19.36$\pm$19.36 &    19.80$\pm$19.80 & CCC &    37.60$\pm$8.04 &  $<$36.87 &  $<$36.94 & PPP &    11.10 &     1.05 &    10.37 \\
J09264137+0447260 &    16.36$\pm$16.36 &    19.36$\pm$19.36 &    19.80$\pm$19.80 & CCC &  $<$35.49 &  $<$36.21 &  $<$36.72 & PPP &    11.38 &     0.91 &   $<$9.70 \\
J09374413+0245394 &  2601.60$\pm$38.65 &  4557.40$\pm$45.05 &  6157.90$\pm$78.87 & AAA &  3101.00$\pm$ 124.44 &  1347.92$\pm$77.60 &   505.41$\pm$50.50 & AAA &    11.16 &     9.83 &    10.37 \\
J10100079+5440198 &   427.64$\pm$8.06 &   969.53$\pm$7.00 &  1404.19$\pm$26.20 & MMM &   735.36$\pm$57.31 &   356.73$\pm$22.99 &   123.55$\pm$13.80 & PPP &    10.89 &     6.40 &    10.35 \\
J10100212+5440279 &   168.22$\pm$8.06 &   312.57$\pm$7.00 &   430.61$\pm$26.20 & MMM &   366.34$\pm$40.03 &   128.19$\pm$14.54 &  $<$57.69 & PPP &    10.84 &     3.30 &    10.08 \\
J10155257+0657330 &  $<$15.62 &    64.94$\pm$5.58 &    85.74$\pm$7.00 & AAA &    95.76$\pm$12.28 &  $<$51.46 &  $<$45.76 & PPP &    10.45 &     0.28 &     9.16 \\
J10205188+4831096 &   138.95$\pm$5.57 &   298.69$\pm$9.27 &   290.63$\pm$16.50 & AAA &   167.91$\pm$13.07 &    70.55$\pm$8.64 &  $<$34.74 & PPP &    10.59 &     2.10 &     9.67 \\
J10225647+3446564 &   244.08$\pm$6.22 &   332.40$\pm$7.06 &   352.31$\pm$18.98 & AAA &   176.55$\pm$10.60 &    85.87$\pm$8.87 &    42.32$\pm$8.71 & PPP &    10.66 &     5.65 &     9.88 \\
J10225655+3446468 &  $<$13.71 &  $<$15.56 &  $<$19.50 & AAA &  $<$42.02 &  $<$28.51 &  $<$36.47 & PPP &    10.99 &   $<$0.27 &   $<$9.37 \\
J10233658+4220477 &   718.84$\pm$8.05 &  1085.12$\pm$8.56 &  1265.47$\pm$21.92 & MMM &   558.50$\pm$21.07 &   239.08$\pm$11.65 &    83.17$\pm$9.48 & PPP &    10.81 &     9.52 &    10.05 \\
J10233684+4221037 &   268.06$\pm$8.05 &   346.87$\pm$8.56 &   290.12$\pm$21.92 & MMM &   195.67$\pm$17.87 &    71.96$\pm$12.90 &  $<$36.43 & PPP &    10.55 &     3.84 &     9.59 \\
J10272950+0114490 &   746.29$\pm$11.60 &  1175.20$\pm$11.27 &  1262.90$\pm$35.01 & AAA &   586.38$\pm$47.10 &   251.36$\pm$15.89 &    70.93$\pm$9.21 & PPP &    10.42 &     1.91 &     9.44 \\
J10325316+5306536 &  $<$17.50 &    47.03$\pm$6.53 &    85.32$\pm$18.46 & AAA &  $<$39.18 &  $<$36.44 &  $<$32.48 & PPP &    11.02 &     1.03 &   $<$9.46 \\
J10332972+4404342 &   679.80$\pm$7.60 &  1143.78$\pm$13.46 &  1000.27$\pm$23.74 & AMM &   561.42$\pm$20.29 &   310.54$\pm$13.69 &   113.35$\pm$10.87 & GPP &    11.12 &    14.35 &    10.39 \\
J10333162+4404212 &   211.61$\pm$8.35 &   443.82$\pm$13.46 &   779.93$\pm$23.74 & AMM &   296.30$\pm$31.67 &   119.29$\pm$17.97 &    41.60$\pm$9.00 & GPP &    10.93 &     4.17 &     9.99 \\
J10364274+5447356 &  $<$15.18 &  $<$14.42 &  $<$15.42 & AAA &  $<$50.46 &  $<$41.84 &  $<$40.66 & PPP &    10.78 &   $<$0.17 &   $<$9.29 \\
J10392338+3904501 &  $<$14.47 &  $<$16.22 &    54.88$\pm$10.71 & AAA &    57.84$\pm$10.40 &    23.99$\pm$3.76 &  $<$32.84 & PPP &    10.72 &   $<$0.16 &     9.29 \\
J10435053+0645466 &   908.74$\pm$6.93 &  1426.00$\pm$9.42 &  1527.20$\pm$30.11 & AAA &   739.62$\pm$55.76 &   268.30$\pm$26.47 &   126.27$\pm$13.54 & APP &    10.53 &     4.46 &     9.78 \\
J10435268+0645256 &    70.02$\pm$5.10 &   128.94$\pm$5.83 &   345.66$\pm$17.53 & AAA &   245.67$\pm$22.10 &   116.31$\pm$15.33 &  $<$76.56 & APP &    10.42 &     0.56 &     9.80 \\
J10452478+3910298 &   112.42$\pm$12.10 &   122.56$\pm$7.02 &   244.20$\pm$16.00 & AAA &   189.69$\pm$18.59 &    81.94$\pm$10.03 &  $<$40.28 & APP &    10.63 &     0.67 &     9.63 \\
J10514450+5101303 &  $<$55.53 &  $<$53.29 &  $<$62.73 & AAA &  $<$35.27 &  $<$35.19 &  $<$32.71 & PPP &    10.80 &   $<$0.16 &   $<$8.66 \\
J10595915+0857357 &  $<$14.47 &  $<$14.63 &  $<$20.26 & AAA &  $<$43.21 &  $<$28.80 &  $<$40.62 & PPP &    10.86 &   $<$0.32 &   $<$9.47 \\
J11014364+5720336 &   $<$9.04 & $<$178.93 &  $<$24.27 & AAA &  $<$58.01 &  $<$52.54 &  $<$64.05 & PPP &    10.54 &   $<$2.01 &   $<$9.37 \\
J11064944+4751119 &  $<$11.86 &  $<$12.65 &  $<$13.89 & AAA &  $<$85.33 &  $<$48.39 &  $<$43.96 & PPP &    11.00 &   $<$0.30 &   $<$9.74 \\
J11065068+4751090 &   259.74$\pm$8.73 &   456.00$\pm$9.30 &   519.09$\pm$21.25 & AAA &   352.17$\pm$24.81 &   134.43$\pm$8.95 &    47.63$\pm$10.59 & PPP &    11.11 &     8.47 &    10.22 \\
J11204657+0028142 &   318.52$\pm$38.33 &   275.79$\pm$36.70 &   220.16$\pm$12.98 & AAA &    87.88$\pm$8.61 &    44.90$\pm$8.95 &  $<$44.09 & GPP &    10.49 &     1.12 &     8.55 \\
J11204801+0028068 &  $<$49.52 &  $<$47.52 &  $<$46.04 & AAA &    67.86$\pm$9.51 &  $<$40.36 &  $<$46.21 & GPP &    10.80 &   $<$0.14 &     8.90 \\
J11251704+0227007 &  $<$13.50 &  $<$17.16 &  $<$19.39 & AAA &    40.95$\pm$10.25 &  $<$43.79 &  $<$51.00 & PPP &    10.72 &   $<$0.23 &     9.26 \\
J11251716+0226488 &    51.04$\pm$5.94 &   138.38$\pm$9.60 &   219.03$\pm$22.83 & AAA &   164.52$\pm$12.36 &    87.73$\pm$14.98 &  $<$45.87 & PPP &    10.97 &     1.13 &     9.96 \\
\enddata
\end{deluxetable*}

\begin{deluxetable*}{cccccccccccc}
\label{tbl:hkpair_table2}
\tabletypesize{\scriptsize}
\setlength{\tabcolsep}{0.05in} 
\tablenum{2}
\tablewidth{0.95\linewidth}
\tablecaption{(Continued)}
\tablehead{
\ch{(1)}   &\ch{(2)}    &\ch{(3)}    &\ch{(4)}&\ch{(5)}&\ch{(6)} &\ch{(7)}
             &\ch{(8)}         &\ch{(9)}  &\ch{(10)}  &\ch{(11)}  &\ch{(12)}       \\
\ch{Galaxy ID} & \ch{F70$\mu$m} & \ch{F100$\mu$m}  & \ch{F160$\mu$m} & \ch{P-PACS} &
\ch{F250$\mu$m} & \ch{F350$\mu$m} & \ch{F500$\mu$m} & \ch{P-SPIRE} &
\ch{logM$_{star}$} & \ch{SFR} & \ch{log$M_{\rm gas}$} \\
\ch{(2MASX)}  & \ch{(mJy)} & \ch{(mJy)} & \ch{(mJy)} & & \ch{(mJy)} & \ch{(mJy)} & \ch{(mJy)} & &
\ch{(M$_{\odot}$)} & \ch{(M$_{\odot}$/yr)} & \ch{(M$_{\odot}$)}
}
\startdata
J11273289+3604168 &    70.87$\pm$5.20 &   154.65$\pm$9.90 &   275.72$\pm$13.37 & AAA &   185.12$\pm$17.20 &    70.08$\pm$9.25 &    46.05$\pm$9.85 & AGG &    10.84 &     0.78 &     9.62 \\
J11273467+3603470 &   624.34$\pm$12.74 &  1417.00$\pm$12.21 &  1887.00$\pm$24.35 & AAA &  1180.76$\pm$79.62 &   482.73$\pm$14.53 &   158.63$\pm$10.97 & AGG &    11.19 &     4.98 &    10.24 \\
J11375801+4728143 &  $<$17.95 &  $<$19.32 &  $<$22.62 & AAA &  $<$76.98 &  $<$38.08 &  $<$51.27 & PPP &    10.75 &   $<$0.12 &   $<$9.20 \\
J11440433+3332339 &   156.58$\pm$6.84 &   334.38$\pm$6.35 &   386.61$\pm$11.86 & AAA &   277.05$\pm$9.74 &   101.84$\pm$20.70 &  $<$51.09 & PPP &    10.35 &     1.11 &     9.58 \\
J11484370+3547002 &   245.71$\pm$12.07 &   562.98$\pm$11.80 &   928.88$\pm$22.28 & MMM &   514.88$\pm$37.78 &   216.07$\pm$11.40 &    78.64$\pm$10.56 & PPP &    10.94 &     7.84 &    10.44 \\
J11484525+3547092 &   267.61$\pm$12.07 &   598.32$\pm$11.80 &   459.22$\pm$22.28 & MMM &   402.27$\pm$54.30 &   189.07$\pm$11.91 &    53.54$\pm$8.82 & PPP &    11.27 &     7.14 &    10.25 \\
J11501333+3746107 &  $<$15.49 &    41.94$\pm$3.82 &   124.04$\pm$13.85 & AAM &    47.70$\pm$7.24 &  $<$33.95 &  $<$33.85 & PPP &    10.84 &     0.66 &     9.40 \\
J11501399+3746306 &   148.80$\pm$5.95 &   320.16$\pm$6.54 &   309.70$\pm$13.85 & AAM &   248.53$\pm$9.61 &   113.71$\pm$7.55 &    44.04$\pm$10.79 & PPP &    10.95 &     3.44 &    10.05 \\
J11505764+1444200 &  $<$14.87 &  $<$14.06 &  $<$17.12 & AAA &  $<$25.48 &  $<$27.27 &  $<$26.19 & PPP &    10.87 &   $<$0.25 &   $<$9.19 \\
J11542299+4932509 &  $<$14.52 &  $<$15.08 &  $<$16.62 & AAA &    35.45$\pm$8.80 &  $<$32.56 &  $<$38.89 & PPP &    10.95 &   $<$0.42 &     9.49 \\
J12020424+5342317 &    41.22$\pm$5.33 &    71.54$\pm$4.85 &   119.44$\pm$10.72 & AAA &    97.15$\pm$8.57 &    47.16$\pm$6.46 &  $<$39.11 & PPP &    10.88 &     1.78 &    10.03 \\
J12054066+0135365 &  $<$34.41 &  $<$42.56 &  $<$32.47 & AAA &  $<$43.02 &  $<$36.57 &  $<$39.41 & AAA &    10.21 &   $<$0.09 &   $<$8.61 \\
J12115507+4039182 &   467.63$\pm$21.33 &   690.98$\pm$20.83 &   851.82$\pm$43.60 & MMM &   264.79$\pm$20.04 &   123.73$\pm$15.98 &  $<$67.02 & PPP &    10.40 &     1.08 &     9.14 \\
J12115648+4039184 &  2413.87$\pm$21.33 &  2090.60$\pm$20.83 &  1031.37$\pm$43.60 & MMM &   467.84$\pm$17.31 &   155.57$\pm$10.70 &    90.97$\pm$11.70 & PPP &    10.36 &     7.14 &     9.12 \\
J12191866+1201054 &   164.66$\pm$7.39 &   237.80$\pm$7.52 &   392.89$\pm$26.58 & AAA &   189.20$\pm$11.91 &    82.43$\pm$14.21 &    35.33$\pm$8.12 & PPP &    10.12 &     0.83 &     9.34 \\
J12433887+4405399 &    93.44$\pm$6.88 &   190.87$\pm$7.06 &   291.07$\pm$16.53 & AAA &   189.84$\pm$11.52 &    82.28$\pm$12.26 &  $<$29.52 & PPP &    10.84 &     1.17 &     9.64 \\
J12525011+4645272 &    51.06$\pm$5.86 &   106.38$\pm$5.43 &   124.40$\pm$4.74 & AAA &   122.60$\pm$9.68 &    58.86$\pm$10.75 &  $<$44.78 & PPP &    11.05 &     1.84 &    10.07 \\
J13011662+4803366 &   922.45$\pm$11.95 &  1206.16$\pm$12.70 &  1144.11$\pm$30.74 & MMM &   504.63$\pm$36.55 &   198.88$\pm$11.70 &    74.39$\pm$10.65 & PPP &    10.51 &     5.62 &     9.62 \\
J13011835+4803304 &   675.05$\pm$11.95 &   975.24$\pm$12.70 &   845.78$\pm$30.74 & MMM &   366.05$\pm$31.51 &   162.10$\pm$11.22 &    47.29$\pm$11.94 & PPP &    10.31 &     2.94 &     9.43 \\
J13082737+0422125 &    54.10$\pm$7.99 &   124.30$\pm$6.83 &   179.01$\pm$14.30 & AAA &   139.97$\pm$14.04 &    75.34$\pm$7.13 &    33.20$\pm$5.76 & GPP &     9.84 &     0.28 &     9.35 \\
J13082964+0422045 &    74.68$\pm$7.92 &   149.92$\pm$8.14 &   222.27$\pm$12.76 & AAA &   171.76$\pm$10.19 &    93.85$\pm$14.69 &    35.38$\pm$6.46 & GPP &    10.22 &     0.38 &     9.43 \\
J13131470+3910382 &  $<$23.14 &  $<$24.15 &  $<$67.02 & AAA &  $<$33.20 &  $<$36.51 &  $<$31.20 & PPP &    10.92 &   $<$0.68 &   $<$9.48 \\
J13151386+4424264 &   256.70$\pm$11.79 &   483.17$\pm$6.95 &   569.02$\pm$13.79 & AAA &   411.36$\pm$35.87 &   164.61$\pm$16.70 &  $<$80.63 & APP &    10.73 &     2.35 &     9.88 \\
J13151726+4424255 &  1105.10$\pm$9.30 &  1094.20$\pm$8.30 &   669.71$\pm$20.69 & AAA &   340.75$\pm$31.14 &   138.57$\pm$11.12 &  $<$61.67 & APP &    11.04 &     8.29 &     9.41 \\
J13153076+6207447 &  1894.62$\pm$24.13 &  2488.20$\pm$21.36 &  2602.03$\pm$63.41 & MMM &   923.01$\pm$78.71 &   320.41$\pm$44.00 &   116.71$\pm$26.82 & APP &    10.62 &     7.59 &     9.80 \\
J13153506+6207287 &  9522.37$\pm$24.13 &  8581.77$\pm$21.36 &  5586.27$\pm$63.41 & MMM &  1615.77$\pm$ 135.74 &   539.46$\pm$41.27 &   167.12$\pm$24.28 & APP &    10.80 &    55.30 &     9.81 \\
J13325525-0301347 &   423.10$\pm$11.84 &   542.11$\pm$12.36 &   416.85$\pm$20.37 & MMM &   300.33$\pm$12.92 &    93.49$\pm$10.37 &  $<$33.69 & PPP &    10.62 &     7.08 &     9.67 \\
J13325655-0301395 &   276.81$\pm$11.84 &   476.79$\pm$12.36 &   612.95$\pm$20.37 & MMM &   281.72$\pm$19.14 &   148.02$\pm$12.66 &    49.38$\pm$9.48 & PPP &    10.91 &     4.37 &     9.91 \\
J13462001-0325407 &    34.85$\pm$4.92 &   202.52$\pm$10.65 &   309.66$\pm$25.46 & AAA &   223.82$\pm$18.24 &   107.98$\pm$12.36 &    42.40$\pm$7.74 & AAP &    10.69 &     0.28 &     9.45 \\
J14003661-0254327 &  $<$41.24 &    25.07$\pm$4.92 &  $<$51.99 & AAA &  $<$32.23 &  $<$24.77 &  $<$42.83 & PPP &    10.60 &     0.08 &   $<$8.63 \\
J14003796-0254227 &  $<$13.46 &  $<$16.89 &  $<$15.99 & AAA &  $<$48.94 &  $<$29.11 &  $<$30.13 & PPP &    10.55 &   $<$0.06 &   $<$8.82 \\
J14005783+4251203 &  1101.70$\pm$9.93 &  1573.70$\pm$11.16 &  1726.93$\pm$39.41 & AAM &   893.18$\pm$69.75 &   352.67$\pm$29.30 &   129.50$\pm$12.79 & APP &    10.71 &     8.10 &    10.01 \\
J14005879+4250427 &  1407.30$\pm$10.47 &  1788.80$\pm$12.41 &  1798.37$\pm$39.41 & AAM &   849.70$\pm$62.63 &   303.06$\pm$19.83 &    99.48$\pm$14.24 & APP &    10.60 &     9.88 &     9.86 \\
J14055079+6542598 &    81.29$\pm$6.76 &   161.01$\pm$6.26 &   221.49$\pm$10.95 & AAA &   182.80$\pm$17.26 &   103.07$\pm$10.63 &    55.79$\pm$9.48 & APP &    10.30 &     0.73 &     9.80 \\
J14062157+5043303 &  4848.30$\pm$25.34 &  8820.20$\pm$24.14 & 10480.00$\pm$58.91 & AAA &  4350.49$\pm$ 166.18 &  1990.01$\pm$76.04 &   695.15$\pm$63.28 & AAA &    10.14 &     0.92 &     9.32 \\
J14070703-0234513 &    44.43$\pm$5.16 &   165.84$\pm$9.56 &   220.38$\pm$10.74 & AAA &   164.48$\pm$15.13 &    86.58$\pm$8.92 &    36.21$\pm$8.79 & PPP &    10.98 &     1.49 &    10.01 \\
J14234238+3400324 &   954.06$\pm$8.13 &  1465.10$\pm$12.39 &  1596.00$\pm$22.30 & AAA &   789.00$\pm$58.13 &   360.55$\pm$28.53 &   124.10$\pm$13.66 & AAP &    10.06 &     0.97 &     9.23 \\
J14234632+3401012 &   237.13$\pm$8.52 &   499.65$\pm$11.25 &   698.09$\pm$17.35 & AAA &   446.27$\pm$42.09 &   224.34$\pm$22.98 &    76.06$\pm$11.45 & AAP &    10.16 &     0.26 &     9.08 \\
J14245831-0303597 &   253.64$\pm$7.85 &   591.09$\pm$11.13 &   857.00$\pm$22.18 & AAA &   325.65$\pm$16.98 &   115.16$\pm$15.83 &  $<$53.15 & GPP &    10.91 &     4.07 &     9.96 \\
J14245913-0304012 &  $<$40.60 &  $<$43.69 &  $<$27.89 & AAA &   222.15$\pm$16.49 &   129.20$\pm$15.86 &    61.39$\pm$13.81 & GPP &    11.12 &   $<$0.62 &     9.92 \\
J14250739+0313560 &  $<$15.68 &  $<$13.71 &  $<$13.43 & AAA &  $<$39.41 &  $<$33.13 &  $<$33.58 & PPP &    10.36 &   $<$0.10 &   $<$9.01 \\
J14294766+3534275 &    18.04$\pm$3.67 & $<$175.35 &    49.51$\pm$12.07 & AAA &    66.45$\pm$10.80 &  $<$65.79 &  $<$46.34 & APP &    10.97 &   $<$0.68 &     9.02 \\
J14295031+3534122 &   180.58$\pm$5.65 &   283.04$\pm$15.59 &   254.06$\pm$9.01 & AAA &   177.90$\pm$16.70 &    61.32$\pm$9.05 &  $<$49.03 & APP &    10.60 &     1.06 &     9.24 \\
J14334683+4004512 &   886.49$\pm$12.00 &  1345.40$\pm$14.96 &  1763.50$\pm$37.31 & AAM &   892.53$\pm$74.85 &   446.44$\pm$41.59 &   206.70$\pm$10.84 & AAG &    10.94 &     4.38 &    10.11 \\
J14334840+4005392 &  1459.00$\pm$12.23 &  2146.80$\pm$10.61 &  2496.30$\pm$37.31 & AAM &  1102.57$\pm$85.57 &   462.39$\pm$36.70 &   136.34$\pm$8.90 & AAG &    10.79 &     5.30 &     9.81 \\
J14442055+1207429 &   258.73$\pm$11.30 &   392.32$\pm$12.00 &   565.22$\pm$29.96 & MMM &   180.05$\pm$20.87 &   164.41$\pm$12.67 &    68.79$\pm$11.10 & GGP &    10.76 &     1.15 &     9.21 \\
J14442079+1207552 &   574.60$\pm$11.30 &  1234.18$\pm$12.00 &  1598.18$\pm$29.96 & MMM &   895.89$\pm$22.73 &   287.70$\pm$10.92 &   117.72$\pm$10.60 & GGP &    11.11 &     3.48 &    10.05 \\
J15002500+4317131 &    27.94$\pm$5.11 &  $<$18.18 &    16.77 & AAA &  $<$45.82 &  $<$52.13 &  $<$48.99 & PPP &    10.73 &   $<$0.10 &   $<$8.96 \\
J15053137+3427534 &  $<$13.13 &  $<$14.13 &  $<$12.73 & AAA &  $<$35.21 &  $<$36.48 &  $<$34.79 & PPP &    11.00 &   $<$0.44 &   $<$9.53 \\
J15064391+0346364 &    33.08$\pm$7.23 &    32.59$\pm$7.23 &  $<$85.47 & AAA &  $<$45.73 &  $<$42.30 &  $<$35.47 & GGP &    10.93 &     0.21 &   $<$9.04 \\
J15064579+0346214 &   232.45$\pm$9.24 &   477.59$\pm$13.24 &   780.68$\pm$66.13 & AAA &   441.27$\pm$36.88 &   202.46$\pm$10.04 &    88.87$\pm$10.08 & GGP &    10.86 &     2.22 &     9.97 \\
J15101587+5810425 &    41.02$\pm$10.00 &  $<$18.51 &  $<$24.37 & MMM &  $<$66.01 &  $<$39.97 &  $<$37.62 & PPP &    10.70 &   $<$0.09 &   $<$9.07 \\
J15101776+5810375 &   304.72$\pm$10.00 &   697.73$\pm$9.03 &   806.48$\pm$65.40 & MMM &   427.43$\pm$21.74 &   176.68$\pm$7.71 &    62.51$\pm$10.13 & PPP &    10.50 &     2.04 &     9.68 \\
\enddata
\end{deluxetable*}

\begin{deluxetable*}{cccccccccccc}
\label{tbl:hkpair_table2}
\tabletypesize{\scriptsize}
\setlength{\tabcolsep}{0.05in} 
\tablenum{2}
\tablewidth{0.95\linewidth}
\tablecaption{(Continued)}
\tablehead{
\ch{(1)}   &\ch{(2)}    &\ch{(3)}    &\ch{(4)}&\ch{(5)}&\ch{(6)} &\ch{(7)}
             &\ch{(8)}         &\ch{(9)}  &\ch{(10)}  &\ch{(11)}  &\ch{(12)}       \\
\ch{Galaxy ID} & \ch{F70$\mu$m} & \ch{F100$\mu$m}  & \ch{F160$\mu$m} & \ch{P-PACS} &
\ch{F250$\mu$m} & \ch{F350$\mu$m} & \ch{F500$\mu$m} & \ch{P-SPIRE} &
\ch{logM$_{star}$} & \ch{SFR} & \ch{log$M_{\rm gas}$} \\
\ch{(2MASX)}  & \ch{(mJy)} & \ch{(mJy)} & \ch{(mJy)} & & \ch{(mJy)} & \ch{(mJy)} & \ch{(mJy)} & &
\ch{(M$_{\odot}$)} & \ch{(M$_{\odot}$/yr)} & \ch{(M$_{\odot}$)}
}
\startdata
J15144544+0403587 &  $<$26.81 &    39.84$\pm$7.79 &  $<$69.46 & AAA &    64.47$\pm$12.30 &    29.82$\pm$7.12 &  $<$35.19 & PPP &    10.83 &     0.29 &     9.21 \\
J15144697+0403576 &  $<$17.54 &  $<$18.90 &  $<$48.40 & AAA &  $<$33.51 &  $<$30.61 &  $<$35.12 & PPP &    10.77 &   $<$0.15 &   $<$8.99 \\
J15233768+3749030 &   130.26$\pm$6.35 &   164.97$\pm$5.60 &   154.33$\pm$17.71 & AAA &   119.51$\pm$12.49 &    47.45$\pm$8.17 &  $<$53.55 & APP &    10.14 &     0.47 &     8.95 \\
J15264774+5915464 &  $<$14.01 &  $<$12.35 &  $<$15.42 & AAA &  $<$28.48 &  $<$23.89 &  $<$29.13 & PPP &    10.77 &   $<$0.14 &   $<$9.06 \\
J15281276+4255474 &  1502.20$\pm$17.05 &  3183.20$\pm$20.78 &  3633.40$\pm$41.97 & AAA &  1966.20$\pm$ 114.11 &   807.69$\pm$54.68 &   294.20$\pm$29.05 & AAA &    10.96 &     2.93 &     9.91 \\
J15281667+4256384 &    84.09$\pm$6.07 &   147.59$\pm$5.43 &   192.57$\pm$21.48 & AAA &    72.08$\pm$10.49 &  $<$58.76 &  $<$49.29 & AAA &    10.70 &     0.14 &     8.46 \\
J15523393+4620237 &   110.88$\pm$6.21 &   261.34$\pm$7.68 &   385.20$\pm$20.21 & AAA &   267.30$\pm$14.16 &   116.31$\pm$8.99 &    38.57$\pm$7.70 & PPP &    10.92 &     3.53 &    10.16 \\
J15562191+4757172 &   522.75$\pm$7.89 &   721.78$\pm$7.39 &   761.40$\pm$17.63 & AAA &   466.39$\pm$34.08 &   198.57$\pm$19.07 &    66.37$\pm$7.78 & AAP &    10.15 &     1.50 &     9.37 \\
J15583749+3227379 &   $<$8.29 &   $<$8.79 &  $<$10.26 & AAA &    78.57$\pm$18.18 &    32.05$\pm$7.55 &  $<$25.68 & PPP &    10.56 &   $<$0.12 &     9.50 \\
J15583784+3227471 &   349.87$\pm$6.76 &   627.04$\pm$7.22 &   702.98$\pm$12.78 & AAA &   361.72$\pm$20.96 &   151.29$\pm$5.88 &    41.59$\pm$7.21 & PPP &    10.89 &     4.81 &     9.89 \\
J16024254+4111499 &  1442.55$\pm$14.88 &  2267.31$\pm$12.52 &  2364.66$\pm$28.23 & MMM &  1056.54$\pm$27.49 &   470.49$\pm$34.98 &   184.42$\pm$11.57 & GPP &    10.81 &    10.66 &    10.13 \\
J16024475+4111589 &   413.45$\pm$14.88 &   691.29$\pm$12.52 &   910.24$\pm$28.23 & MMM &   380.22$\pm$16.91 &   182.97$\pm$18.21 &    67.52$\pm$9.14 & GPP &    10.48 &     2.84 &     9.71 \\
J16080559+2529091 &  $<$92.56 &   214.78$\pm$9.41 &   558.62$\pm$20.08 & MMM &   181.27$\pm$14.99 &  $<$67.98 &  $<$58.70 & PPP &    10.90 &     1.78 &     9.67 \\
J16080648+2529066 &   131.04$\pm$14.51 &   240.68$\pm$9.41 &   122.49$\pm$20.08 & MMM &   342.93$\pm$19.66 &   198.37$\pm$18.06 &  $<$56.18 & PPP &    11.16 &     2.14 &    10.08 \\
J16082261+2328459 &   138.82$\pm$9.96 &   267.16$\pm$8.47 &   669.17$\pm$33.34 & AAM &   326.25$\pm$13.67 &    99.03$\pm$12.32 &    57.11$\pm$8.58 & GPP &    10.38 &     1.82 &     9.93 \\
J16082354+2328240 &   951.20$\pm$6.74 &  1278.60$\pm$16.78 &  1109.63$\pm$33.34 & AAM &   488.88$\pm$18.52 &   263.92$\pm$11.27 &    72.67$\pm$7.70 & GPP &    10.67 &    10.37 &     9.95 \\
J16145418+3711064 &  $<$22.82 &    24.92$\pm$4.90 &  $<$22.74 & AAA &  $<$30.49 &  $<$46.44 &  $<$47.06 & PPP &    11.12 &     0.46 &   $<$9.29 \\
J16282497+4110064 &  $<$51.50 &  $<$52.55 &  $<$45.85 & AAA &  $<$38.44 &  $<$40.42 &  $<$37.49 & APP &    10.89 &   $<$0.29 &   $<$8.93 \\
J16282756+4109395 &    18.13$\pm$3.50 &    57.43$\pm$5.41 &    40.50$\pm$9.25 & AAA &  $<$38.44 &  $<$31.81 &  $<$37.19 & APP &    10.83 &     0.30 &   $<$8.90 \\
J16354293+2630494 &    25.36$\pm$5.11 &    33.53$\pm$5.36 &    59.02$\pm$12.87 & AAA &  $<$38.60 &  $<$41.14 &  $<$36.54 & PPP &    11.23 &     0.90 &   $<$9.52 \\
J16372583+4650161 &    41.58$\pm$6.93 &   107.55$\pm$6.02 &   308.13$\pm$24.89 & AAM &   140.86$\pm$9.72 &   106.33$\pm$18.11 &    53.50$\pm$9.39 & GPP &    11.26 &     1.64 &    10.20 \\
J16372754+4650054 &    69.34$\pm$7.52 &   179.90$\pm$8.12 &   205.43$\pm$24.89 & AAM &   267.81$\pm$7.56 & $<$109.99 &  $<$59.72 & GPP &    10.98 &     2.09 &    10.19 \\
J17020378+1859495 &  $<$20.26 &  $<$18.22 &  $<$23.73 & AAA &  $<$28.63 &  $<$37.35 &  $<$31.44 & PPP &    10.67 &   $<$0.31 &   $<$9.23 \\
J17045089+3448530 &   303.07$\pm$8.38 &   479.89$\pm$9.62 &   156.54$\pm$30.62 & MMM &   300.39$\pm$34.50 & $<$117.59 &  $<$66.97 & PPP &    10.74 &     7.42 &     9.88 \\
J17045097+3449020 &  1310.33$\pm$8.38 &  1628.11$\pm$9.62 &  1808.06$\pm$30.62 & MMM &   654.28$\pm$36.22 &   298.36$\pm$29.33 &    88.79$\pm$16.17 & PPP &    10.99 &    24.71 &    10.21 \\
J20471908+0019150 &  1002.60$\pm$63.17 &  1419.10$\pm$54.45 &  3087.20$\pm$74.43 & AAA &  2383.84$\pm$ 105.02 &  1214.49$\pm$76.78 &   481.05$\pm$32.56 & AAA &    11.09 &     1.83 &    10.18 \\
\enddata
\tablecomments{Descriptions of Columns: (1) Galaxy ID, taken from 2MASS. (2) Herschel PACS
70$\mu$m flux (mJy). (3) Herschel PACS 100$\mu$m flux (mJy). (4) Herschel PACS 160$\mu$m flux
(mJy). (5) Herschel PACS photometric methods, 'A': aperture, 'M': model fitting, 'C': compact.
(6) Herschel SPIRE 250$\mu$m flux (mJy). (7) Herschel SPIRE 350$\mu$m flux (mJy). (8) Herschel
SPIRE 500$\mu$m flux (mJy). (9) Herschel SPIRE photometric methods, 'A': aperture, 'P': PSF fitting,
'G': Gaussian fitting. (10) Stellar mass (log(M$_{\odot}$)). (11) Star formation rate (M$_{\odot}$/yr).
(12) Total gas mass (log(M$_{\odot}$)).}
\end{deluxetable*}
\end{turnpage}}

{\begin{turnpage}
\begin{deluxetable*}{cccccccccccc}
\label{tbl:hkpair_table3}
\tabletypesize{\scriptsize}
\setlength{\tabcolsep}{0.05in} 
\tablenum{3}
\tablewidth{0.95\linewidth}
\tablecaption{Herschel 6-band fluxes and physical parameters of H-KPAIR ellipticals}
\tablehead{
\ch{(1)}   &\ch{(2)}    &\ch{(3)}    &\ch{(4)}&\ch{(5)}&\ch{(6)} &\ch{(7)}
             &\ch{(8)}         &\ch{(9)}  &\ch{(10)}  &\ch{(11)}  &\ch{(12)}       \\
\ch{Galaxy ID} & \ch{F70$\mu$m} & \ch{F100$\mu$m}  & \ch{F160$\mu$m} & \ch{P-PACS} &
\ch{F250$\mu$m} & \ch{F350$\mu$m} & \ch{F500$\mu$m} & \ch{P-SPIRE} &
\ch{logM$_{star}$} & \ch{SFR} & \ch{log$ M_{\rm gas}$} \\
\ch{(2MASX)}  & \ch{(mJy)} & \ch{(mJy)} & \ch{(mJy)} & & \ch{(mJy)} & \ch{(mJy)} & \ch{(mJy)} & &
\ch{(M$_{\odot}$)} & \ch{(M$_{\odot}$/yr)} & \ch{(M$_{\odot}$)}
}
\startdata
J00202748+0050009 &  $<$25.37  &    32.41$\pm$7.31 &    91.59$\pm$12.06 & AAA &    55.55$\pm$10.24 &  $<$48.28  &  $<$46.92  & APP &    10.75 &     0.06 &     8.62 \\
J03381299+0109414 &   326.73$\pm$12.60 &   647.89$\pm$9.68 &   837.69$\pm$22.90 & AAA &  $<$32.20  &  $<$67.56  &  $<$62.95  & APP &    10.55 &     4.85 &   $<$9.03 \\
J08083563+3854522 &  $<$18.58  &  $<$16.62  &  $<$18.30  & AAA &  $<$29.87  &  $<$37.62  &  $<$51.16  & PPP &    10.89 &   $<$0.15 &   $<$8.99 \\
J08385973+3613164 &  $<$59.26  &    31.69$\pm$5.30 &  $<$29.82  & AAA &  $<$41.85  &  $<$47.14  &  $<$37.81  & PPP &    11.12 &     0.53 &   $<$9.37 \\
J08415054+2642475 &  $<$12.21  &  $<$14.45  &  $<$12.85  & AAA &  $<$34.25  &  $<$20.46  &  $<$29.21  & PPP &    11.13 &   $<$0.62 &   $<$9.64 \\
J09060283+5144411 &  $<$20.84  &  $<$21.48  &  $<$21.92  & AAA &  $<$33.19  &  $<$24.55  &  $<$51.83  & AAP &    10.70 &   $<$0.10 &   $<$8.80 \\
J09123636+3547180 &  $<$20.22  &  $<$18.51  &  $<$21.32  & AAA &  $<$31.78  &  $<$45.99  &  $<$44.77  & PPP &    10.56 &   $<$0.06 &   $<$8.61 \\
J09134461+4742165 &  $<$10.53  &    17.28$\pm$3.35 &  $<$25.01  & AAA &    64.80$\pm$11.55 &    44.85$\pm$8.38 &  $<$32.64  & GGP &    11.10 &     0.25 &     9.46 \\
J09374506+0244504 &  $<$49.45  &  $<$57.59  &  $<$65.53  & AAA &   129.50$\pm$15.27 &   107.02$\pm$16.36 &    80.00$\pm$15.11 & AAA &    10.96 &   $<$0.15 &     9.09 \\
J10155338+0657495 &    25.62$\pm$4.03 &    33.36$\pm$4.06 &    26.67$\pm$5.34 & AAA &  $<$61.30  &  $<$50.34  &  $<$57.53  & PPP &    10.81 &     0.14 &   $<$8.98 \\
J10205369+4831246 &  $<$18.30  &    20.07$\pm$3.82 &    48.91$\pm$6.01 & AAA &    46.45$\pm$10.62 &  $<$33.10  &  $<$36.51  & PPP &    10.99 &     0.31 &     9.37 \\
J10272970+0115170 &  $<$46.39  &  $<$45.08  &  $<$58.29  & AAA &  $<$76.12  &  $<$42.78  &  $<$49.04  & PPP &    10.76 &   $<$0.12 &   $<$8.88 \\
J10325321+5306477 &  $<$17.50  &  $<$16.11  &  $<$20.65  & AAA &    50.93$\pm$10.39 &  $<$35.66  &  $<$36.89  & PPP &    11.20 &   $<$0.37 &     9.55 \\
J10364400+5447489 &  $<$15.18  &    27.10$\pm$4.76 &    57.72$\pm$11.39 & AAA &  $<$30.85  &  $<$23.60  &  $<$41.35  & PPP &    11.12 &     0.31 &   $<$9.11 \\
J10392515+3904573 &  $<$14.47  &  $<$16.22  &  $<$21.51  & AAA &  $<$38.05  &  $<$21.70  &  $<$33.19  & PPP &    10.79 &   $<$0.16 &   $<$9.13 \\
J10452496+3909499 &  $<$28.51  &  $<$38.90  &  $<$34.46  & AAA &  $<$35.64  &  $<$36.29  &  $<$31.78  & APP &    10.68 &   $<$0.13 &   $<$8.71 \\
J10514368+5101195 &    47.68$\pm$6.11 &    99.25$\pm$6.01 &   138.40$\pm$17.17 & AAA &    55.87$\pm$10.04 &  $<$33.47  &  $<$29.42  & PPP &    11.12 &     0.18 &     8.71 \\
J10595869+0857215 &  $<$14.47  &  $<$14.63  &  $<$20.26  & AAA &  $<$47.95  &  $<$27.39  &  $<$44.41  & PPP &    11.23 &   $<$0.32 &   $<$9.50 \\
J11014357+5720058 &   $<$9.04  & $<$178.93  &  $<$24.27  & AAA &  $<$55.69  &  $<$66.14  &  $<$85.02  & PPP &    10.80 &   $<$1.93 &   $<$9.34 \\
J11375476+4727588 &  $<$24.93  &  $<$26.84  &  $<$72.58  & AAA &  $<$37.68  &  $<$30.10  &  $<$45.17  & PPP &    10.99 &   $<$0.17 &   $<$8.95 \\
J11440335+3332062 &    89.03$\pm$6.78 &   192.15$\pm$7.59 &   193.80$\pm$8.79 & AAA &   105.13$\pm$11.00 &  $<$62.13  &  $<$34.15  & PPP &    10.73 &     0.48 &     9.04 \\
J11505844+1444124 &  $<$20.46  &  $<$19.35  &  $<$22.64  & AAA &  $<$25.34  &  $<$28.65  &  $<$33.72  & PPP &    11.24 &   $<$0.32 &   $<$9.18 \\
J11542307+4932456 &  $<$14.52  &  $<$15.08  &  $<$16.62  & AAA &  $<$33.47  &  $<$33.82  &  $<$32.59  & PPP &    11.35 &   $<$0.43 &   $<$9.48 \\
J12020537+5342487 &  $<$14.04  &  $<$12.79  &  $<$13.48  & AAA &  $<$34.01  &  $<$33.20  &  $<$43.16  & PPP &    11.09 &   $<$0.30 &   $<$9.40 \\
J12054073+0134302 &  $<$16.63  &  $<$20.56  &  $<$17.76  & AAA &  $<$43.02  &  $<$36.57  &  $<$39.41  & AAA &    10.52 &   $<$0.04 &   $<$8.57 \\
J12191719+1200582 &  $<$24.65  &  $<$25.10  &  $<$33.60  & AAA &  $<$35.82  &  $<$39.65  &  $<$28.11  & PPP &    10.53 &   $<$0.09 &   $<$8.73 \\
J12433936+4406046 &  $<$14.76  &  $<$15.11  &  $<$16.20  & AAA &  $<$46.90  &  $<$38.94  &  $<$27.56  & PPP &    10.88 &   $<$0.14 &   $<$9.16 \\
J12525212+4645294 &    47.54$\pm$4.72 &    59.27$\pm$5.71 &    80.75$\pm$8.80 & AAA &    47.91$\pm$10.15 &  $<$29.37  &  $<$50.12  & PPP &    11.14 &     1.47 &     9.47 \\
J13131429+3910360 &  $<$23.14  &  $<$24.15  &  $<$67.02  & AAA &  $<$24.10  &  $<$33.87  &  $<$28.42  & PPP &    11.26 &   $<$0.68 &   $<$9.36 \\
J13462215-0325057 &    30.10$\pm$4.24 &    54.45$\pm$5.69 &    65.60$\pm$7.45 & AAA &  $<$37.54  &  $<$27.94  &  $<$27.52  & AAP &    10.53 &     0.16 &   $<$8.68 \\
J14055334+6542277 &    26.07$\pm$4.78 &  $<$20.03  &  $<$31.41  & AAA &  $<$46.59  &  $<$26.44  &  $<$37.84  & APP &    10.62 &   $<$0.10 &   $<$8.96 \\
J14064127+5043239 & $<$204.35  & $<$138.04  & $<$124.38  & AAA &  $<$45.17  &  $<$51.68  &  $<$70.05  & AAA &    10.30 &   $<$0.04 &   $<$7.85 \\
J14070720-0234402 &  $<$22.89  &  $<$24.17  &  $<$20.88  & AAA &  $<$53.97  &  $<$36.13  &  $<$35.11  & PPP &    10.76 &   $<$0.41 &   $<$9.46 \\
J14250552+0313590 &    59.56$\pm$6.02 &   101.37$\pm$7.03 &    97.44$\pm$14.59 & AAA &    61.35$\pm$9.52 &    37.92$\pm$7.03 &  $<$35.97  & PPP &    10.75 &     0.58 &     9.14 \\
J15002374+4316559 &    37.39$\pm$7.52 &    19.76$\pm$3.59 &  $<$23.67  & AAA &  $<$32.89  &  $<$36.62  &  $<$45.18  & PPP &    10.88 &     0.10 &   $<$8.82 \\
J15053183+3427526 &  $<$13.13  &  $<$14.13  &  $<$12.73  & AAA &  $<$34.70  &  $<$33.02  &  $<$31.14  & PPP &    11.24 &   $<$0.43 &   $<$9.52 \\
J15233899+3748254 &  $<$20.50  &  $<$18.09  &  $<$24.72  & AAA &    38.32$\pm$9.46 &  $<$36.84  &  $<$38.41  & APP &    10.19 &   $<$0.05 &     8.67 \\
J15264892+5915478 &  $<$14.01  &  $<$12.35  &  $<$15.42  & AAA &  $<$25.85  &  $<$25.94  &  $<$28.87  & PPP &    10.91 &   $<$0.14 &   $<$9.04 \\
J15523258+4620180 &  $<$15.23  &  $<$14.07  &  $<$17.84  & AAA &  $<$35.36  &  $<$40.27  &  $<$32.18  & PPP &    11.17 &   $<$0.28 &   $<$9.36 \\
J15562738+4757302 &  $<$16.61  &  $<$15.53  &  $<$16.63  & AAA &  $<$40.91  &  $<$47.06  &  $<$39.00  & AAP &    10.17 &   $<$0.03 &   $<$8.58 \\
J16145421+3711136 &  $<$22.82  &  $<$19.59  &  $<$22.74  & AAA &  $<$31.27  &  $<$43.27  &  $<$50.60  & PPP &    11.16 &   $<$0.36 &   $<$9.30 \\
J16354366+2630505 &  $<$14.64  &  $<$15.37  &  $<$16.04  & AAA &  $<$40.94  &  $<$40.90  &  $<$41.63  & PPP &    11.27 &   $<$0.44 &   $<$9.55 \\
J17020320+1900006 &  $<$20.26  &  $<$18.22  &    55.42$\pm$13.02 & AAA &  $<$28.45  &  $<$38.78  &  $<$31.68  & PPP &    11.00 &   $<$0.33 &   $<$9.25 \\
J20472428+0018030 &  $<$61.19  &  $<$54.39  &  $<$61.46  & AAA &  $<$48.59  &  $<$51.33  &  $<$52.71  & AAA &    10.74 &   $<$0.05 &   $<$8.30 \\
\enddata
\tablecomments{Descriptions of Columns: (1) Galaxy ID, taken from 2MASS. (2) Herschel PACS
70$\mu$m flux (mJy). (3) Herschel PACS 100$\mu$m flux (mJy). (4) Herschel PACS 160$\mu$m flux
(mJy). (5) Herschel PACS photometric methods, 'A': aperture, 'M': model fitting, 'C': compact.
(6) Herschel SPIRE 250$\mu$m flux (mJy). (7) Herschel SPIRE 350$\mu$m flux (mJy). (8) Herschel
SPIRE 500$\mu$m flux (mJy). (9) Herschel SPIRE photometric methods, 'A': aperture, 'P': PSF fitting,
'G': Gaussian fitting. (10) Stellar mass (log(M$_{\odot}$)). (11) Star formation rate (M$_{\odot}$/yr).
(12) Total gas mass (log(M$_{\odot}$)).}
\end{deluxetable*}
\end{turnpage}}

\citet{Chang2015} recently published stellar masses and SFRs for
1M galaxies from SDSS$+$WISE, estimated based on 0.4 to 22$\mu$m
SEDs using MAGPHYS \citep{daCunha2008} energy balance SED fitting
technique. 85 spirals in pairs and all 132 spirals in controls in
our samples are also included in Chang et al.'s work. Comparisons
on stellar masses and SFRs estimated in our work and in \citet{Chang2015}
are shown in Figure~\ref{fig:comp_mstar} $\&$ Figure~\ref{fig:comp_SFR},
respectively.

From the comparisons, we found for most of the spirals with higher stellar
masses, using 2MASS Ks band luminosities (our work) and SED model fitting
\citep{Chang2015} will results in very similar M$_{star}$ values. While only
for spirals with the lowest stellar masses, \citet{Chang2015} estimated lower
M$_{star}$ ($\sim$0.3dex) than ours. For comparison on SFRs, \citet{Chang2015}'s
values are systematically lower than ours, especially for spirals in the control
sample. This may be caused by: (1) in our work, larger beams of Herschel FIR-to-submm
bands resulting in more background contaminations, (2) in \citet{Chang2015},
under-estimation of dust obscurations for optical-to-MIR bands using in MAGPHYS
SED fittings. (3) Due to the fact that the sizes of our local paired and control
galaxies are generally much larger than the PSFs of SDSS and WISE, therefore,
using MODELFLUX for SDSS 5-bands and PSF model flux for WISE 4-bands (although
applied a correction using R$_{e}$) may still cause some flux loss in \citet{Chang2015},
especially for the estimations of SFRs using WISE bands.

\begin{figure}[!htb]
\includegraphics[width=0.45\textwidth,angle=90]{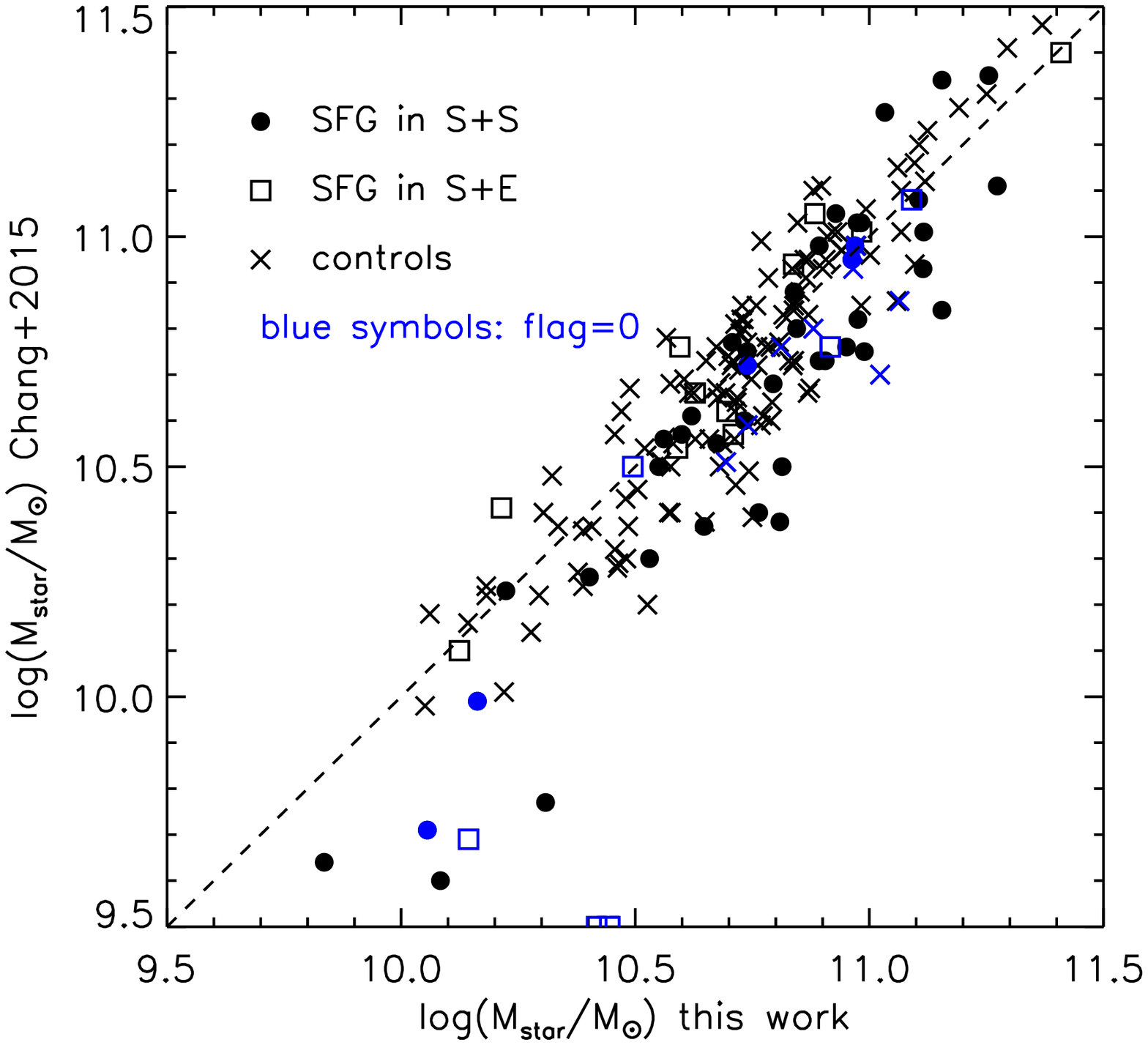}
\caption{Comparison of stellar masses estimated in our work and in
\citet{Chang2015}. Star-forming spirals in S$+$S pairs are shown
as black dots, those in S$+$E pairs are shown as open squares, and
SFGs in the control sample are shown as crosses.
}
\label{fig:comp_mstar}
\setcounter{figure}{5}
\end{figure}

\begin{figure}[!htb]
\includegraphics[width=0.45\textwidth,angle=90]{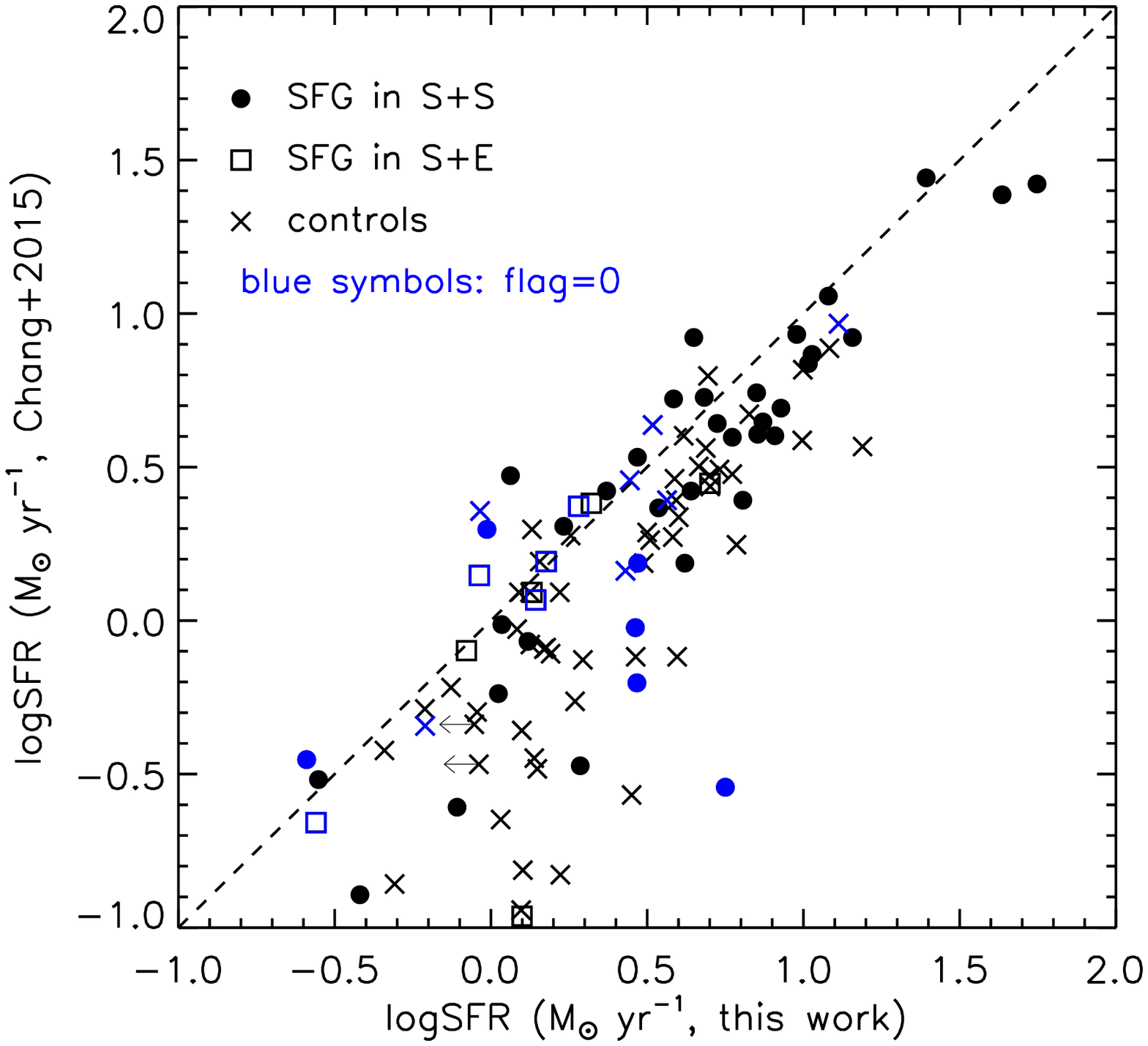}
\caption{Comparison of SFRs estimated in our work and in
\citet{Chang2015}. Star-forming spirals in S$+$S pairs are shown
as black dots, those in S$+$E pairs are shown as open squares, and
SFGs in the control sample are shown as crosses.
}
\label{fig:comp_SFR}
\setcounter{figure}{6}
\end{figure}

\section{Control sample}\label{section:Control}
The control sample was selected among 2MASS galaxies of redshift
$z <0.1$ (from SDSS), found in four Level-5 HerMES fields:
Bootes HerMES, EGS HerMES, ELAIS N1 HerMES, and Lockman SWIRE.
Details of the Herschel PACS and SPIRE observations in
these fields can be found in \citet{Oliver2012}.
Morphologies of these galaxies (S or E) were obtained from visual
classifications by Galaxy Zoo (data release v1,
\citealt{Lintott2011}). For those galaxies labeled as 'UNCERTAIN' in
the Galaxy Zoo catalog, two of us (C.C. $\&$ C.K.X) visually re-classified
them into S or E categories using SDSS optical images. For those
galaxies with controversial classifications between us, we used an
automatic classification algorithm as described in \citet{Xu2010}
(u$-$r color and R$_{50}$/R$_{90}$ ratio) as the third party.
Then close pairs (have companions with
projected distance $<$ 70kpc and dM$_{star}$ $<$ 0.4 dex), peculiars
(visually selected by C.C. $\&$ C.K.X.), and objects at the edge of
PACS or SPIRE images (with low coverage) were rejected from the
sample. The resulting parent control sample was then 1-to-1 matched
with H-KPAIRs. The matched control galaxies have:
(1) the same morphology (S or E);
(2) similar stellar mass: $\delta M_{\rm star} < 0.1$ dex except for the
match of J13082737+0422125, which is $\delta M_{\rm star} = 0.15$;
(3) with the closest z to that of corresponding H-KPAIR galaxy.

The Herschel PACS images at 100 $\&$ 160$\mu$m (the 70$\mu$m band was
not included in the HerMES survey) and SPIRE images at 250, 350, 500$\mu$m
were taken from the HerMES data release (v2) for the 176 control sample
galaxies. PACS $\&$ SPIRE aperture photometries were made using IDL/phot
and circular apertures. Fixed circular aperture sizes were used for most
galaxies, with aperture radii of 12", 18" for PACS 100 $\&$ 160$\mu$m,
24", 33.33", 48" for SPIRE 250, 350, 500$\mu$m, respectively. Larger
apertures were used for a few extended galaxies. The background and error
estimates are similar to those in the aperture photometry of H-KPAIR
galaxies (Section~3). The total dust masses, SFRs and stellar masses of
control sample galaxies were calculated using the same method as for
the H-KPAIR galaxies (Section~4). In Table~4 we listed the galaxy ID (name),
RA $\&$ Dec, redshifts, the Herschel fluxes, stellar mass M$_{star}$,
the SFR, and total gas masses $M_{\rm gas}$ ($M_{\rm gas} = 100 \times M_{\rm dust}$)
of spiral galaxies in the control sample.

It should be noted in the following comparisonal analysis with the
control sample, each galaxy in the pairs of our H-KPAIR sample was
included independently, rather than the pair as whole, which would
introduce a bias.

{\begin{turnpage}
\begin{deluxetable*}{ccccccccccccc}
\centering
\label{tbl:hkpair_table4}
\tabletypesize{\scriptsize}
\setlength{\tabcolsep}{0.005in} 
\tablenum{4}
\tablewidth{0.999\linewidth}
\tablecaption{Herschel 5-band fluxes and physical parameters of spirals in the control sample}
\tablehead{
\ch{(1)}   &\ch{(2)}    &\ch{(3)}    &\ch{(4)}&\ch{(5)}&\ch{(6)} &\ch{(7)}
             &\ch{(8)}         &\ch{(9)}  &\ch{(10)}  &\ch{(11)}  &\ch{(12)} &\ch{(13)}       \\
\ch{Galaxy ID} & \ch{RA} & \ch{Dec} & \ch{z} & \ch{F100$\mu$m}  & \ch{F160$\mu$m} &
\ch{F250$\mu$m} & \ch{F350$\mu$m} & \ch{F500$\mu$m} &
\ch{logM$_{star}$} & \ch{SFR} & \ch{log$M_{\rm gas}$} & \ch{Galaxy ID} \\
\ch{(Control)} & \ch{(J2000)} & \ch{(J2000)} & \ch{redshift} &
\ch{(mJy)} & \ch{(mJy)} & \ch{(mJy)} & \ch{(mJy)} & \ch{(mJy)} &
\ch{(logM$_{\odot}$)} & \ch{(M$_{\odot}$/yr)} & \ch{(logM$_{\odot}$)} & \ch{(Pair)}
}
\startdata
lk-096   & 164.74153 & 58.133747 & 0.0319167 &   146.43$\pm$7.43 &   253.13$\pm$9.49 &   168.30$\pm$11.17 &    90.81$\pm$15.02 &  $<$65.03  &    10.58 &     0.49 &     9.61 & J00202580+0049350 \\
lk-416   & 158.55316 & 57.079830 & 0.0482240 &   113.78$\pm$6.87 &   244.52$\pm$10.39 &   212.03$\pm$16.14 &   107.13$\pm$22.57 & $<$104.06  &    10.88 &     1.17 &    10.20 & J01183417-0013416 \\
lk-075   & 158.30994 & 56.741673 & 0.0468842 &  $<$43.80  &  $<$53.66  & $<$107.36  & $<$100.03  & $<$135.08  &    10.71 &   $<$0.50 &   $<$9.57 & J01183556-0013594 \\
egs-009  & 213.96425 & 51.856659 & 0.0456582 &   171.32$\pm$6.56 &   291.75$\pm$7.25 &   156.07$\pm$12.94 &    79.33$\pm$13.09 &  $<$49.66  &    10.46 &     1.23 &     9.76 & J02110638-0039191 \\
en1-045  & 242.20769 & 53.997375 & 0.0627264 &    34.02$\pm$7.18 &    56.46$\pm$8.34 &    60.07$\pm$10.28 &  $<$67.36  &  $<$58.72  &    10.58 &     0.72 &     9.59 & J02110832-0039171 \\
lk-244   & 161.47475 & 56.506535 & 0.0461360 &    83.70$\pm$8.18 &   161.28$\pm$8.62 &   112.31$\pm$9.62 &    49.64$\pm$12.35 &  $<$54.88  &    10.73 &     0.90 &     9.58 & J03381222+0110088 \\
lk-108   & 159.86114 & 57.662411 & 0.0719180 &  $<$26.53  &  $<$36.79  &  $<$43.59  &  $<$86.05  &  $<$52.88  &    10.86 &   $<$0.76 &   $<$9.59 & J07543194+1648214 \\
lk-368   & 163.04863 & 58.438011 & 0.0315815 &  1061.65$\pm$14.82 &  1822.36$\pm$13.97 &  1049.99$\pm$13.65 &   500.07$\pm$9.39 &   194.86$\pm$12.86 &    11.06 &     3.29 &    10.28 & J07543221+1648349 \\
en1-016  & 242.03619 & 53.878147 & 0.0641164 &   135.56$\pm$6.36 &   238.76$\pm$7.81 &   118.15$\pm$12.24 &    60.87$\pm$12.12 &  $<$90.64  &    10.69 &     1.80 &     9.92 & J08083377+3854534 \\
lk-290   & 165.13409 & 57.497932 & 0.0275216 &  $<$26.97  &  $<$41.11  &  $<$53.30  &  $<$41.20  &  $<$84.54  &    10.18 &   $<$0.11 &   $<$8.91 & J08233266+2120171 \\
lk-148   & 164.15407 & 59.215527 & 0.0325229 &    40.95$\pm$7.85 &   114.71$\pm$12.27 &  $<$60.81  &    66.83$\pm$11.68 &  $<$64.11  &    10.38 &     0.22 &   $<$9.09 & J08233421+2120515 \\
lk-390   & 160.60169 & 58.458820 & 0.0451383 &  $<$25.89  &    65.22$\pm$8.39 &  $<$36.63  &  $<$55.28  &  $<$57.79  &    10.65 &   $<$0.28 &   $<$9.16 & J08291491+5531227 \\
lk-248   & 162.23958 & 56.620098 & 0.0468625 &    79.00$\pm$7.22 &   175.56$\pm$11.18 &   163.64$\pm$13.32 &   102.59$\pm$14.25 &  $<$66.15  &    10.71 &     1.01 &    10.26 & J08292083+5531081 \\
lk-282   & 164.20754 & 57.412735 & 0.0468024 &    37.15$\pm$7.00 &  $<$38.31  &  $<$61.42  &  $<$73.00  &  $<$51.33  &    10.91 &     0.43 &   $<$9.37 & J08364482+4722188 \\
lk-171   & 163.19229 & 60.094456 & 0.0697656 &   289.83$\pm$7.96 &   463.92$\pm$11.13 &   279.43$\pm$28.53 &    97.37$\pm$20.75 & $<$126.28  &    11.06 &     4.95 &    10.24 & J08364588+4722100 \\
lk-405   & 159.23137 & 58.560928 & 0.0715059 &  $<$31.49  &    78.56$\pm$8.87 &  $<$54.77  &  $<$63.28  &  $<$55.17  &    10.71 &   $<$0.88 &   $<$9.66 & J08381759+3054534 \\
lk-271   & 161.58758 & 56.765396 & 0.0672263 &   142.12$\pm$7.21 &   177.08$\pm$8.76 &   145.22$\pm$9.67 &    89.74$\pm$9.71 &  $<$53.52  &    10.94 &     3.52 &    10.32 & J08381795+3055011 \\
lk-009   & 163.78018 & 59.654129 & 0.0451426 &   858.23$\pm$13.86 &   948.72$\pm$17.93 &   531.34$\pm$30.12 &   185.74$\pm$36.85 & $<$156.50  &    10.81 &     5.39 &    10.03 & J08390125+3613042 \\
btsh-212 & 218.07066 & 33.590675 & 0.0847390 &   101.22$\pm$7.29 &   280.92$\pm$8.35 &   214.35$\pm$12.92 &   112.67$\pm$12.23 &  $<$54.57  &    11.46 &     3.22 &    10.65 & J08414959+2642578 \\
lk-327   & 158.39201 & 57.133083 & 0.0466682 &   123.93$\pm$7.39 &   260.04$\pm$9.79 &   143.05$\pm$21.06 & $<$112.46  & $<$120.05  &    10.66 &     1.34 &     9.67 & J09060498+5144071 \\
btsh-111 & 218.37558 & 34.561275 & 0.0287766 &   133.23$\pm$6.08 &   179.24$\pm$9.26 &   139.30$\pm$8.88 &  $<$65.03  &  $<$60.62  &    10.34 &     0.54 &     9.29 & J09123676+3547462 \\
egs-043  & 214.41007 & 52.693192 & 0.0631147 &  $<$24.61  &    34.02$\pm$7.12 &  $<$39.34  &  $<$47.90  &  $<$56.15  &    10.97 &   $<$0.54 &   $<$9.45 & J09134606+4742001 \\
lk-276   & 162.26814 & 56.911495 & 0.0718246 &  $<$25.25  &  $<$35.21  &    57.55$\pm$13.64 &    88.90$\pm$11.89 &    83.98$\pm$16.52 &    10.73 &   $<$0.72 &     9.69 & J09155467+4419510 \\
btsh-146 & 218.82675 & 35.118813 & 0.0284341 &  5395.36$\pm$35.44 &  5274.42$\pm$52.53 &  1902.35$\pm$32.67 &   725.94$\pm$33.25 &   254.42$\pm$24.96 &    11.02 &    12.96 &    10.05 & J09155552+4419580 \\
lk-056   & 161.98213 & 57.188320 & 0.0898139 &  $<$26.02  &  $<$45.15  &  $<$42.39  &  $<$48.52  &  $<$58.04  &    11.00 &   $<$1.20 &   $<$9.76 & J09264111+0447247 \\
lk-175   & 158.62358 & 59.784828 & 0.0905889 &   122.43$\pm$15.91 &   109.51$\pm$23.54 &   153.42$\pm$24.42 &   111.23$\pm$23.32 & $<$138.69  &    11.37 &     5.27 &    10.78 & J09264137+0447260 \\
en1-018  & 242.25880 & 53.743439 & 0.0643115 &    43.30$\pm$7.09 &    44.96$\pm$9.26 &  $<$76.72  &  $<$54.51  &  $<$73.00  &    11.12 &     0.96 &   $<$9.70 & J09374413+0245394 \\
en1-048  & 242.60513 & 53.779949 & 0.0658474 &   148.85$\pm$6.95 &   251.71$\pm$9.55 &   197.29$\pm$17.72 &   107.12$\pm$16.14 &  $<$58.32  &    10.79 &     3.87 &    10.42 & J10100079+5440198 \\
lk-150   & 164.46387 & 59.365425 & 0.0461592 &   786.03$\pm$11.98 &   863.67$\pm$21.83 &   423.96$\pm$19.93 &   200.22$\pm$21.31 & $<$103.25  &    10.77 &     6.11 &    10.01 & J10100212+5440279 \\
lk-256   & 162.47282 & 56.647766 & 0.0459531 &   103.97$\pm$6.84 &   181.74$\pm$10.33 &   112.72$\pm$9.52 &    51.06$\pm$10.78 &  $<$59.50  &    10.48 &     0.75 &     9.64 & J10155257+0657330 \\
btsh-148 & 218.99033 & 35.172455 & 0.0540843 &    74.98$\pm$11.19 &   120.83$\pm$11.79 &    85.83$\pm$11.99 &  $<$61.30  &  $<$53.69  &    10.52 &     1.12 &     9.60 & J10205188+4831096 \\
lk-323   & 158.20479 & 56.945797 & 0.0569093 &  $<$54.04  &   122.49$\pm$15.06 & $<$105.94  & $<$122.96  & $<$111.23  &    10.61 &   $<$0.91 &   $<$9.72 & J10225647+3446564 \\
lk-155   & 158.16884 & 58.766632 & 0.0735668 &  $<$28.58  &  $<$40.19  &  $<$57.93  &  $<$62.76  &  $<$55.66  &    10.93 &   $<$0.86 &   $<$9.71 & J10225655+3446468 \\
lk-014   & 161.36885 & 59.496750 & 0.0724664 &    42.32$\pm$7.49 &    97.26$\pm$9.76 &    75.35$\pm$8.81 &    54.89$\pm$12.54 &  $<$65.91  &    10.75 &     1.25 &    10.17 & J10233658+4220477 \\
lk-259   & 163.53919 & 56.821007 & 0.0464892 &   444.33$\pm$6.80 &   545.14$\pm$10.77 &   293.35$\pm$13.75 &   132.02$\pm$16.50 &    62.02$\pm$12.72 &    10.65 &     4.87 &     9.98 & J10233684+4221037 \\
lk-049   & 158.95834 & 56.568207 & 0.0437173 &    88.81$\pm$7.24 &   213.97$\pm$9.91 &   133.30$\pm$17.40 &    99.05$\pm$18.54 &  $<$88.23  &    10.50 &     0.64 &     9.95 & J10272950+0114490 \\
lk-203   & 162.58281 & 56.347740 & 0.0683074 &  $<$32.69  &    56.70$\pm$12.84 &   104.47$\pm$23.80 &    83.08$\pm$13.14 &  $<$50.25  &    10.92 &   $<$0.83 &     9.86 & J10325316+5306536 \\
lk-082   & 159.70738 & 57.004242 & 0.0725609 &   100.16$\pm$7.13 &   145.67$\pm$9.50 &    74.33$\pm$8.73 &  $<$57.07  &  $<$66.15  &    11.07 &     2.74 &     9.79 & J10332972+4404342 \\
en1-046  & 242.37892 & 53.863384 & 0.0623371 &    92.44$\pm$7.53 &   150.81$\pm$9.32 & $<$109.83  &  $<$63.50  &  $<$88.18  &    10.84 &     1.84 &   $<$9.81 & J10333162+4404212 \\
lk-311   & 163.57295 & 57.726276 & 0.0751972 &  $<$30.52  &   102.61$\pm$10.27 &   109.41$\pm$12.00 &    74.62$\pm$16.78 &    58.92$\pm$14.05 &    10.85 &     1.21 &    10.75 & J10364274+5447356 \\
lk-048   & 158.79388 & 56.524883 & 0.0648007 &  $<$35.42  &  $<$46.16  &  $<$97.36  & $<$135.89  & $<$100.23  &    10.72 &   $<$0.80 &   $<$9.79 & J10392338+3904501 \\
lk-210   & 164.78847 & 56.918480 & 0.0467608 &   116.47$\pm$9.11 &   125.71$\pm$11.82 &   104.12$\pm$10.38 &  $<$59.94  &  $<$56.94  &    10.46 &     1.27 &     9.56 & J10435053+0645466 \\
lk-006   & 161.48135 & 59.154591 & 0.0443921 &   115.51$\pm$7.66 &   144.93$\pm$9.13 &   140.13$\pm$21.92 &    97.94$\pm$23.19 &  $<$71.20  &    10.46 &     1.26 &    10.04 & J10435268+0645256 \\
lk-294   & 158.20099 & 56.595795 & 0.0468146 &   125.28$\pm$13.13 &   136.48$\pm$23.07 &   129.06$\pm$24.26 & $<$125.81  & $<$133.61  &    10.72 &     1.36 &     9.64 & J10452478+3910298 \\
lk-355   & 159.07225 & 57.611805 & 0.0734656 &    52.54$\pm$7.38 &    93.05$\pm$9.02 &  $<$70.37  &  $<$47.40  &  $<$54.86  &    10.71 &     1.52 &   $<$9.78 & J10514450+5101303 \\
egs-017  & 213.85736 & 52.332111 & 0.0738315 &  $<$22.04  &  $<$30.92  &  $<$44.65  &  $<$41.72  &  $<$61.61  &    10.84 &   $<$0.67 &   $<$9.62 & J10595915+0857357 \\
lk-217   & 164.84148 & 56.605415 & 0.0478231 &   437.11$\pm$15.59 &   288.91$\pm$20.60 &   180.96$\pm$17.65 &    79.12$\pm$16.90 &  $<$84.74  &    10.53 &     5.01 &     9.53 & J11014364+5720336 \\
lk-379   & 159.88892 & 58.362122 & 0.0711125 &    59.99$\pm$7.35 &    47.98$\pm$8.99 &  $<$78.01  &  $<$72.30  &  $<$51.81  &    10.99 &     1.61 &   $<$9.79 & J11064944+4751119 \\
lk-357   & 159.91113 & 57.933495 & 0.0743553 &  $<$32.79  &    61.94$\pm$8.95 &    96.29$\pm$10.65 &    60.19$\pm$12.21 &    62.14$\pm$11.12 &    11.12 &     0.76 &    10.73 & J11065068+4751090 \\
lk-030   & 162.42171 & 60.293438 & 0.0444752 &   130.94$\pm$7.59 &   231.78$\pm$12.74 &   192.14$\pm$12.74 &   118.29$\pm$13.71 &  $<$59.04  &    10.56 &     1.50 &    10.24 & J11204657+0028142 \\
lk-273   & 161.65997 & 56.893730 & 0.0739821 &  $<$28.15  &  $<$41.51  &  $<$39.40  &  $<$70.40  &  $<$62.29  &    10.74 &   $<$0.85 &   $<$9.57 & J11204801+0028068 \\
lk-235   & 163.38641 & 56.303272 & 0.0719963 &  $<$66.31  &  $<$78.02  &  $<$53.87  &  $<$54.18  &  $<$63.11  &    10.63 &   $<$1.82 &   $<$9.66 & J11251704+0227007 \\
lk-403   & 159.10716 & 58.556229 & 0.0271570 &   448.17$\pm$14.58 &   844.33$\pm$19.78 &   562.00$\pm$34.14 &   335.35$\pm$44.40 &   223.49$\pm$32.19 &    10.88 &     1.87 &    10.33 & J11251716+0226488 \\
\enddata
\end{deluxetable*}

\begin{deluxetable*}{ccccccccccccc}
\centering
\label{tbl:hkpair_table4}
\tabletypesize{\scriptsize}
\setlength{\tabcolsep}{0.005in} 
\tablenum{4}
\tablewidth{0.999\linewidth}
\tablecaption{(Continued)}
\tablehead{
\ch{(1)}   &\ch{(2)}    &\ch{(3)}    &\ch{(4)}&\ch{(5)}&\ch{(6)} &\ch{(7)}
             &\ch{(8)}         &\ch{(9)}  &\ch{(10)}  &\ch{(11)}  &\ch{(12)} &\ch{(13)}       \\
\ch{Galaxy ID} & \ch{RA} & \ch{Dec} & \ch{z} & \ch{F100$\mu$m}  & \ch{F160$\mu$m} &
\ch{F250$\mu$m} & \ch{F350$\mu$m} & \ch{F500$\mu$m} &
\ch{logM$_{star}$} & \ch{SFR} & \ch{log$ M_{\rm gas}$} & \ch{Galaxy ID} \\
\ch{(Control)} & \ch{(J2000)} & \ch{(J2000)} & \ch{redshift} &
\ch{(mJy)} & \ch{(mJy)} & \ch{(mJy)} & \ch{(mJy)} & \ch{(mJy)} &
\ch{(logM$_{\odot}$)} & \ch{(M$_{\odot}$/yr)} & \ch{(logM$_{\odot}$)} & \ch{(Pair)}
}
\startdata
lk-200   & 162.21944 & 56.336193 & 0.0457280 &   272.88$\pm$8.58 &   445.66$\pm$12.10 &   232.15$\pm$10.65 &   120.48$\pm$13.54 &    64.99$\pm$13.83 &    10.77 &     1.86 &     9.94 & J11273289+3604168 \\
lk-411   & 164.21436 & 59.534958 & 0.0564417 &   271.46$\pm$9.06 &   394.70$\pm$15.80 &   233.74$\pm$20.27 &   136.18$\pm$25.48 & $<$104.61  &    11.10 &     3.99 &    10.19 & J11273467+3603470 \\
egs-005  & 213.76653 & 52.056580 & 0.0730016 &   140.23$\pm$6.04 &   237.95$\pm$7.26 &   124.57$\pm$9.91 &  $<$54.80  &  $<$64.39  &    10.76 &     3.82 &     9.98 & J11375801+4728143 \\
egs-064  & 215.64249 & 53.585861 & 0.0391728 &  $<$23.97  &  $<$26.64  &    65.03$\pm$14.33 &    67.53$\pm$14.67 &  $<$56.35  &    10.30 &   $<$0.20 &     9.25 & J11440433+3332339 \\
lk-143   & 161.00293 & 58.760281 & 0.0727644 &   409.63$\pm$7.16 &   501.07$\pm$9.47 &   266.59$\pm$9.86 &   128.54$\pm$8.99 &  $<$54.95  &    10.84 &     9.91 &    10.25 & J11484370+3547002 \\
lk-054   & 160.78687 & 56.820522 & 0.0669273 &    52.30$\pm$6.56 &    79.20$\pm$10.29 &    42.51$\pm$9.46 &  $<$41.99  &  $<$61.66  &    11.25 &     1.25 &     9.52 & J11484525+3547092 \\
lk-257   & 162.54466 & 56.727203 & 0.0480569 &    29.78$\pm$6.97 &    82.68$\pm$9.78 &  $<$52.06  &  $<$56.41  &  $<$38.82  &    10.76 &     0.37 &   $<$9.33 & J11501333+3746107 \\
lk-196   & 161.76175 & 56.254833 & 0.0460456 &    47.67$\pm$8.59 &  $<$48.78  &  $<$46.06  &  $<$40.77  &  $<$72.34  &    10.87 &     0.52 &   $<$9.25 & J11501399+3746306 \\
en1-024  & 242.69853 & 53.422432 & 0.0630939 &    63.42$\pm$14.58 &   116.81$\pm$26.63 &  $<$51.28  &  $<$49.94  &  $<$41.89  &    10.78 &     1.32 &   $<$9.54 & J11505764+1444200 \\
lk-396   & 164.28052 & 59.031815 & 0.0738622 &    81.36$\pm$8.56 &   143.36$\pm$10.04 &   122.00$\pm$11.69 &    82.89$\pm$14.02 &  $<$77.19  &    10.87 &     3.08 &    10.57 & J11542299+4932509 \\
lk-267   & 161.13715 & 56.678345 & 0.0671536 &   493.37$\pm$7.34 &   454.31$\pm$10.58 &   224.42$\pm$9.07 &    97.70$\pm$14.05 &  $<$49.95  &    10.79 &    12.11 &     9.99 & J12020424+5342317 \\
lk-306   & 162.71820 & 57.585117 & 0.0268691 &   188.75$\pm$6.88 &   194.89$\pm$9.89 &   116.89$\pm$12.05 &    69.83$\pm$9.63 &  $<$53.31  &    10.29 &     0.69 &     9.21 & J12054066+0135365 \\
egs-050  & 216.14351 & 53.639969 & 0.0305066 &   556.26$\pm$5.28 &   635.53$\pm$7.29 &   289.99$\pm$14.49 &   133.67$\pm$13.69 &  $<$67.75  &    10.48 &     1.35 &     9.48 & J12115507+4039182 \\
egs-066  & 215.92018 & 53.915695 & 0.0337332 &   158.61$\pm$6.09 &   297.18$\pm$8.51 &   184.69$\pm$16.18 &   113.68$\pm$16.65 &  $<$56.01  &    10.46 &     0.61 &     9.73 & J12115648+4039184 \\
lk-007   & 162.01831 & 59.344875 & 0.0285956 &    35.20$\pm$7.66 &  $<$38.28  &  $<$73.25  & $<$112.69  &  $<$97.98  &    10.18 &     0.15 &   $<$9.06 & J12191866+1201054 \\
lk-334   & 160.10893 & 57.439854 & 0.0471860 &   156.15$\pm$6.90 &   200.26$\pm$9.94 &   118.32$\pm$13.26 &    64.13$\pm$12.23 &  $<$51.23  &    10.78 &     1.84 &     9.73 & J12433887+4405399 \\
lk-406   & 159.30702 & 58.571445 & 0.0710580 &  $<$29.39  &    72.43$\pm$8.91 &  $<$77.49  &  $<$70.40  &  $<$64.35  &    10.97 &   $<$0.82 &   $<$9.78 & J12525011+4645272 \\
lk-079   & 158.74356 & 56.912376 & 0.0467561 &    50.86$\pm$8.57 &    83.89$\pm$9.61 &  $<$86.29  &  $<$74.66  & $<$110.59  &    10.41 &     0.58 &   $<$9.49 & J13011662+4803366 \\
en1-015  & 241.90225 & 53.958252 & 0.0297046 &   500.88$\pm$10.97 &   563.42$\pm$14.85 &   262.97$\pm$19.24 &   107.80$\pm$24.03 &  $<$89.74  &    10.39 &     1.33 &     9.37 & J13011835+4803304 \\
lk-242   & 160.95529 & 56.281364 & 0.0240286 &   162.87$\pm$6.55 &   107.87$\pm$10.17 &    71.16$\pm$13.78 &  $<$51.34  &  $<$57.17  &     9.71 &     0.46 &     8.92 & J13082737+0422125 \\
lk-015   & 161.59250 & 59.595676 & 0.0309583 &  $<$26.33  &    87.37$\pm$9.83 &  $<$63.95  &  $<$54.83  &  $<$54.81  &    10.28 &   $<$0.13 &   $<$9.07 & J13082964+0422045 \\
btsh-094 & 216.88402 & 34.510860 & 0.0684431 &    56.61$\pm$7.59 &   144.69$\pm$10.43 &   118.59$\pm$13.21 &    78.80$\pm$19.06 &  $<$68.62  &    10.83 &     1.41 &     9.91 & J13131470+3910382 \\
en1-003  & 240.67156 & 54.337185 & 0.0653331 &  $<$42.65  &  $<$58.05  &  $<$53.20  &  $<$58.65  &  $<$74.50  &    10.69 &   $<$0.98 &   $<$9.58 & J13151386+4424264 \\
btsh-049 & 218.32701 & 34.734619 & 0.0343026 &   160.07$\pm$6.58 &   259.88$\pm$8.81 &   142.83$\pm$9.14 &    70.35$\pm$11.08 &  $<$55.22  &    10.97 &     0.62 &     9.46 & J13151726+4424255 \\
lk-035   & 159.37373 & 60.002872 & 0.0281504 &   657.91$\pm$30.36 &  1118.77$\pm$39.53 &   552.61$\pm$31.50 &   261.23$\pm$35.17 & $<$179.53  &    10.71 &     1.66 &     9.89 & J13153076+6207447 \\
egs-040  & 214.24640 & 52.730427 & 0.0742157 &  $<$18.28  &  $<$24.81  &  $<$43.90  &  $<$60.95  &  $<$57.34  &    10.84 &   $<$0.57 &   $<$9.61 & J13153506+6207287 \\
lk-332   & 159.63516 & 57.400383 & 0.0470770 &   132.21$\pm$7.70 &   101.67$\pm$9.57 &    85.55$\pm$13.16 &  $<$65.46  &  $<$53.79  &    10.67 &     1.45 &     9.49 & J13325525-0301347 \\
lk-279   & 163.63444 & 57.159027 & 0.0676157 &   170.72$\pm$7.05 &   132.56$\pm$10.32 &    69.30$\pm$14.74 &  $<$69.46  &  $<$64.70  &    10.81 &     3.92 &     9.70 & J13325655-0301395 \\
lk-127   & 161.54971 & 58.449306 & 0.0496845 &    81.77$\pm$6.99 &   126.44$\pm$9.94 &    96.08$\pm$8.88 &    63.09$\pm$9.53 &  $<$59.96  &    10.72 &     1.19 &    10.02 & J13462001-0325407 \\
btsh-063 & 217.60631 & 35.321106 & 0.0104899 &  7971.50$\pm$34.10 & 10153.50$\pm$49.49 &  4483.15$\pm$51.59 &  2000.98$\pm$40.83 &   778.11$\pm$50.94 &    10.69 &     2.69 &     9.80 & J14003661-0254327 \\
lk-078   & 158.71724 & 56.837643 & 0.0438288 &   112.75$\pm$7.45 &   134.45$\pm$9.45 &   104.62$\pm$19.58 &  $<$72.55  &  $<$73.73  &    10.62 &     1.07 &     9.51 & J14003796-0254227 \\
btsh-123 & 217.08795 & 35.405602 & 0.0292952 &   271.58$\pm$13.39 &   417.54$\pm$12.38 &   285.34$\pm$16.88 &   143.71$\pm$17.97 &  $<$84.96  &    10.73 &     1.06 &     9.76 & J14005783+4251203 \\
lk-338   & 160.79364 & 57.647472 & 0.0467886 &   161.47$\pm$7.16 &   237.07$\pm$9.62 &   165.98$\pm$10.06 &    66.66$\pm$13.05 &  $<$90.31  &    10.57 &     1.46 &     9.79 & J14005879+4250427 \\
lk-130   & 163.01991 & 58.599899 & 0.0317531 &    54.19$\pm$6.54 &    54.17$\pm$8.89 &    85.58$\pm$9.70 &    64.20$\pm$12.10 &  $<$50.77  &    10.32 &     0.34 &     9.78 & J14055079+6542598 \\
lk-409   & 161.11449 & 58.903076 & 0.0312686 &   185.49$\pm$7.64 &   266.49$\pm$10.36 &   177.81$\pm$9.91 &    93.54$\pm$14.95 &  $<$55.48  &    10.15 &     0.92 &     9.65 & J14062157+5043303 \\
lk-170   & 161.25214 & 59.736507 & 0.0443880 &    74.04$\pm$6.60 &   130.72$\pm$8.45 &    93.14$\pm$9.62 &  $<$60.19  &  $<$55.10  &    10.91 &     0.74 &     9.48 & J14070703-0234513 \\
lk-119   & 159.54161 & 58.017307 & 0.0237380 &    77.94$\pm$6.04 &   153.66$\pm$10.70 &    97.03$\pm$11.75 &    76.17$\pm$17.04 &  $<$68.58  &    10.05 &     0.15 &     9.25 & J14234238+3400324 \\
btsh-004 & 217.07809 & 34.284489 & 0.0333809 &   178.35$\pm$7.68 &   213.48$\pm$9.38 &   134.03$\pm$7.29 &    72.16$\pm$9.97 &  $<$54.91  &    10.06 &     1.15 &     9.51 & J14234632+3401012 \\
lk-270   & 161.44302 & 56.836800 & 0.0726175 &  $<$27.74  &  $<$40.26  &  $<$46.23  &  $<$57.17  &  $<$65.56  &    10.82 &   $<$0.81 &   $<$9.62 & J14245831-0303597 \\
lk-140   & 159.77400 & 58.361572 & 0.0733310 &    60.41$\pm$7.80 &   186.05$\pm$11.47 &   140.70$\pm$10.95 &    72.19$\pm$11.29 &  $<$53.95  &    11.07 &     1.67 &    10.37 & J14245913-0304012 \\
lk-321   & 158.19514 & 57.005798 & 0.0435024 &    95.40$\pm$7.95 &   185.53$\pm$10.26 & $<$163.27  & $<$116.77  & $<$120.75  &    10.39 &     0.90 &   $<$9.67 & J14250739+0313560 \\
lk-383   & 160.20712 & 58.256981 & 0.0744073 &  $<$30.14  &  $<$39.98  &  $<$57.00  &  $<$72.24  &  $<$58.32  &    10.87 &   $<$0.92 &   $<$9.71 & J14294766+3534275 \\
lk-221   & 164.12077 & 56.910526 & 0.0469288 &    33.46$\pm$7.51 &   129.71$\pm$9.38 &    95.80$\pm$10.13 &  $<$53.62  &  $<$93.02  &    10.57 &     0.39 &     9.53 & J14295031+3534122 \\
lk-245   & 161.70120 & 56.567902 & 0.0449944 &    49.80$\pm$7.16 &    82.05$\pm$10.14 &    65.79$\pm$9.69 &  $<$53.06  &  $<$48.52  &    10.85 &     0.52 &     9.36 & J14334683+4004512 \\
lk-115   & 161.89583 & 58.101940 & 0.0742169 &  $<$33.16  &  $<$41.34  &  $<$74.73  &  $<$70.31  &  $<$57.00  &    10.86 &   $<$1.01 &   $<$9.81 & J14334840+4005392 \\
lk-055   & 161.28499 & 56.976227 & 0.0722381 &   175.59$\pm$6.84 &   211.69$\pm$9.60 &   114.39$\pm$7.90 &  $<$50.18  &  $<$83.80  &    10.71 &     4.63 &     9.94 & J14442055+1207429 \\
egs-006  & 213.82188 & 51.831017 & 0.0745795 &    46.67$\pm$5.32 &    54.83$\pm$7.23 &    72.37$\pm$10.70 &  $<$94.89  &  $<$61.91  &    11.06 &     1.41 &     9.80 & J14442079+1207552 \\
btsh-110 & 217.33472 & 34.753334 & 0.0688272 &    63.96$\pm$6.98 &    78.00$\pm$9.44 &    79.16$\pm$8.47 &  $<$45.69  &  $<$64.54  &    10.68 &     1.60 &     9.77 & J15002500+4317131 \\
lk-156   & 158.50607 & 58.807034 & 0.0739712 &  $<$31.01  &  $<$41.82  &  $<$49.94  &  $<$48.82  &  $<$72.75  &    10.94 &   $<$0.94 &   $<$9.66 & J15053137+3427534 \\
lk-392   & 161.18837 & 58.454979 & 0.0313002 &   711.58$\pm$15.32 &  1074.62$\pm$19.74 &   690.00$\pm$22.86 &   386.12$\pm$25.36 &   193.31$\pm$24.17 &    10.84 &     3.94 &    10.37 & J15064391+0346364 \\
btsh-200 & 217.19626 & 33.387711 & 0.0424442 &   817.96$\pm$19.21 &   767.78$\pm$21.78 &   376.05$\pm$9.11 &   159.56$\pm$12.17 &    46.20$\pm$10.36 &    10.77 &     5.92 &     9.78 & J15064579+0346214 \\
btsh-102 & 218.83510 & 33.846039 & 0.0581498 &   168.61$\pm$6.19 &   209.07$\pm$8.98 &   127.37$\pm$14.89 &    67.08$\pm$16.65 &    54.19$\pm$13.48 &    10.68 &     2.82 &     9.80 & J15101587+5810425 \\
lk-365   & 162.09343 & 58.195721 & 0.0465291 &    37.22$\pm$7.18 &   138.01$\pm$9.67 &    89.73$\pm$15.62 &    48.23$\pm$10.88 &  $<$58.27  &    10.49 &     0.45 &     9.90 & J15101776+5810375 \\
\enddata
\end{deluxetable*}

\begin{deluxetable*}{ccccccccccccc}
\label{tbl:hkpair_table4}
\tabletypesize{\scriptsize}
\setlength{\tabcolsep}{0.005in} 
\tablenum{4}
\tablewidth{0.999\linewidth}
\tablecaption{(Continued)}
\tablehead{
\ch{(1)}   &\ch{(2)}    &\ch{(3)}    &\ch{(4)}&\ch{(5)}&\ch{(6)} &\ch{(7)}
             &\ch{(8)}         &\ch{(9)}  &\ch{(10)}  &\ch{(11)}  &\ch{(12)} &\ch{(13)}       \\
\ch{Galaxy ID} & \ch{RA} & \ch{Dec} & \ch{z} & \ch{F100$\mu$m}  & \ch{F160$\mu$m} &
\ch{F250$\mu$m} & \ch{F350$\mu$m} & \ch{F500$\mu$m} &
\ch{logM$_{star}$} & \ch{SFR} & \ch{log$M_{\rm gas}$} & \ch{Galaxy ID} \\
\ch{(Control)} & \ch{(J2000)} & \ch{(J2000)} & \ch{redshift} &
\ch{(mJy)} & \ch{(mJy)} & \ch{(mJy)} & \ch{(mJy)} & \ch{(mJy)} &
\ch{(logM$_{\odot}$)} & \ch{(M$_{\odot}$/yr)} & \ch{(logM$_{\odot}$)} & \ch{(Pair)}
}
\startdata
egs-044  & 214.52928 & 52.697277 & 0.0659781 &    61.25$\pm$5.59 &    95.40$\pm$8.69 &  $<$47.40  &  $<$50.54  &  $<$63.38  &    10.73 &     1.40 &   $<$9.55 & J15144544+0403587 \\
lk-169   & 159.02901 & 59.379044 & 0.0732209 &   104.47$\pm$7.13 &   168.32$\pm$8.29 &    80.07$\pm$13.21 &  $<$51.76  &  $<$60.20  &    10.67 &     2.91 &     9.82 & J15144697+0403576 \\
btsh-050 & 218.26311 & 34.744881 & 0.0296272 &    75.02$\pm$8.03 &   187.08$\pm$10.74 &   190.83$\pm$7.02 &   113.02$\pm$14.13 &    54.16$\pm$11.13 &    10.14 &     0.40 &     9.97 & J15233768+3749030 \\
lk-091   & 163.20827 & 57.747452 & 0.0730649 &   175.22$\pm$7.75 &   151.40$\pm$9.59 &    77.31$\pm$9.95 &    51.25$\pm$11.66 &  $<$36.18  &    10.70 &     5.03 &     9.61 & J15264774+5915464 \\
lk-106   & 159.34660 & 57.520683 & 0.0718920 &   397.72$\pm$7.14 &   376.18$\pm$10.43 &   168.52$\pm$16.53 &  $<$64.12  &  $<$61.08  &    10.87 &     9.96 &    10.07 & J15281276+4255474 \\
lk-164   & 163.10425 & 59.685532 & 0.0277227 &  1146.73$\pm$12.16 &  1605.01$\pm$17.47 &   875.35$\pm$19.63 &   416.53$\pm$35.14 &   132.22$\pm$26.56 &    10.74 &     2.78 &     9.99 & J15281667+4256384 \\
lk-269   & 161.45438 & 56.701805 & 0.0672907 &    57.40$\pm$6.92 &   103.00$\pm$8.95 &    48.54$\pm$10.61 &  $<$54.29  &  $<$58.99  &    10.84 &     1.38 &     9.57 & J15523393+4620237 \\
lk-146   & 162.65512 & 59.095657 & 0.0323721 &   290.86$\pm$7.64 &   259.00$\pm$10.45 &   145.66$\pm$13.88 &  $<$64.38  &  $<$61.28  &    10.22 &     1.43 &     9.40 & J15562191+4757172 \\
lk-040   & 162.04411 & 56.704330 & 0.0477595 &    74.63$\pm$6.81 &   125.26$\pm$10.66 &    74.89$\pm$7.77 &  $<$37.45  &  $<$47.58  &    10.48 &     0.87 &     9.46 & J15583749+3227379 \\
lk-028   & 161.42595 & 60.067890 & 0.0719979 &    53.20$\pm$7.18 &   103.07$\pm$11.01 &    71.96$\pm$10.60 &  $<$53.86  &  $<$68.80  &    10.79 &     1.48 &     9.77 & J15583784+3227471 \\
egs-026  & 213.78456 & 51.613804 & 0.0742349 &   244.41$\pm$10.83 &   287.85$\pm$12.13 &   127.77$\pm$14.48 &  $<$71.02  &  $<$62.44  &    10.75 &     6.71 &    10.00 & J16024254+4111499 \\
lk-224   & 165.37244 & 57.117775 & 0.0451835 &   303.19$\pm$7.74 &   467.53$\pm$12.26 &   229.45$\pm$9.54 &   108.68$\pm$9.88 &  $<$60.84  &    10.58 &     1.97 &     9.84 & J16024475+4111589 \\
lk-366   & 162.32683 & 58.345104 & 0.0280867 &    78.43$\pm$8.30 &   237.33$\pm$9.72 &   215.25$\pm$10.28 &   112.63$\pm$13.33 &  $<$87.83  &    10.81 &     0.32 &     9.87 & J16080559+2529091 \\
lk-281   & 164.13226 & 57.347748 & 0.0675492 &    51.07$\pm$7.74 &   127.73$\pm$9.47 &   120.37$\pm$12.37 &    65.86$\pm$13.76 &  $<$75.45  &    11.11 &     1.06 &    10.31 & J16080648+2529066 \\
btsh-028 & 218.14639 & 34.364929 & 0.0422711 &    30.39$\pm$6.33 &    68.86$\pm$8.07 &    57.41$\pm$9.14 &  $<$53.69  &  $<$70.53  &    10.47 &     0.29 &     9.27 & J16082261+2328459 \\
lk-345   & 162.32899 & 57.821774 & 0.0702557 &   128.20$\pm$7.22 &   176.90$\pm$9.49 &    66.96$\pm$13.36 &  $<$68.14  &  $<$53.16  &    10.57 &     3.23 &     9.72 & J16082354+2328240 \\
lk-387   & 160.41603 & 58.317230 & 0.0722775 &    65.57$\pm$6.61 &    96.26$\pm$10.94 &    84.75$\pm$12.78 &  $<$63.62  &  $<$61.34  &    11.10 &     1.82 &     9.83 & J16145418+3711064 \\
lk-243   & 161.15924 & 56.369724 & 0.0240542 &  1293.33$\pm$11.99 &  1304.87$\pm$16.83 &   588.23$\pm$21.66 &   286.68$\pm$21.50 & $<$135.42  &    10.81 &     3.66 &     9.63 & J16282497+4110064 \\
lk-139   & 159.74431 & 58.376003 & 0.0679610 &   172.97$\pm$7.39 &   165.39$\pm$9.12 &    84.48$\pm$10.98 &    62.33$\pm$12.82 &  $<$97.77  &    10.74 &     4.14 &     9.73 & J16282756+4109395 \\
lk-278   & 163.33688 & 57.242729 & 0.0802541 &    75.46$\pm$6.70 &    91.48$\pm$8.92 &    53.06$\pm$13.16 &  $<$71.33  &  $<$54.35  &    11.19 &     2.60 &     9.75 & J16354293+2630494 \\
egs-038  & 213.91566 & 52.543880 & 0.0737351 &    61.64$\pm$5.41 &   143.09$\pm$7.66 &   164.18$\pm$11.88 &    66.53$\pm$14.99 &  $<$48.53  &    11.29 &     2.04 &    10.52 & J16372583+4650161 \\
lk-227   & 162.26764 & 56.223995 & 0.0719589 &   549.85$\pm$8.26 &   538.53$\pm$11.64 &   269.88$\pm$10.05 &   115.67$\pm$12.98 &    61.87$\pm$10.89 &    10.90 &    15.46 &    10.19 & J16372754+4650054 \\
en1-041  & 241.28255 & 54.887733 & 0.0632211 &  $<$27.27  &  $<$45.90  &    58.78$\pm$14.37 &  $<$54.11  & $<$113.47  &    10.60 &   $<$0.60 &     9.59 & J17020378+1859495 \\
lk-122   & 160.10104 & 58.155136 & 0.0720025 &   174.47$\pm$6.81 &   321.35$\pm$10.52 &   175.95$\pm$9.86 &    73.90$\pm$15.11 &  $<$55.78  &    10.72 &     3.16 &    10.17 & J17045089+3448530 \\
lk-376   & 159.78526 & 58.231022 & 0.0729322 &    57.36$\pm$7.32 &  $<$45.37  &    83.03$\pm$8.45 &  $<$68.12  &  $<$66.86  &    10.90 &     1.63 &     9.83 & J17045097+3449020 \\
lk-240   & 160.43297 & 56.250572 & 0.0704346 &    58.85$\pm$6.97 &    80.12$\pm$9.71 &    64.40$\pm$11.86 &  $<$43.86  &  $<$55.05  &    10.98 &     1.55 &     9.71 & J20471908+0019150 \\
\enddata
\tablecomments{Descriptions of Columns: (1) Control Galaxy ID. (2) Control Galaxy RA (deg, J2000).
(3) Control Galaxy Dec (deg, J2000). (4) Control Galaxy redshift z taken from SDSS. (5) Herschel
PACS 100$\mu$m flux (mJy). (6) Herschel PACS 160$\mu$m flux (mJy). (7) Herschel SPIRE 250$\mu$m
flux (mJy). (8) Herschel SPIRE 350$\mu$m flux (mJy). (9) Herschel SPIRE 500$\mu$m flux (mJy).
(10) Stellar mass (log(M$_{\odot}$)). (11) Star formation rate (M$_{\odot}$/yr). (12) Total gas mass
(log(M$_{\odot}$)). (13) ID of matched H-KPAIR galaxy.}
\end{deluxetable*}
\end{turnpage}}

\section{Specific Star Formation Rate Enhancement}

Figure~\ref{fig:m_ssfr_scatter} is a scatter plot for the
relation between M$_{star}$ and the sSFR for spirals in H-KPAIR and in
the control sample.  The maximum likelihood Kaplan-Meier (K-M)
estimator (\citet{KM58}, see also \citet{Xu1998}), which exploits
information in the upper-limits, was used to calculate the means of the
sSFR of different populations. For spirals in H-KPAIR the mean is $\rm
log(sSFR/yr^{-1}) = -10.66 \pm 0.06$, and for spirals in the control
sample it is nearly identical: $\rm log(sSFR/yr^{-1}) = -10.65 \pm
0.04$. However, detailed inspections of
Figure~\ref{fig:m_ssfr_scatter} reveal a mismatch between the H-KPAIR
and the control sample: while none of the control galaxies has a value
or upper-limit of $\rm log(sSFR/yr^{-1})$ below $-11.3$, many H-KPAIR
spirals have their $\rm log(sSFR/yr^{-1})$ values/upper-limits below
this threshold. This is due to the fact that control galaxies are more
distant (median $z = 0.056$) than H-KPAIR galaxies (median $z =
0.040$), and their Herschel observations in the HerMES Level-5 survey
\citep{Oliver2012} are shallower than our H-KPAIR observations.
Therefore their upper-limits are in general significantly higher than
those of H-KPAIR galaxies, and none of them was detected below $\rm
log(sSFR/yr^{-1}) = -11.3$. This mismatch can bias the mean of the
control sample toward higher value even though the K-M estimator is
used, for the algorithm requires that the detections cover the entire
data range and the detections and upper-limits are well mixed
\citep{Feigelson1985}. In order to confirm that this is indeed the
case, the following test was carried out: a hybrid sample was
constructed by replacing in the control sample those galaxies matching
to the low sSFR H-KPAIR galaxies ($\rm log(sSFR/yr^{-1})$
values/upper-limits below -11.3) by the H-KPAIR galaxies
themselves. Using again the K-M estimator, we found a mean $\rm
log(sSFR/yr^{-1}) = -10.94\pm 0.05$ for such a sample, significantly
lower than the result for the control sample.  The implicit assumption
in this test is that below $\rm log(sSFR/yr^{-1}) = -11.3$ the control
and H-KPAIR samples have nearly identical sSFR distributions, which is
reasonable if the red spirals with low sSFR are similar to early type
galaxies that show no interaction induced star formation enhancement
\citep{Sulentic1989}.

In Figure~\ref{fig:m_ssfr_scatter} we plotted the lines representing
the blue sequence and the red sequence, taken from Figure~7 of
\citet{Schiminovich2007}, where it also shows that galaxies with
$\rm log(sSFR/yr^{-1}) < -11.3$ belong to the red sequence.

\begin{figure}[!htb]
\includegraphics[width=0.35\textwidth,angle=90]{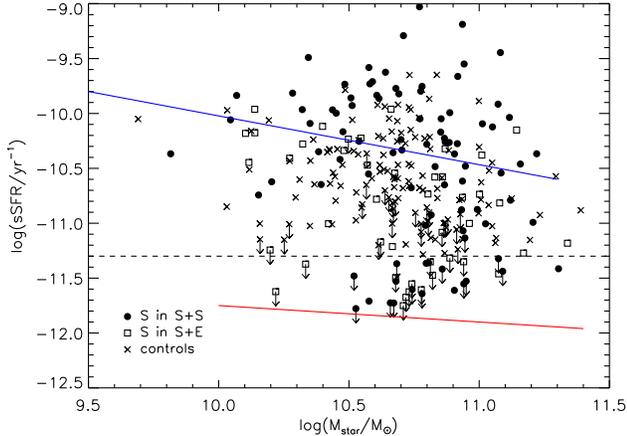}
\caption{Stellar mass (M$_{star}$) vs. sSFR scatter plot, for S in S$+$S (filled
circles) $\&$ S$+$E (open squares) pairs, and control samples (crosses). The blue
and red lines represent the blue and red sequences, taken from \citet{Schiminovich2007}.}
\label{fig:m_ssfr_scatter}
\setcounter{figure}{7}
\end{figure}

In order to minimize the bias due to the mismatch, we confined the
star formation enhancement analysis to H-KPAIR galaxies with
values/upper-limits of $\rm log(sSFR/yr^{-1}) \geq -11.3$ and the
control galaxies matched to them. Among these 101 paired SFG galaxies,
the sSFR of 5 galaxies are upper-limits; for their matches in the
control sample, 19 have the sSFR in upper-limits. It is possible that
the true values of some of the upper-limits are below the $\rm
log(sSFR)$ cutoff. However, they are only small fractions of both
samples. Furthermore, because in the K-S estimator the probability
distribution of the true values of the upper-limits are estimated using
data points included in the analysis (none of them is below the
cutoff), the possible contamination by low sSFR red spirals should not
affect significantly the mean sSFR of SFGs in both H-KPAIR and control
samples.

We found a mean of $\rm log(sSFR/yr^{-1}) = -10.34\pm 0.05$
for SFGs in the H-KPAIR, and $\rm log(sSFR/yr^{-1}) = -10.67\pm 0.04$
for their matches in the control sample. The corresponding sSFR
enhancement \citep{Xu2010} is $\rm enh = <log(sSFR)>_{pair} -
<log(sSFR)>_{cont} = 0.33\pm 0.06$.  This indicates that the sSFR
of paired SFGs is enhanced on average
by a factor of $\sim 2$ compared to control
galaxies, and the enhancement is significant at $> 5\sigma$ level.
When paired SFGs are divided into subsamples of galaxies in S$+$S
pairs and in S$+$E pairs, we found a mean of $\rm log(sSFR/yr^{-1}) =
-10.21\pm 0.06$ for the S$+$S subsample and $\rm log(sSFR/yr^{-1}) =
-10.64\pm 0.07$ for the S$+$E subsample.  This shows that SFGs in
S$+$S pairs have strong sSFR enhancement (with $\rm enh = 0.46 \pm
0.08$) and those in S$+$E have no significant sSFR enhancement. The
result is in very good agreement with that of \citep{Xu2010}
based on Spitzer observations of a smaller subsample of KPAIR.
We further divided the S$+$S and S$+$E subsamples by the interaction
morphology (INT/MER vs JUS), and found that the mean values
for SFGs in S$+$S pairs of INT/MER types and of JUS type are
$\rm log(sSFR/yr^{-1}) = -10.14\pm 0.07$ and
$\rm log(sSFR/yr^{-1}) = -10.31\pm 0.08$, respectively.
This suggests that SFGs in S$+$S pairs of INT and MER types are slightly more
enhanced (by 0.17 dex) than those in JUS pairs, though the significance
of the difference is only at $1.5\sigma$ level.
No significant sSFR enhancement is found
for SFGs in S$+$E pairs of either the INT/MER types
(mean $\rm log(sSFR/yr^{-1}) = -10.58\pm 0.11$) or the
JUS type (mean $\rm log(sSFR/yr^{-1}) = -10.71\pm 0.11$).

In Figure~\ref{fig:plot_mass-ssfr-statis_sfg} we compared the sSFR of SFGs
in S$+$S pairs and S$+$E pairs with the controls in four $M_{\rm star}$ bins
(log(M$_{star}$/M$_{\odot}$) $<$ 10.4, 10.4 $\leq$ log(M$_{star}$/M$_{\odot}$)
$<$ 10.7, and 10.7 $\leq$ log(M$_{star}$/M$_{\odot}$). For SFGs in S$+$S pairs
we found a similar amount of sSFR enhancement in different $M_{\rm dust}$ bins.
For SFGs in S$+$E pairs no significant sSFR enhancement is found in any
$M_{\rm dust}$ bin. It is worth noting that \citet{Xu2010} found that SFGs
in S$+$S pairs in their lowest $M_{\rm star}$ bin (corresponding
to $\rm 9.3<log(M_{star}/M_{\odot}) < 9.8$ for the $M_{\rm star}$ calibration used
in this paper) do not show any significant sSFR enhancement.
These low mass galaxies are not included in H-KPAIR because
of the exclusion of pairs of $\rm v < 2000\; km/s$.

\begin{figure}[!htb]
\includegraphics[width=0.35\textwidth,angle=90]{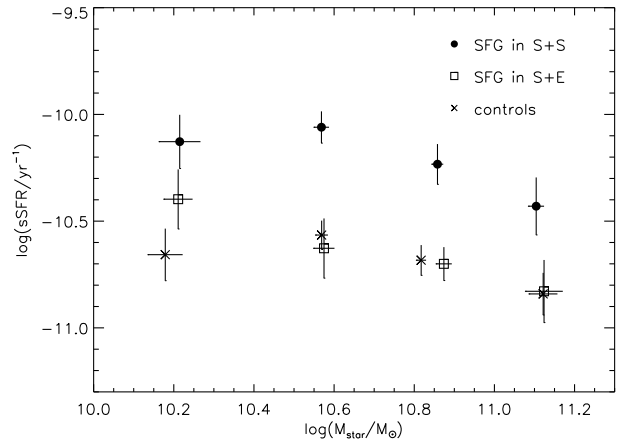}
\caption{Plot of the mean values of sSFRs in four stellar mass (M$_{star}$)
bins: for SFGs in S$+$S
(black filled circles) $\&$ S$+$E (open squares) pairs, and control samples (crosses).
}
\label{fig:plot_mass-ssfr-statis_sfg}
\setcounter{figure}{8}
\end{figure}

\section{Star Formation Efficiency Enhancement}

The total gas mass ($M_{\rm gas}$) can be estimated from $M_{\rm dust}$
according \citet{James2002}: $M_{\rm gas}/M_{\rm dust} = 1/(Z\times
e\times f) = 114/Z$, where Z is the metallicity in solar units, e $=$
0.456 the fraction of metals incorporated in dust in the ISM, and f
$=$ 0.019 the metal mass fraction in gas of solar
metallicity. \citet{Draine2007} found a very similar gas-to-dust ratio
($M_{\rm gas}$/$M_{\rm dust}$ $\sim$ 100) for SINGS galaxies, including
interacting galaxies. Studies on gas-to-dust mass ratios in local
galaxies (e.g., \citealt{Remy-Ruyer2014}) show that for metal rich
galaxies the ratios are uniformally distributed, and only very metal
poor galaxies have significantly higher $M_{\rm gas}$/$M_{\rm dust}$ ratios.
In the local universe, only low mass galaxies have such low metalicities
\citep{Tremonti2004}.  All SFGs in H-KPAIR have $M_{\rm star} >
10^{9.8} M_{\rm \odot}$, and the control galaxies are 1-to-1 matched to
paired galaxies according to $M_{\rm star}$.  Therefore we adopted a
fixed gas-to-dust ratio of 100 for the conversion from $M_{\rm dust}$ to
$M_{\rm gas}$.
Using the K-M estimator, we found that means of $log (M_{\rm gas}/M_{\rm star})$
of SFGs in S$+$S pairs, in S$+$E pairs, and in the control sample
are $-0.93\pm 0.05$, $-1.07\pm 0.07$, and $-1.00\pm 0.04$, respectively.
For both SFGs in S$+$S pairs and in S$+$E pairs, the mean
$log (M_{\rm gas}/M_{\rm star})$ differs from that of control galaxies
only at $\sim 1\sigma$ level. In Figure~\ref{fig:plot_mass-mgas-statis_sfg} the
means of $log (M_{\rm gas}/M_{\rm star})$ of different samples in different
mass bins are compared. No clear trend is found in the plot.

The SFE is defined as the ratio between the SFR and total gas mass:
$SFE = SFR/M_{\rm gas}$. The SFE analysis was confined to SFGs with
$M_{\rm gas}$ detections. The means of $\rm log (SFE/yr^{-1})$ of SFGs
in S$+$S pairs, in S$+$E pairs, and in the control sample are found to
be $-9.26\pm 0.05$, $-9.56\pm 0.05$, and $-9.67\pm 0.05$,
respectively.  Figure~\ref{fig:plot_mass-sfe-statis_sfg}
shows that the means of
$\rm log (SFE)$ of SFGs in S$+$S pairs are consistently above that of
control galaxies at a $\sim $ 0.4 dex level in all mass bins. On the
other hand, the means of $\rm log (SFE)$ of SFGs in S$+$E pairs are
seen both above and below that of control galaxies, consistent with no
significant SFE enhancement.

In Figure~\ref{fig:plot_fgas-sfe-scatter-statis} we plotted the mean
$\rm log(SFE)$ against the gas fraction $f_{\rm gas} =
M_{\rm gas}/(M_{\rm star} + M_{\rm gas})$ for the S$+$S pair and control
samples. Within the range of $f_{\rm gas}$
covered by SFGs in our samples, which is [0,0.25], there is no significant
dependence of the SFE enhancement on $f_{\rm gas}$. Taken at the
face value, this does not confirm the results of theoretical simulations showing
weaker merger-induced star formation enhancement
for high $f_{\rm gas}$ \citep{Hopkins2009a, Perret2014, Scudder2015}.
However, it appears that the
effect of $f_{\rm gas}$ on star formation enhancement
is significant only when $f_{\rm gas} \gsim 0.3$ \citep{Scudder2015},
which is beyond the range covered by our samples.

\begin{figure}[!htb]
\includegraphics[width=0.35\textwidth,angle=90]{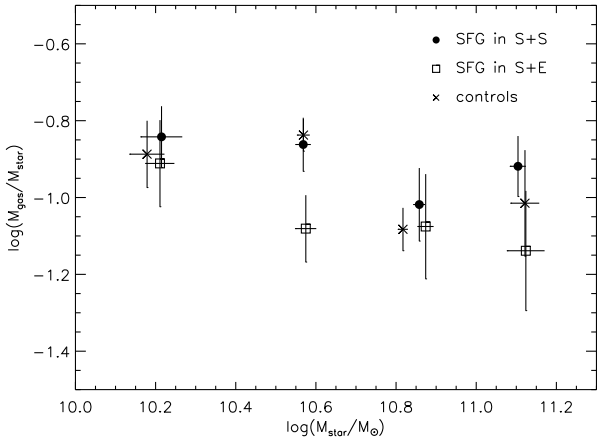}
\caption{Plot of means of $log(M_{\rm gas}/M_{\rm star})$) in four stellar mass
(M$_{star}$) bins for SFGs in S$+$S (filled circles) $\&$ S$+$E (open squares)
pairs, and control samples (crosses).
}
\label{fig:plot_mass-mgas-statis_sfg}
\setcounter{figure}{9}
\end{figure}

\begin{figure}[!htb]
\includegraphics[width=0.35\textwidth,angle=90]{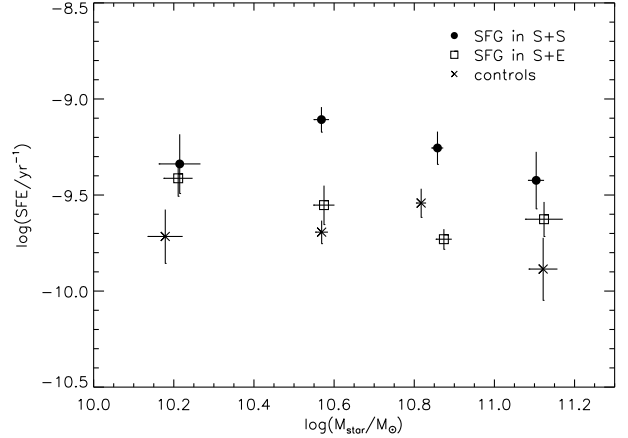}
\caption{Statistical comparison of star formation efficiencies (SFE = SFR/$M_{\rm gas}$)
in four stellar mass (M$_{star}$) bins for SFGs in S$+$S (filled circles) $\&$
S$+$E (open squares) pairs, and control samples (crosses).
}
\label{fig:plot_mass-sfe-statis_sfg}
\setcounter{figure}{10}
\end{figure}

\begin{figure}[!htb]
\includegraphics[width=0.35\textwidth,angle=90]{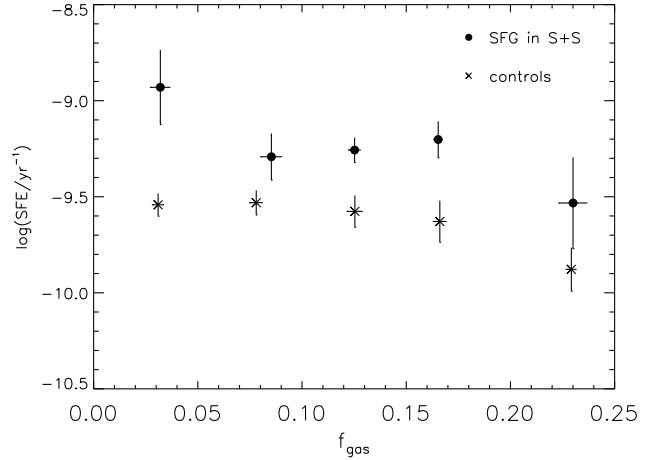}
\caption{Statistical comparison of SFEs in five gas mass fraction (f$_{gas}$ =
$M_{\rm gas}$/($M_{\rm gas}+M_{\rm star}$)) bins, for SFGs in S$+$S pairs (filled circles)
and control samples (crosses).
}
\label{fig:plot_fgas-sfe-scatter-statis}
\setcounter{figure}{11}
\end{figure}

\begin{deluxetable}{lccc}
\label{tbl:sfenhance}
\tabletypesize{\scriptsize}
\setlength{\tabcolsep}{0.05in} 
\tablenum{5}
\tablewidth{0.95\linewidth}
\tablecaption{sSFR and SFE enhancement}
\tablehead{
\ch{(1)}   &\ch{(2)}    &\ch{(3)}    &\ch{4 }\\
\ch{Type} & \ch{sSFR enhancement} & \ch{SFE enhancement} & \ch{$\delta log(M_{\rm gas}/M_{\rm star})$} \\
          & \ch{(dex)} & \ch{(dex)}    & \ch{(dex)}
}
\startdata
S in S$+$S & 0.46$\pm$0.08 & 0.41$\pm$0.07 & 0.07$\pm$0.06  \\
S in S$+$E & 0.03$\pm$0.08 & 0.11$\pm$0.07 & $-0.07\pm$0.08 \\
\enddata
\tablecomments{Mean values of sSFR \& SFE enhancements (dex)
and of the difference in $\log(M_{\rm gas}/M_{\rm star})$,
for SFGs in S$+$S $\&$ S$+$E pairs.}
\end{deluxetable}

\section{Discussion}

\subsection{AGNs in H-KPAIR}
In the H-KPAIR sample, some spiral galaxies are classified as Active Galactic
Nuclei (AGNs) based on their optical spectra. J08364588$+$4722100,
J11251716$+$0226488, J11484525$+$3547092, J13151386$+$4424264,
J13325655$-$0301395, J14234632$+$3401012, J14294766$+$3534275,
J15281276$+$4255474, J20471908$+$0019150 are sub-classified (SUBCLASS)
as AGN or AGN Broadline in SDSS archive (using line ratios).
J09134606$+$4742001 and J13151726$+$4424255 are classified as AGN in
``Quasars and Active Galactic Nuclei'' (13th Ed.) \citep{Veron2010}.
J13462001$-$0325407 is classified as Sy2 in \citet{Veron2010}.

Identification of AGN has been successful through the use of WISE photometry
\citep{Jarrett2011, Stern2012, Mateos2012, Secrest2015}. To apply the corresponding
AGN color criteria of \citet{Mateos2012}, the W1, W2 and W3 magnitudes are taken
from the unWISE forced photometry catalog \citep{Lang2014a, Lang2014b}. The catalog
apertures are chosen from SDSS photometry and are held constant for each identified
object across the three WISE bands. The forced photometry should be sufficient to
measure the magnitude differences W1-W2 and W2-W3 for the placement of the H-KPAIRs
in the WISE color-color diagram. A comparison to manual aperture photometry is
examined in Domingue et al. (2015, in prep.). From the full set of spirals in the
H-KPAIRs, only J13151726+4424255 falls within the color-color diagram area associated
with AGN in \citet{Mateos2012} and therefore this may be our only AGN candidate
in addition to the optically determined AGNs. More details on discussions on WISE
AGNs will be in Domingue et al. (2015, in prep.).

\citet{Lam2013} found that the FIR/sub-mm properties of galaxies with
AGNs are similar to those of star-forming galaxies, and the AGN contribution
to the FIR/sub-mm luminosity is in general insignificant. Even in the MIR
band where the AGN contribution is much stronger, \citet{Nordon2012} found
that only in two out of 18 AGN-hosting galaxies the emission is significantly
enhanced by the AGN. Therefore, different from \citet{Xu2010}, we choose
to keep the AGN candidates in our analyses. Indeed we found that keeping or
removing AGN candidates will not affect the major results of this paper.

\subsection{Why is SFR enhancement absent in S+E pairs?}
The significant sSFR enhancements of SFGs in S$+$S pairs found in
\citet{Xu2010} and in this work are due to a significant SFE enhancement
in them, while they have the same mean gas content as the control
galaxies. On the other hand, the lack of sSFR enhancement in SFGs in
S$+$E pairs can not be attributed to a deficiency of gas content,
as speculated by \citet{Xu2010}, but to a significantly lower SFE than
that of SFGS in S$+$S pairs (Table~5). It is difficult to understand
the SFE difference between SFGs in S$+$S pairs and those in S$+$E pairs,
if the enhancement is triggered by tidal effects (the tidal effects
caused by a spiral companion and by a elliptical companion should
be similar). It has been suggested that cloud-cloud collisions
\citep{Scoville1986, Tan2006} or cloud-squeezing by shock-heated
diffuse gas in two colliding disks \citep{Jog1992, Saitoh2009,
Hayward2014, Renaud2015} can trigger starbursts. Our results indicate
that these mechanisms, associated with the collision between the ISM
in two gas rich galaxies (which is absent in S+E pairs), may indeed
be the major mode for the SFR enhancement in paired SFGs (mostly early
stage mergers). \citet{Hwang2011} also found in the GOODS-Herschel fields,
the SFR and sSFR increase as a late-type galaxy approaches a late-type
neighbor (S$+$S pairs), while the SFR and sSFR decrease or do not
change much as the galaxy approaches an early-type neighbor (S$+$E
pairs). They argued for X-rays from the Ellipticals as the reason for
the suppression of star formations in the spirals in S$+$E pairs.

\subsection{Separation and SFR Enhancement}
In Section 5.7 of \citet{Xu2010}, they discussed how the star formation
enhancement is related to the projected separation between the two galaxies
in pairs. Here we made a similar analysis using the normalized separation
parameter: SEP $=$ s $/$ (r1$+$r2) . Here s is the projected separation,
and r1 and r2 are the K-band Kron radii (taken from 2MASS) of the primary
and the secondary, respectively, in the same units as those of s (arcsec).
Figures~\ref{fig:plot_sep_ssfr} and~\ref{fig:plot_sep_sfe} are plots of
mean log(SFR/M$_{star}$) and log(SFR/M$_{gas}$) vs. log(M$_{star}$) of
star-forming spirals (SFGs) in S$+$S pairs, separated into two subsamples
of SEP < 1 (39 SFGs) and SEP > 1 (30 SFGs). Similar to \citet{Xu2010}, we found for both
subsamples, the log(SFR/M$_{star}$) and log(SFR/M$_{gas}$) vs. log(M$_{star}$)
relations scatter around the means of the total sample without any obvious trend,
and no significant differences between the two subsamples are detected.
This confirms \citet{Xu2010}'s finding and suggests that, for early stage mergers
in H-KPAIR, the separation is not an important parameter any more once the two galaxies
are close enough. This is different from FIR selected late stage mergers for which
\citet{Gao1999} found an anti-correlation between SFE and the pair separation.

\begin{figure}[!htb]
\includegraphics[width=0.35\textwidth,angle=90]{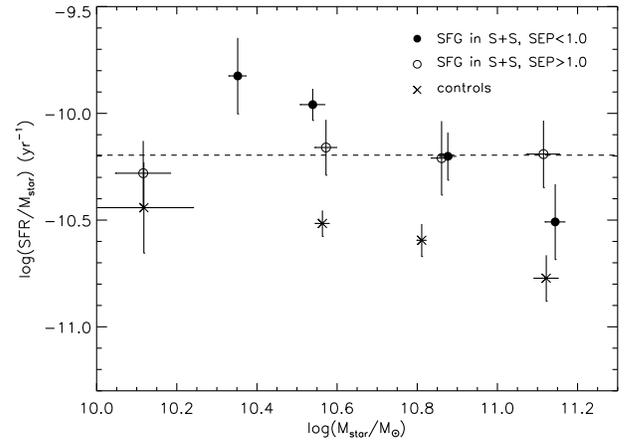}
\caption{Similar to Figure~17 in \citet{Xu2010}, average sSFRs of SFGs in
S$+$S pairs, separated into two subsamples of SEP < 1 (filled dots) and
SEP > 1 (opened circles), respectively. The dotted line marks the mean
log(SFR/M$_{star}$) of all SFGs in the H-KPAIR sample.
}
\label{fig:plot_sep_ssfr}
\setcounter{figure}{12}
\end{figure}

\begin{figure}[!htb]
\includegraphics[width=0.35\textwidth,angle=90]{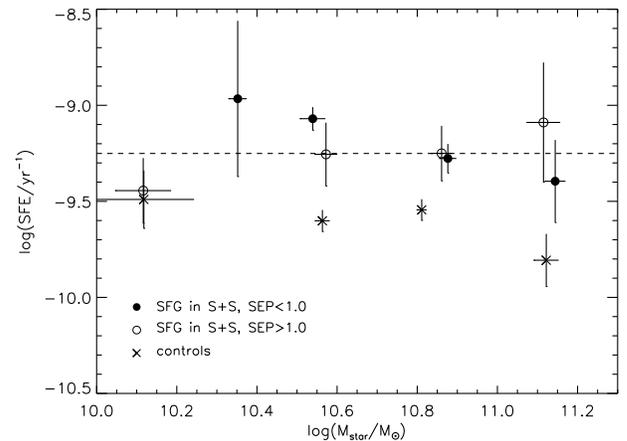}
\caption{Average log(SFE) (SFR/M$_{gas}$) of SFGs in S$+$S pairs, separated
into two subsamples of SEP < 1 (filled dots) and SEP > 1 (opened circles),
respectively. The dotted line marks the mean log(SFR/M$_{gas}$) of all SFGs
in the H-KPAIR sample.
}
\label{fig:plot_sep_sfe}
\setcounter{figure}{13}
\end{figure}

\subsection{"Holmberg effect"}
Figure~\ref{fig:plot_holmberg_ssfr} shows a correlation between the 
sSFR of primary galaxies and that of secondary galaxies in S$+$S
pairs with the Spearman's rank correlation coefficient equal to 0.56,
while for corresponding control galaxies there is no such correlation.
This reconfirms the "Holmberg effect" on the sSFR previously found by
\citet{Xu2010} using Spitzer observations of a small subsample of KPAIR.

\begin{figure}[!htb]
\includegraphics[width=0.45\textwidth,angle=90]{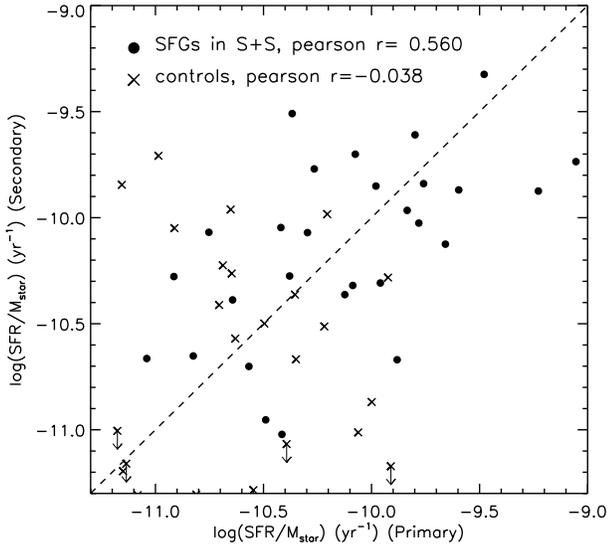}
\caption{Plot of sSFRs of primary $\&$ secondary star-forming spiral galaxies
in S$+$S pairs (filled circles) and control samples (crosses).
}
\label{fig:plot_holmberg_ssfr}
\setcounter{figure}{14}
\end{figure}

\section{Summary $\&$ future plan}

In this paper we present Herschel observations for a large
and complete sample of close major-merger pairs of galaxies
selected from a 2MASS/SDSS-DR5 cross-match. The H-KPAIR sample
includes 176 galaxies (132 spirals and 44
ellipticals) in 88 pairs (44 S$+$S and 44 S$+$E).
Herschel maps, in the three PACS bands at 70,
100 and 160 $\mu m$ and the three SPIRE bands at 250, 350 and 500
$\mu m$, show very diversified FIR-submm emission properties
among H-KPAIR galaxies. The SFR (estimated using the IR luminosity),
the sSFR, the total gas mass $M_{\rm gas}$
(estimated using the SED fitting resulted
dust mass), and the SFE of the spiral galaxies in H-KPAIR are
compared to those of single spirals in a control sample in order
to study the interaction-induced enhancement.
The control sample is selected from 2MASS galaxies with Herschel data
obtained in HerMES Level-5 surveys. In order to minimize a bias due to
different depths of the Herschel data of H-KPAIR and of
the control sample, the enhancement analyses
are confined to SFGs with $\rm log(sSFR/yr^{-1}) > -11.3$.
The following results are found:
\begin{description}
\item{(1)} The mean $\rm log(sSFR)$ of the SFGs in in
  S$+$S pairs is significantly higher than that of the SFGs in the
  control sample, but that of SFGs in the S$+$E pairs is not. The sSFR
  enhancement level for SFGs in S$+$S pairs is $\sim$ 0.5 dex.
  This result is in very good agreement with \citet{Xu2010}.
\item{(2)} SFGs in S$+$S pairs of INT and MER types have enhancement
  of log(sSFR) (by 0.17 dex) above those in JUS pairs, though the
  significance of the difference is only at $1. 5\sigma$ level. No
  significant sSFR enhancement is found for SFGs in S$+$E pairs of
  either the INT/MER types or the JUS type .
\item{(3)} There is no significant difference among
  the means of $log(M_{\rm gas}/M_{\rm star})$ of SFGs in  S$+$S pairs,
  in  S$+$E pairs, and in the control sample.
\item{(4)} The mean $\rm log(SFE)$ of the SFGs in
  S$+$S pairs is higher than that of the SFGs
  in S$+$E pairs and in the control
  sample. This indicates that
  star formation triggered by disc-disc collision may
  play an important role in the
  interaction-induced star formation activity.
\item{(5)} There is no dependence of the SFE enhancement on the gas
fraction, fgas, in the range of fgas of our sample ($0<f_{\rm gas} <0.25$).
\end{description}

In future works we will study the atomic and molecular gas contents in
H-KPAIR galaxies using GBT HI and IRAM-30m CO observations (Lisenfeld et al.,
in preparatio). Using WISE mid-infrared images, we will analyze the PAH
properties and AGN activities of paired galaxies (Domingue et al.,
in preparation). Analysis on the
stellar populations and metallicity gradients in paired galaxies and
detailed studies on SFGs in S$+$E pairs will be carried out by using
optical spectroscopic observations with the 2.16m telescope of National
Astronomical Observatory of China (NAOC), and narrow-band H$_{\alpha}$
imaging observations with the 2.4m telescope of Yunnan Astronomical
Observatory.

\vskip1truecm
\noindent{\it Acknowledgments}:
We thank the anonymous referee for helpful comments and suggestions.
We acknowledge useful discussions with Dr. Yinghe Zhao, and thank
Dr. Yuyen Chang for the help on data tables of 1M galaxies from
SDSS$+$WISE. C.C. and Y.G. are supported by NSFC-11503013, NSFC-11420101002,
NSFC-10978014, NSFC-11173059, NSFC-11390373, and CAS-XDB09000000.
UL acknowledges support by the research projects AYA2011-24728 and
AYA2014-53506-P from the Spanish Ministerio de Econom\'ia y Competividad
and the Junta de Andaluc\'ia (Spain). The Herschel spacecraft was
designed, built, tested, and launched under a contract to ESA managed
by the Herschel/Planck Project team by an industrial consortium under
the overall responsibility of the prime contractor Thales Alenia Space
(Cannes), and including Astrium (Friedrichshafen) responsible for the
payload module and for system testing at spacecraft level, Thales
Alenia Space (Turin) responsible for the service module, and Astrium
(Toulouse) responsible for the telescope, with in excess of a hundred
subcontractors. HCSS / HSpot / HIPE is a joint development (are joint
developments) by the Herschel Science Ground Segment Consortium,
consisting of ESA, the NASA Herschel Science Center, and the HIFI,
PACS and SPIRE consortia.  This research has made use of the NASA/IPAC
Extragalactic Database (NED), which is operated by the Jet Propulsion
Laboratory, California Institute of Technology, under contract with
the National Aeronautics and Space Administration.

\bibliographystyle{hapj}
\bibliography{papers_biblio}

\end{document}